\documentclass[3p,amsmath,amssymb,compress]{elsarticle}
\usepackage{amsmath}
\usepackage{graphicx}
\usepackage{latexsym}
\usepackage{amsmath,amssymb}
\usepackage{bm} 
\usepackage{xcolor}
\usepackage{stackrel}
\usepackage{subcaption}
\usepackage{accents}
\usepackage{ulem}
\usepackage{bbm}
\usepackage{comment}
\usepackage{float}
\journal{Annals of Physics}

\numberwithin{equation}{section}

\DeclareMathOperator{\rms}{rms}
\DeclareMathOperator{\var}{var}
\DeclareMathOperator{\Tr}{Tr}

\begin{document}
\begin{frontmatter}	

\title{Mesoscopic fluctuations in superconductor-topological insulator \\  Josephson junctions}

\author[Rio]{Marcus Marinho}
\author[Rio]{Guilherme Vieira}
\author[Rio]{Tobias Micklitz}
\author[UA]{Georg Schwiete}
\author[UW]{Alex Levchenko}

\address[Rio]{Centro Brasileiro de Pesquisas F\'isicas, Rua Xavier Sigaud 150, 22290-180, Rio de Janeiro, Brazil}
\address[UA]{Department of Physics and Astronomy, The University of Alabama, Alabama 35487, USA}
\address[UW]{Department of Physics, University of Wisconsin-Madison, Madison, Wisconsin 53706, USA}

\date{June 12, 2022}

\begin{abstract}
We study mesoscopic fluctuations in the supercurrent of a Josephson junction consisting of a topological insulator microbridge between two conventional superconductors. In the model, we account for the strong proximity effect when superconductors induce a gap in the spectrum of surface states as well as a magnetic field piercing the junction area that causes depairing and gap filling. The overall magnitude and functional form of the Josephson current fluctuations are determined analytically, and found to sensitively depend on the coupling strength to surface states, Thouless energy, and pair-breaking energy scales in the problem. We also study the density of states that can be measured by scanning probes. Technically, mesoscopic fluctuations on top of the mean field description of the proximity effect in the topological region are described by a field theory approach, the replica nonlinear $\sigma$-model in the class-D of the extended symmetry classification.
\end{abstract}


\end{frontmatter}



\section{Introduction}\label{sec:intro}

The signature phenomenon of mesoscopic quantum transport is universality of conductance fluctuations (UCF) \cite{BLA-UCF,Lee-Stone}, see also \cite{Lee-Stone-Fukuyama,Beenakker-RMP} and references therein.  The variance of the conductance, $\var G=\langle G^2\rangle-\langle G\rangle^2$, where angular brackets denote ensemble average (or equivalently average over the impurity configurations) is found to be expressed universally through the quantum of conductance, $\var G\sim (e^2/h)^2$. The prefactor in this relation depends on the symmetries of the system and the dimensionality, but not on the disorder strength or the sample size. For a wire geometry, for example, the variance is found as
\begin{equation}\label{eq:UCF}
\var G=\frac{2}{15\beta}G^2_0,\qquad G_0=\frac{2e^2}{h}, \qquad \beta=1,2,4,    \end{equation}
where the values of parameter $\beta$, respectively, correspond to standard Dyson symmetry classes of the orthogonal, unitary, and symplectic ensembles in random matrix theory \cite{Mehta}. The physical origin of this effect can be traced back to the quantum interference that leads to reproducible sample-to-sample fluctuations in the conductance at low temperatures. Experimentally, these fluctuations can be observed in a single
sample as a function of magnetic field (or gate voltage changing the chemical potential), since a small
change in field (carrier density) has a similar effect on the interference
pattern as a change in impurity configuration. In contrast to the sample average conductance, $\langle G\rangle\sim G_0(Nl/L)$, universality of fluctuations is manifested by the fact that there is no dependence on the disorder mean free path $l$, the number of transverse modes $N$, and the system size $L$
provided $l\ll L\ll Nl$ (the second inequality insures that the wire length is shorter than the localization length). Universality is also robust against interaction effects provided that the system size is smaller than the dephasing length, $L<L_\phi(T)$, although interactions determine the typical scale of $L_\phi(T)$ and its temperature dependence \cite{Aleiner-Blanter}.  

There are two complementary explanations for the universality of conductance
fluctuations that can be given either in terms of the distribution of transmission eigenvalues \cite{Imry} or in terms of the level statistics in disordered conductors \cite{BLA-BIS}.  
From the theory of localization it is known that the distribution function density $\rho(\mathcal{T})$ of transmission eigenvalues $\mathcal{T}_n$, through a disordered region is bimodal \cite{Beenakker-RMP,Dorokhov}, $\rho(\mathcal{T})=(Nl/2L)(\mathcal{T}\sqrt{1-\mathcal{T}})^{-1}$,
with a peak at unit transmission and a peak at exponentially
small transmission. The fact that it is non-normalizable at small transmissions is the manifestation of localization -- most channels are closed as most transmission eigenvalues in a disordered conductor are exponentially small. Only a small fraction $l/L$ of the total number $N$ of transmission eigenvalues is of order unity and effectively contributes to the total conductance: $\langle G\rangle=N_{\text{eff}}G_0\approx (Nl/L)G_0$. Then the fluctuations in the conductance can be interpreted as fluctuations in the effective number of open channels $N_{\text{eff}}$. An alternative argument explores the relationship of $N_{\text{eff}}=E_{\text{Th}}/\delta$ to the Thouless energy $E_{\text{Th}}$ and the
mean level spacing $\delta$ in the system. In this language conductance fluctuations can be interpreted as fluctuations in the number of energy levels
in an energy strip of width $\sim E_{\text{Th}}$. For the uncorrelated statistics one would naturally estimate that fluctuations in $N_{\text{eff}}$ would be of order $\sqrt{N_{\text{eff}}}$, however strong level repulsion dictates that in fact $N_{\text{eff}}\sim 1$ and thus $\var G/G_0\sim 1$ \cite{BLA-BIS}.  

When superconductivity is introduced as a boundary effect, the processes of Andreev reflection determine the conductance of the junction, while the conductance fluctuations remain universal. This was verified both by direct numerical simulation \cite{Marmorkos,Brunn} and diagrammatic calculation \cite{Takane-Ebisawa}, and subsequently confirmed experimentally \cite{Exp-UCF-SN-1,Exp-UCF-SN-2}. For instance,  the variance of conductance in a normal-superconductor (NS) junction with ideal NS interface is found to be     
\begin{equation}
\var G_{\rm NS}=\frac{16}{15\beta}\left[1-\frac{45}{\pi^4}\right]G^2_0,\qquad \text{for} \qquad \beta=1,4 
\end{equation}
which differs from Eq. \eqref{eq:UCF} only in the numerical prefactor (the case of $\beta=2$ is somewhat special as discussed in \cite{Brouwer-Beenakker}). 

In contrast, when superconducting correlations are present in the bulk of the sample, one would expect global properties of the system to be affected by mesoscopic effects. For instance, this concerns supercurrents in a superconductor-normal-superconductor (SNS) Josephson junctions. For a short wire geometry with transparent NS interfaces, for example, the variance in the current-phase relation is known to be as the following series \cite{Chalker-Macedo,Beenakker-PRL91}
\begin{equation}\label{eq:UJF}
\var I(\phi)=\frac{2\pi^2}{15}I^2_0\sin^2\phi\left[1+\frac{62}{63}\sin^2(\phi/2)+\frac{3631}{3780}\sin^4(\phi/2)+\ldots\right],\qquad I_0=\frac{e\Delta}{h},\qquad \beta=1.    
\end{equation}
In complete analogy with Eq. \eqref{eq:UCF} these mesoscopic fluctuations are universal in the sense that they do not depend on the size of the junction or on the degree of disorder, as long as the criteria $l\ll L\ll Nl$ and $L\ll\xi$ for the diffusive, short-junction regime are satisfied. Here $\xi$ is the superconducting coherence length. In the diffusive limit $\xi=\sqrt{\xi_0l}$ with $\xi_0=v_F/\pi\Delta$. The overall scale of the Josephson current fluctuations is set by the  energy gap $\Delta$ in a superconductor. The same is true for the root-mean-square value of the critical current $I_c=\mathrm{max}[I(\phi)]$, for which $\rms I_c\approx 1.8\,e\Delta/h$ \cite{Chalker-Macedo}. The numerical factor in this relation for $\rms I_c$ does not immediately follow from Eq. \eqref{eq:UJF}. Indeed, $I_c$ is not a simple linear statistics of transmission eigenvalues since phases at which the maximum supercurrent is reached depends  itself on all the transmission eigenvalues. 

The short-junction limit $L\ll\xi$ is essential for universality in Eq. \eqref{eq:UJF}. The opposite long-junction limit was considered subsequently in Ref. \cite{BLA-BZS}. It was shown that fluctuations are no longer universal and the variance of the critical current scales with the Thouless energy, namely $\var I_c\simeq(eE_{\text{Th}}/h)^2$. Nevertheless, this result captures the remarkable property that the entire critical current through the SNS junction is determined by the mesoscopic contribution. Indeed, in the long junction limit, the sample average critical current decays exponentially with the junction length, $\langle I_c\rangle=(\Delta\langle G\rangle/e)\exp(-L/\xi)$, whereas the mesoscopic term decays only algebraically. Therefore, one could reach the regime where $\langle I^2_c\rangle\gg \langle I_c\rangle^2$. The sensitivity of these results to the transparency of NS interfaces was investigated in Ref. \cite{Micklitz}. Additionally, these results were extended to various geometries, including chaotic quantum dots, and different temperature regimes in Ref. \cite{Houzet-Skvortsov}, where weak localization corrections to the supercurrent in Josephson junctions with coherent diffusive electron dynamics in the normal part were also computed. Mesoscopic fluctuations of supercurrents were addressed in the special limit of the single-channel multiterminal devices based on the scattering matrix formalism \cite{HXAL}.

The simplicity of the universal limit for the supercurrent fluctuations given by Eq. \eqref{eq:UJF} is the result of an approximation that neglects the complexity of the proximity effect induced by a superconductor into the normal region. This concerns the spectral gap, an energy scale seemingly missing in Eq. \eqref{eq:UJF}. In the nonuniversal regime of a long-junction, the spectral gap in the normal region is of the order of the Thouless energy $E_{\text{Th}}$ \cite{BBS,Hammer}. When the junction size is made smaller, the spectral gap grows. One would naturally expect it to reach the full superconducting gap $\Delta$ in the regime of strong proximity effect, where the universal limit of mesoscopic fluctuations is realized. However, this scenario occurs only in the limit of a point-contact junction $L/\xi\to0$, which alternatively can be reformulated as a limit of energy scale ratio $\Delta/E_{\text{Th}}\to0$. For large but finite Thouless energy, the spectral gap does not reach the full superconducting gap $\Delta$, as there remains a strip of energies $\sim \Delta^3/E^2_{\text{Th}}$ with the finite density of states in the normal region \cite{AL-DOS}. Remarkably, depending on the quality and properties of the interfaces a secondary minigap may develop near the spectral edge \cite{Reutlinger-1,Reutlinger-2}. The mesoscopic fluctuations of the secondary gap follows the Tracy-Widom distribution \cite{Reutlinger-3}, the same as found in Ref. \cite{Vavilov} for the distribution of the minigap edge in the opposite limit $E_{\text{Th}}\ll\Delta$. However, the implication of these interesting features on the supercurrent fluctuations has not been addressed, only an average current-phase relation was calculated \cite{LKG,Whisler}. Furthermore, Josephson junctions are typically operated in an external magnetic field, which introduces yet another energy scale $E_\Phi$ into the problem. This scale quantifies the strength of field-induced depairing effects, which in part lead to a population of sub-gap states and ultimately gap closure. It should be then expected that  $\var I(\phi)$ must depend sensitively on both the spectral gap and $E_\Phi$. 

In this work we explore mesoscopic fluctuations in hybrid proximity circuits of topological insulator (TI) thin films and conventional superconductors (S) deposited on their surfaces. There is a wealth of transport data on these systems including the STIS Josephson junctions \cite{Sacepe,Brinkman,Lu,Goldhaber-Gordon,Mason,Kurter-PRB,Kurter-NC,Sochnikov,Stehno,Finck,Bobkova16, Alidoust17, Zyuzin18}. In part our study is motivated by proposals that an architecture of networks of lateral S-TI Josephson junctions holds promise for realizing quantum computing hardware with topological states of matter \cite{Vishveshwara}. Thus understanding mesoscopic effects in these system is important for establishing fundamental limits of their transport functionality. The rest of the paper is organized as follows. In Sec. \ref{sec:model} we introduce the model that includes the geometry of the junction, relevant energy scales, and the Hamiltonian of the system. In Sec. \ref{sec:action} we reformulate the problem in the language of the effective field theory of the nonlinear $\sigma$-model. This approach is convenient as disorder averaging is performed at the first step explicitly and the formalism enables computation of both observables of interest and their higher order correlation functions. To benchmark this approach, we derive the Usadel equation in Sec. \ref{sec:Usadel} and apply it to study density of states in Sec. \ref{sec:DOS}. Here we contrast our results to the previous computations in similar geometries and settings. In Sec. \ref{sec:Z} we introduce the semiclassical partition function. Knowledge of this function enables us to compute the Josephson current-phase relation and to derive our central results for the variance of the Josephson current fluctuations. The average current and the current fluctuations in the absence of a magnetic field are discussed in Sec. \ref{sec:Josephson-Fluctuations}. The influence of a finite magnetic field is subsequently studied in Sec. \ref{sec:Josephson-Fluctuations-FiniteB}. We conclude in Sec. \ref{sec:discussion}, where we also summarize our main results in a compact form in Table \ref{table1}. The main text of the paper is accompanied by several appendices where we provide additional technical details of the presented analysis.


\section{Model}\label{sec:model}

\subsection{Setup}

\begin{figure}
\centering
\includegraphics[scale=0.4]{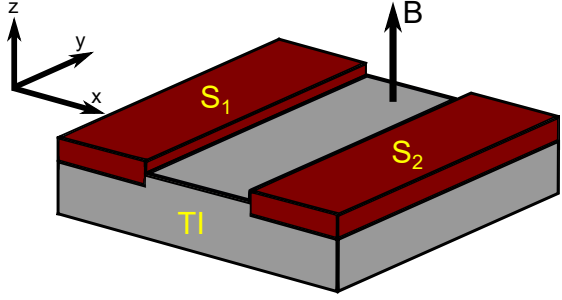}
\caption{A schematic of the planar STIS Josephson junction. Two superconductors, $S_1$ and $S_2$, are deposited onto the top surface of the topological insulator (TI) thin film marked by a gray slab. We chose TI surface as $xy$-plane of the coordinate system with magnetic field $\mathbf{B}$ pointing in $z$-direction. The length of the junction along the $x$-direction is $L$, whereas its width along the $y$-direction is $W$.}
\label{Fig-STIS}
\end{figure}

We consider a Josephson junction formed by a topological insulator (TI) in contact between two superconductors (SC) with the phase difference $\phi$. The system geometry we investigate is that of a line junction of width $W$ along the $y$-direction. In the $x$-direction we assume rigid boundary conditions that the order parameter varies as $\Delta_1(x) = \Delta e^{i\phi/2}$ for $x > L/2$ and $\Delta_2(x) = \Delta e^{-i\phi/2}$ for $x < -L/2$, with $L$ being the length of the junction between the superconducting electrodes and $\Delta$ being the superconducting energy gap. Furthermore, the junction is pierced by a perpendicular magnetic field $\mathbf{B}$, see Fig.~\ref{Fig-STIS} for an illustration.

The superconductors are coupled to the TI surface via tunneling contacts leading to the Fermi golden rule level broadening $E_t=\pi \nu w^2$ of surface states, where $w$ sets the strength of the coupling between the topological insulator and the superconducting lead. The presence of disorder in the TI surface is characterized by the elastic mean scattering time $\tau$. We focus on the most relevant case of a strong proximity effect and weak disorder in the short junction limit, in which the Thouless energy 
$E_{\rm Th}=D/L^2$, where $D$ is the diffusion coefficient, and the inverse elastic scattering time set the largest energy scales. The external magnetic field is a parameter which provides a gateway for us to access the physics of gapless surface states in the topological insulator. It will be used as a tuning parameter to close the induced minigap for the TI surface states. Therefore, the focus of our interest throughout this work shall remain mainly on the hierarchy of energy scales satisfying the following inequality
\begin{equation}\label{eq:energy-scales}
\delta \ll \{E_t, E_\Phi, \Delta\} \ll \{E_{\rm Th},1/\tau\}
\end{equation} 
where $\delta=1/\nu$ is the level spacing (with the density of states $\nu$), and the characteristic magnetic energy is defined as $E_\Phi=\frac{\pi^2n_\Phi^2}{3 \tau_{\rm tr}}$. For convenience we introduced the number of superconducting flux quanta $n_\Phi=\Phi / \Phi_0$, where $\Phi_0=\pi/e$ is the (superconducting) magnetic flux quantum. The magnetic flux is determined by the strength of the magnetic field piercing an area defined by the transport mean free path times the transverse length of the line junction, that is, $\Phi=Ll_{\rm tr}B$ is the flux through an area $Ll_{\rm tr}$ with
$l_{\rm tr}=v\tau_{\rm tr}$, where $\tau_{\rm tr}=2\tau$ is the transport mean free time. 

\subsection{Hamiltonian}

The Hamiltonian describing such a system has several distinct contributions
\begin{align}
&H = H_S + H_{S_1} + H_{S_2} + H_T,\\ 
&H_S = \int_{S} d^2x\left[\psi^\dagger_{\bold{x}}
\sigma^{\rm ph}_3\otimes \left(
-i\bm{\sigma}\cdot\partial_{\bf A}+V_{\bold{x}}\right)\psi_{\bold{x}}\right],\label{eq:HS}
\\
&H_{S_i} = \int_{S_i} d^3x\left[\Psi^\dagger_{\bold{x}}
\sigma^{\rm ph}_3\otimes \left(\epsilon_{\hat{\bold{p}}-e\bold{A}}
+i \Delta \sigma_2^{\rm ph} e^{\frac{i}{2}\phi_i\sigma_3^{\rm ph}}
\right)\Psi_{\bold{x}}\right],
\\
&H_T = \sum_{i=1,2}\int d^3x\left[\bar{w}_i(y)\Psi^\dagger_{\bold{x}}\,
\sigma^{\rm ph}_3 \psi_{\bold{x}}\delta_{x-x_i}\delta_{z-z_0}+
{\rm h.c.}\right]. 
\end{align}
Here, $S$ and $S_{1,2}$ are the topological insulator surface in the central region, and the bulk superconductors $1$ and $2$, respectively. The local coupling term is described by $H_T$ 
and we define $x_i$ as the position of the $i$-th topological insulator-superconductor interface. 
We assume that the tunneling barrier between the superconductor and the topological insulator varies in thickness along the $y$ direction, so that it can be effectively modelled as a collection of randomly distributed tunneling centers with short range correlations. Correspondingly, we introduce the ensemble average of the tunneling amplitudes as $\langle \bar{w}_i(y)\bar{w}_j(y')\rangle\propto \delta_{ij} w^2 \delta(y-y')$. The dispersion relation in $S_{1,2}$ is $\epsilon_{\bold{p}}$, ${\bold{p}}$ is the canonical momentum, $x$ and $y$ are the coordinates along and perpendicular to the junction 
(where an infinitesimal shift to left and right of the interfaces is implicit), and $z_0$ is the $z$-coordinate of the TI surface. The magnetic field is given by $\bold{B}={\rm rot}\, \bold{A}$. Its presence promotes the spatial gradient term in the Hamiltonian $\hat H_S$ to a long covariant derivative $\partial_{\bf A}$ (we consider a constant magnetic field). Making use of gauge invariance, we choose for the vector potential $\bold{A}= Bx\bold{e}_y$. Besides providing a simple and efficient way to represent the magnetic field, this form also preserves the translational invariance in the $y$-direction parallel to the interfaces. 
We further introduced Nambu spinors
$\psi^\dagger
=\frac{1}{\sqrt{2}}
(c^\dagger_{\uparrow},
c^\dagger_\downarrow,
-c_\downarrow,
c_\uparrow)$ and $\psi$ for the central region, and their counterparts $\Psi^\dagger$ and $\Psi$ for the surfaces $S_{1,2}$ below the superconductors. Finally, $V_{\bold{x}}$ is
a Gaussian distributed disorder with vanishing mean and 
second moment: 
$\langle  V_{\bold{x}} V_{\bold{x}'}\rangle=\frac{1}{\pi\nu\tau}
\delta_{\bold{x}-\bold{x}'}$.
It is important to distinguish the single particle scattering time $\tau$, which appears in this formula, from the transport scattering time $\tau_{\mathrm{tr}}=2\tau$, which is the relevant time scale entering the diffusion coefficient $D=v\tau_{\rm tr}/2$.

\subsection{Effective channel Hamiltonian}

We assume that the superconducting leads are much larger than the topological insulator thin film. Then, we can neglect the inverse proximity effect as well as depairing effects due to a finite current density or the magnetic field in the leads and integrate them out \cite{Sedlmayr}. This generates the effective channel Hamiltonian ${\cal H} = H_S + H_\Gamma$
for the tunneling junction $|x| \leq L/2$, with
\begin{align}
\label{H_Gam}
H_\Gamma
&=
-
{E_t L\over \sqrt{\epsilon^2_n + \Delta^2}}
\sum_{k=1,2}
\begin{pmatrix}
i\epsilon_n & -\Delta_k
\\
-\Delta^*_k & i\epsilon_n
\end{pmatrix}
\delta(x-x_k),
\end{align}
where $\epsilon_n$ represents the Fermionic Matsubara frequencies. In the short junction limit the tunneling contribution $H_\Gamma$ imposes so-called rigid boundary conditions \cite{Golubov}.

\subsection{Symmetries}

The effective channel Hamiltonian satisfies the particle-hole symmetry, 
\begin{align}
\mathcal{H}(\bold{k})
&=
-(i\sigma_2)\otimes (i\sigma^{\rm ph}_2) 
\mathcal{H}^t(-\bold{k})
(-i\sigma_2)\otimes(-i\sigma^{\rm ph}_2),\label{eq:phsymm}
\end{align}
where $\sigma^{\rm ph}_2$ and $\sigma_2$ operate in Nambu and spin space, respectively. Noting that the particle-hole symmetry involves 
$\Xi\equiv (i\sigma_2)\otimes (i\sigma^{\rm ph}_2)K$ 
with $K$ the complex conjugation and $\Xi^2=\mathbbm{1}_4$ this defines class-${\rm D}$ in the Altland-Zirnbauer symmetry classification \cite{Altland97}.

\subsection{Josephson current and fluctuations}\label{sec:Josephson}

The phase-difference $\phi$ between the superconductor pair potentials $\Delta_1$, $\Delta_2$ induces a stationary current 
\begin{equation}
I(\phi)
=
2e\partial_\phi\langle F\rangle,
\end{equation}
where $e$ is the charge of the electron, 
$F= -T \ln Z$ the free energy in a given disorder realization, and 
$\langle...\rangle$ refers to the disorder average. 
For the calculation of the fluctuations of the Josephson supercurrent it is convenient to introduce two sample copies subject to the same realization of the disorder potential. We introduce the correlator
\begin{equation}
K(\phi_1,\phi_2)=4e^2\partial^2_{\phi_1 \phi_2} \langle F(\phi_1)
F(\phi_2)\rangle_c, 
\end{equation}
where $\langle...\rangle_c$ is the connected disorder average and indices in $\phi_{1,2}$ refer to the sample. Then, the variance of the current fluctuations is given as
\begin{equation}
{\rm var}\, I(\phi)=
K(\phi,\phi).
\end{equation}


\section{Effective action}\label{sec:action}

Following the standard approach to disordered systems, we employ the replica trick to express the disorder averaged free energy in terms of a replicated partition function \cite{Wegner,Efetov}. We then derive for the latter an effective field theory representation  
${\cal Z}=\int DQ e^{-S[Q]}$ with the nonlinear sigma model action
\begin{align}
S[Q]=
\frac{\pi\nu}{8} \int d^2x\; 
{\rm tr}\left(D\partial_{\bf A} Q_\bold{x}\partial_{\bf A} Q_\bold{x}
-
4(\epsilon + iH_\Gamma)\sigma_3^{\rm ph} Q_\bold{x}\right).\label{eq:firstaction}
\end{align}
In this expression, $\nu=\mu / (2\pi v^2)$ is the density of states of the TI surface at the Fermi level, $\partial_{\bf A} O=\partial_\bold{x} O + ie[\bold{A}\sigma_3^{\rm ph}, O]$ is the covariant 
derivative accounting for the presence of 
a magnetic field, with the standard notation for the commutator of two matrices $[A,B]$. The notation $\epsilon$ in here denotes a diagonal matrix of Fermionic Matsubara frequencies with diagonal elements 
 $(\epsilon)_n=\epsilon_n$. In these conventions $Q_\bold{x}$ is a $4MR$ dimensional matrix, where $R$ is the number of replicas 
(send to zero at the end of the calculation, see below) and $M\sim 1/\tau T$ is the number of Matsubara frequencies kept 
in the low energy description. The additional $4=2^2$-dimensional structure 
is the tensor product of the two-dimensional Nambu space 
and the two-dimensional `sample degree of freedom', introduced to accommodate the calculation of sample-to-sample fluctuations in Sec.~\ref{sec:Josephson}. More specifically, doubling of the junction Hamiltonian 
$\mathcal H(\phi)
\mapsto
{\rm diag}(\mathcal H(\phi_1), \mathcal H(\phi_2))$,
allows us to simultaneously 
account for the Josephson currents in the same disorder realizations of the system 
at different superconductor phase differences $\phi_1$, $\phi_2$, 
 and their correlations. 
The Pauli matrices operating in the Nambu space 
are indicated by the index `${\rm ph}$'. 
The matrix degree of freedom obeys the symmetry constraint
\begin{align}
Q(\bold{x};\tau,\tau') 
&=
\sigma_1^{\rm ph}
Q^t(\bold{x};\tau',\tau)
\sigma_1^{\rm ph},\label{eq:symmetry}
\end{align}
inherited from the particle-hole symmetry Eq.~\eqref{eq:phsymm} of the junction Hamiltonian.
In the above $\tau,\tau'$ are the imaginary time arguments related to Matsubara frequencies 
by the Fourier transformation 
$Q_{\epsilon\epsilon'}=\int d\tau d\tau' Q_{\tau\tau'}e^{i\epsilon\tau-i\epsilon'\tau'}$.
A general derivation of the effective action can be found in Ref.~\cite{Altland10}, and its adaptation to the 
topological insulator surface is discussed in \ref{app:replica}. Here, we sketch the main steps.

Starting out from the replicated partition function for the junction Hamiltonian (doubled to account for two realizations of the same sample, as discussed above), the disorder average induces an effective `interaction' between fermions in different replicas, Nambu  and sample sectors. This interaction is decoupled via a Hubbard-Stratonovich 
transformation in terms of a Hermitian 
  matrix field $Q$, satisfying the symmetry constraint induced by the particle-hole symmetry 
 of the junction Hamiltonian. Integration over 
fermionic fields leads to a representation of the averaged partition function entirely in terms 
of the matrix field, which is then exposed to a saddle point analysis.
The latter is stabilized by $1/\tau$ which defines the largest energy scale in the problem,  
and is much larger than the scales $\{\Delta, E_t, E_\Phi\}\ll1/\tau$ of interest.
We employ the ansatz $Q_0=q\sigma^{\rm ph}\otimes \Lambda$ of a homogeneous saddle point  
with the structure in the Nambu and Matsubara spaces dictated by causality, where, here and in the following,
$\Lambda$ is a diagonal matrix in the Matsubara space with the elements 
$(\Lambda)_n={\rm sgn}(\epsilon_n)$.  
For this ansatz, the saddle point equation reads
\begin{align}
q=-\frac{2i}{\pi \mu}\int_0^\rho d\varepsilon \varepsilon \frac{\mu+\frac{i}{2\tau} q}{\varepsilon^2-(\mu+\frac{i}{2\tau} q)^2},
\end{align}
where we used the relation $\nu(\mu)/\nu(\epsilon)=\mu/\epsilon$ for the density of states $\nu(\epsilon)=\epsilon/(2\pi v^2)$, 
the upper cut-off $\rho\gg \mu$ was introduced to regularize the logarithmic ultraviolet divergence, and 
contributions from low energy scales 
$\hat \epsilon$ and $H_\Gamma$ have been neglected. 
As usual, the saddle point equation corresponds to the self-consistent Born approximation (SCBA), and 
upon integration becomes
\begin{align}
q=-\frac{i\alpha}{\pi \mu}\ln\frac{\alpha^2-\rho^2}{\alpha^2}\approx -\frac{i\alpha}{\pi \mu}\ln\frac{-\rho^2}{\alpha^2},\label{eq:qsaddle}
\end{align}
where $\alpha=\mu+\frac{i}{2\tau} q$. 
Decomposing $q$ into the real and imaginary parts, $q=q_1+iq_2$, and using $q_1/\mu\tau$, and $q_2/\mu\tau$ as small parameters, 
the equation can be solved iteratively, with the leading solution 
$q_1=1$ and $q_2=-({2}/{\pi})\ln\rho/\mu$ in the limit $1/\mu\tau\rightarrow \infty$. 
In systems with quadratic dispersion, one usually absorbs the logarithmic divergence 
in $q_2$ into a redefinition of $\mu$. 
The case of a linear dispersion is, however, different and including subleading terms 
in $1/\mu\tau$, one finds
$q_1\approx 1+\frac{2}{\pi\mu \tau}\ln\frac{\rho}{\mu}$, and 
$q_2\approx -\frac{2}{\pi}\ln\frac{\rho}{\mu}-\frac{2}{\pi^2\mu\tau}\ln^2\frac{\rho}{\mu}$.
The key point here is that the real part $q_1$ also acquires logarithmic corrections 
($\ln(\rho/{\mu})\ll \mu\tau$ is implied). While this problem is visible in the SCBA, 
the latter is not sufficient to account for the logarithmic corrections systematically. 
Indeed, it is known from studies of the closely related problem of disordered graphene that terms of the same order arise from contributions to 
the self-energy that are not included in the SCBA scheme \cite{Aleiner06,Ostrovsky06}. It was found that the renormalization group (RG) procedure can be implemented on the level of the fermionic action to sum the logarithmic divergencies and that these renormalizations can be absorbed into effective parameters of the diffusive nonlinear sigma model \cite{Aleiner06}. As is clear from the considerations summarized above, the renormalizations remain weak as long as $1/(\mu\tau)\ln(\rho/\mu)\ll 1$, where $\rho$ 
may be considered as the scale at which the dispersion deviates from being linear. Here we assume that either the ultraviolet renormalizations are irrelevant, or that the parameters of the above effective action  
are effective scale-dependent parameters, and proceed outlining the final step in the derivation of the soft mode action. 

This consists in the inclusion of soft mode fluctuations around the saddle point.
As usual, the fluctuations can be parametrized as slowly varying (local) rotations of  
the saddle point solution, $Q_\bold{x}=T_\bold{x}Q_0T_\bold{x}^{-1}$, where $T_\bold{x}$ rotate in replica, Matsubara, 
Nambu and sample space. Fluctuations in the spin sector are suppressed by spin orbit interaction (see \cite{Garate,Velkov} for a detailed discussion of the surface soft modes), and their role in the derivation of the sigma model action in the spin singlet sector is a renormalization of the charge diffusion coefficient. Essentially, this amounts to a projection onto the spin singlet sector 
of the particle-hole degrees of freedom (the `Diffuson' and `Cooperon' modes, see below), stabilized by the `mass' $\sim \nu/\tau$ of the triplet modes. 
Inserting the soft mode ansatz, a gradient expansion detailed in~\ref{app:replica} then results in the effective soft mode action Eq.~\eqref{eq:firstaction}.

Finally, with the replica field theory at hand we can calculate the 
average Josephson current and its sample-to-sample fluctuations as
\begin{align}
I(\phi)
=
\lim_{R\to 0}
\frac{2e T}{R}
\partial_{\phi_{1}}\left.\mathcal Z\right|_{\phi_1=\phi},\qquad
K(\phi_1,\phi_2)
&=
 \lim_{R\to 0}
\frac{4 e^2 T^2}{R^2}
\partial^2_{\phi_1\phi_2}
{\cal Z}.\label{eq:IKbasic}
\end{align}


\section{Usadel equation}\label{sec:Usadel}

From now on we focus our attention on the short junction limit characterized by the inequality $E_{\rm Th}\gg \Delta$. For this geometry, we can assume that $Q$ is approximately constant as a function of the $x$-coordinate perpendicular to the interfaces and integrate the action in this direction. Next we subject the action \eqref{eq:firstaction} to a second saddle point analysis. The first saddle point analysis was used for the derivation of the nonlinear sigma model and did not account for the presence of $\hat{H}_\Gamma$ nor for the magnetic field. The second saddle point analysis occurs within the manifold of the first saddle point $Q^2=1$. We therefore look for matrices $Q$ for which the condition 
$d_\alpha S[\exp(\alpha W)Q\exp(-\alpha W)]=0$ holds for arbitrary generators $W$. Due to the translational invariance parallel to the interfaces we further restrict ourselves to matrices $Q$ that are $y$-independent. This procedure leads directly to the Usadel equation~\cite{Usadel70} in the form
\begin{align}
\left[ -E_\Phi Q_{\perp} + v_i\sigma^{\rm ph}_i, Q\right]
=0,\label{eq:Usadel}
\end{align}
where $Q_\perp=\frac{1}{2}(Q-\sigma_3^{\rm ph} Q\sigma_3^{\rm ph})$. Furthermore, this equation is similar to equations that describe the effects of spin-flip processes or pair-breaking mechanisms, See Refs.~\cite{Hammer,Bergeret}. The vector components $v_i$ appear as a result of the integration in the transverse direction, $\int dx (\epsilon+iH_\Gamma)=Lv_i\sigma^{\rm ph}_i$, and read as follows 
\begin{align}
v_1(i)
=
0,\quad
v_2(i)
=
\frac{2E_t\Delta\cos\left(\phi_i/2\right)}{\sqrt{\Delta^2 + \epsilon_i^2}},\quad
v_3(i)
=
\epsilon_i + \frac{2E_t \epsilon_i}{\sqrt{\Delta^2 + \epsilon_i^2}},
\label{eq:vcomponents}
\end{align}
where $(i)=(\epsilon_i,\phi_i)$ is a convenient multi-index notation. To this end, inserting the ansatz 
\begin{align}
Q_\Delta
=
m_i\sigma_i^{\rm ph}\label{eq:QDelta}
\end{align}
with the unit vector ${\bm m}$ into Eq.~\eqref{eq:Usadel}, we can express the saddle point condition as a geometric constraint
\begin{align}
(
{\bf v}
+
E_\Phi{\bf m_3}
)
\times
{\bf m}
=
0.\label{eq:MFv}
\end{align}
Due to the nonlinear normalization condition ${\bf m}^2 = 1$ inherited from the $Q$ matrix, the general solution of this equation is rather complicated. For analytical calculations, we will therefore mainly concentrate on the limiting cases of strong and vanishing magnetic fields, where analytical solutions can be constructed straightforwardly. For general magnetic fields, we will use its    numerical solution.

\subsection{Zero magnetic field}

In the absence of a magnetic field, 
${E_\Phi} = 0$, we arrive at the homogeneous Usadel equation 
$[v_i\sigma_i^{\mathrm{ph}}, \bar Q] = 0$. 
In this limit, the solution is readily obtained by a mean 
field parallel to ${\bf v}$, that is, ${\bf m} = {\bf n}\parallel {\bf v}$, where ${\bf n}$ is given by
\begin{align}
{\bf n}
=\frac{1}{v}
\begin{pmatrix}
0\\
v_2\\
v_3
\end{pmatrix}
,
\end{align}
where $v = \sqrt{v_2^2 + v_3^2}$.

\subsection{Finite magnetic field}

For the finite magnetic field we use the saddle point condition to write $m_3$ in terms of $m_2$, namely
$
m_3
(
v_2 - E_\Phi m_2
)
=
v_3 m_2$. If the solution of interest has $m_2\neq 0$, then we can further state
\begin{align}
{\bf m}=\left( \begin{array}{ccc} 0\\ m_2\\ \frac{v_3 m_2}{v_2-E_\Phi m_2}\end{array}\right),\label{eq:mthroughm2}
\end{align}
where $m_2$ is to be fixed by the normalization condition ${\bf m}^2=1$. Substituting the components of the mean field yields
\begin{equation}
(s - m_2)^2(m_2^2 - 1) + s^2\beta^2m_2^2 = 0,\label{eq:mf}
\end{equation}
where $s = v_2 / E_\Phi$ and $\beta = v_3 / v_2$. 

The general solution to this equation can be found in closed form. However the result is complex and it is cumbersome to 
extract meaningful information from it. Progress can be made in the limit of large magnetic fields
where the solution can be constructed in terms of a power series in the small parameter $s$,
\begin{align}
m_2(s)=\sum_{l = 0}^\infty m_{2, l}s^l.
\end{align}
The leading contributions to the components of $m$ read as
\begin{align}
\label{eq:SM-mean-field2}
&m_2(s)=\frac{s}{1 + |\beta| s},\qquad m_3(s)=1.
\end{align}
In constructing this solution, we considered the product $s\beta$ in Eq.~\eqref{eq:mf} as an independent parameter. The strong magnetic field limit imposes that $s$ is small. However, for the calculation of the current fluctuations we will need to work with the solution for a wide range of frequencies $\epsilon$. Under these circumstances, the product $s\beta$ is not necessarily small, since $\beta$ scales with the frequency. We confirmed numerically that the solution stated above provides an excellent approximation for a broad interval of $\beta$ values as long as $s< 0.5$.

Note that for $v_2=0$ (i.e. no superconductor, $\Delta=0$ or $E_t=0$), the normalizable solution has $m_2=0$ and, as a consequence of this result, the solution for the third component of the mean field collapses to $|m_3|=1$. The correct solution in this case is chosen by further demanding ${\rm sgn}(m_3)={\rm sgn}(\epsilon)$. This is consistent with the conventional structure for a normal conductor.

\subsection{Rotation of the $Q$-field}

The solution of the improved saddle point equation, the Usadel equation \eqref{eq:Usadel}, is sufficient for finding the average Josephson current in the short junction limit. The calculation of the current fluctuations, in turn, requires us to go beyond the saddle point approximation and to include fluctuations. A possible strategy to achieve this goal would be to decompose the $Q$ matrix as $Q=\tilde{T} Q_\Delta \tilde{T}^{-1}$, to parametrize $\tilde{T}=\mbox{e}^{\tilde{W}}$ in terms of generators $\tilde{W}$ with constraint $\{Q_\Delta,\tilde{W}\}=0$, where $\{A,B\}$ denotes the anti-commutator of two matrices, and to proceed with an expansion in powers of $\tilde{W}$. Unfortunately, the nontrivial structure of the saddle point solution complicates the constraints for $\tilde{W}$. Following Ref.~\cite{Micklitz}, we therefore introduce a rotation of the $Q$ matrix that will simplify the perturbative expansion. With this goal in mind, we first define rotation matrices $T_\Delta$ and $T_\Delta^{-1}$ by the condition 
\begin{align}
Q_\Delta=T_\Delta Q_0 T_\Delta^{-1}.\label{eq:TDelta}
\end{align}
The rationale behind this change of variables is that $\hat{Q}$ can be parametrized as $\hat{Q}=\hat{T}Q_0\hat{T}^{-1}$, where $\hat{T}=T_\Delta^{-1}\tilde{T}T_\Delta$ can be written as $\hat{T}=\exp(\hat{W})$ and the constraint for $\hat{W}$ is much simplified compared to $\tilde{W}$. Indeed, $\hat{W}=T_{\Delta}^{-1}\tilde{W}T_\Delta$ fulfills the condition $\{\hat{W}, Q_0\}=0$. 

In the next step, we express the action in terms of the field $\hat{Q}=T_\Delta^{-1} Q T_\Delta$. The change of variables from $Q$ to $\hat{Q}$ in the action can be performed with the help of the cyclic property of the trace. In order to write the result in a compact form we introduce the notation
\begin{align}
\hat{\sigma}^{\rm ph}_i=T_{\Delta}^{-1} \sigma_i^{\rm ph} T_\Delta. \label{eq:hatsigma}
\end{align} 
In this way, we arrive at the final form of the action in the presence of a vector potential,
\begin{align}
S
=
\frac{\pi\nu}{8}
\int d^2 x\;
{\rm tr}
\left[
D(\partial_{\bf A}\hat{Q})^2 - 4v_i\hat{\sigma}_i^{\rm ph}\hat{Q}
\right].
\label{eq:smodelaction}
\end{align}
We remind that this form is valid in the small junction limit, when $\hat{Q}=\hat{Q}_{y}$. Calculations with the action \eqref{eq:smodelaction} require explicit knowledge of the rotated Pauli matrices $\hat{\sigma}_i^{\rm ph}$ defined in Eq.~\eqref{eq:hatsigma}. To find them, we insert the ansatz $T_\Delta=\exp(i\theta \Lambda \sigma^{\rm ph}_1/2)$ into Eq.~\eqref{eq:TDelta} and further use relation \eqref{eq:QDelta} to fix the rotation angle $\theta$. This leads us to the two conditions $\cos\theta \Lambda = m_3$ and $\sin\theta = m_2$, which imply $
\hat{\sigma}_2^{\rm ph}=(m_3\sigma_2^{\rm ph} + m_2\sigma_3^{\rm ph})\Lambda$ and $\hat{\sigma}^{\rm ph}_3=(-m_2\sigma^{\rm ph}_2 + m_3 \sigma^{\rm ph}_3)\Lambda$.

The action in the form given in \eqref{eq:smodelaction} forms the basis for our studies of the average Josephson current and the current fluctuations. We will exclusively work with the rotated matrix $\hat{Q}$ and from now on denote it as $Q$ in order to simplify the notation.


\section{Density of states}\label{sec:DOS}

With the solution of the Usadel equation at hand, we can now study the 
influence of the magnetic field on the proximity induced minigap in the
TI film. To this end, we first recall that within the field theory approach 
the density of states (DoS) follows from
\begin{align}
\frac{\nu(\epsilon)}{\nu}
&=
\frac{1}{4}
{\rm Re}\,
\left[
{\rm tr}
\left(
Q_0
(\epsilon\to -i\epsilon_+)
\sigma_3^{\rm ph}
\right)
\right],
\end{align}
where $Q_0$ is the solution of the Usadel equation analytically continued from the discrete set of Matsubara frequencies to the axis of real energies, where  $\epsilon_+=\epsilon+i\eta$ includes a positive small imaginary part, and $\nu$ denotes the density of states at energy $\epsilon=\mu$ in absence of superconducting leads. Building then on the discussion of the previous section, the DoS reads
\begin{align}
\nu(\epsilon)
&=\nu
\mathrm{Re}
\left[
m_3(-i\epsilon_+)
\right],
\end{align}
with $m_3$ specified through Eqs.~\eqref{eq:mthroughm2} and \eqref{eq:mf}.

\begin{figure}
\centering
\includegraphics[scale=0.3]{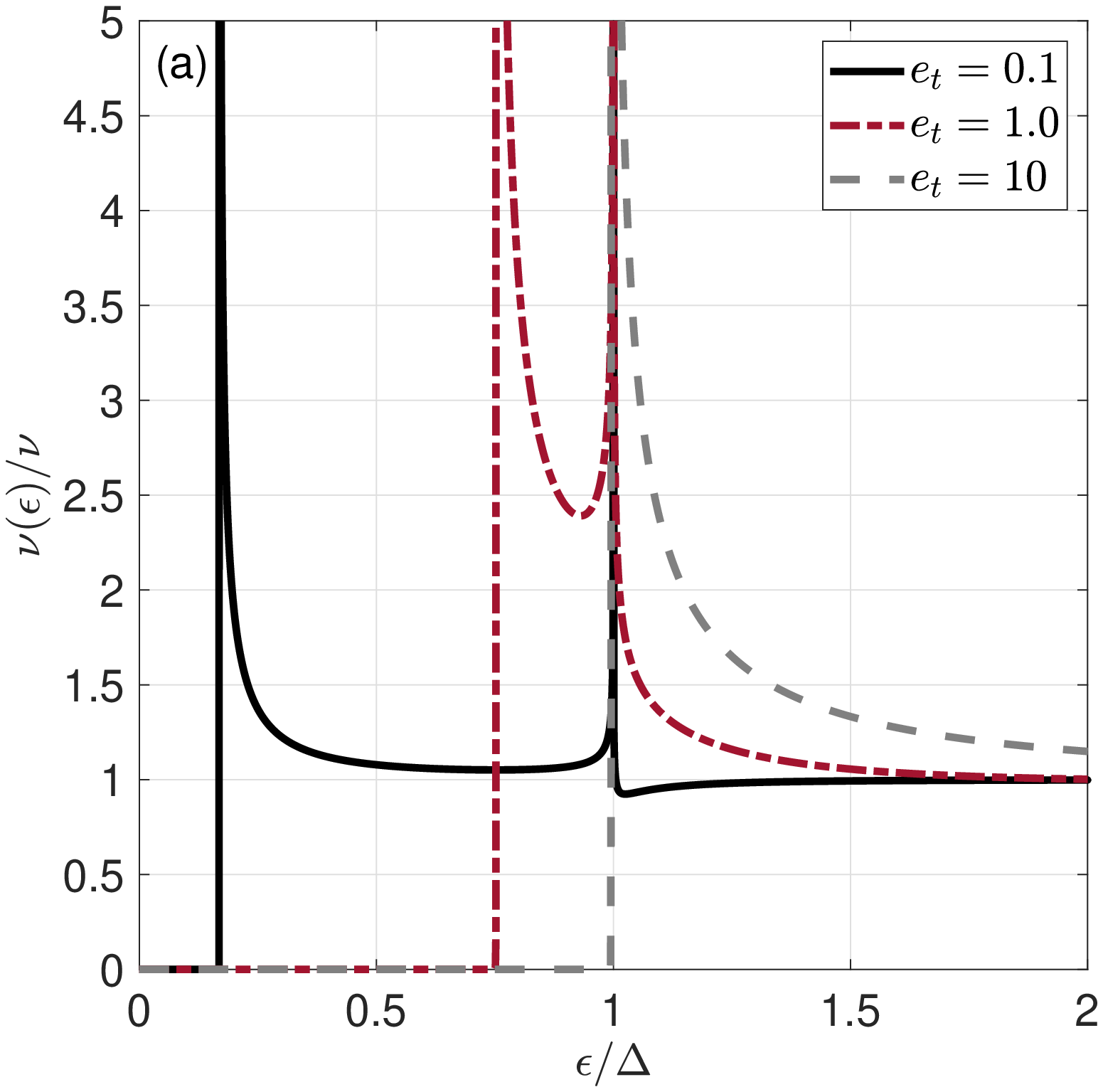} 
\includegraphics[scale=0.3]{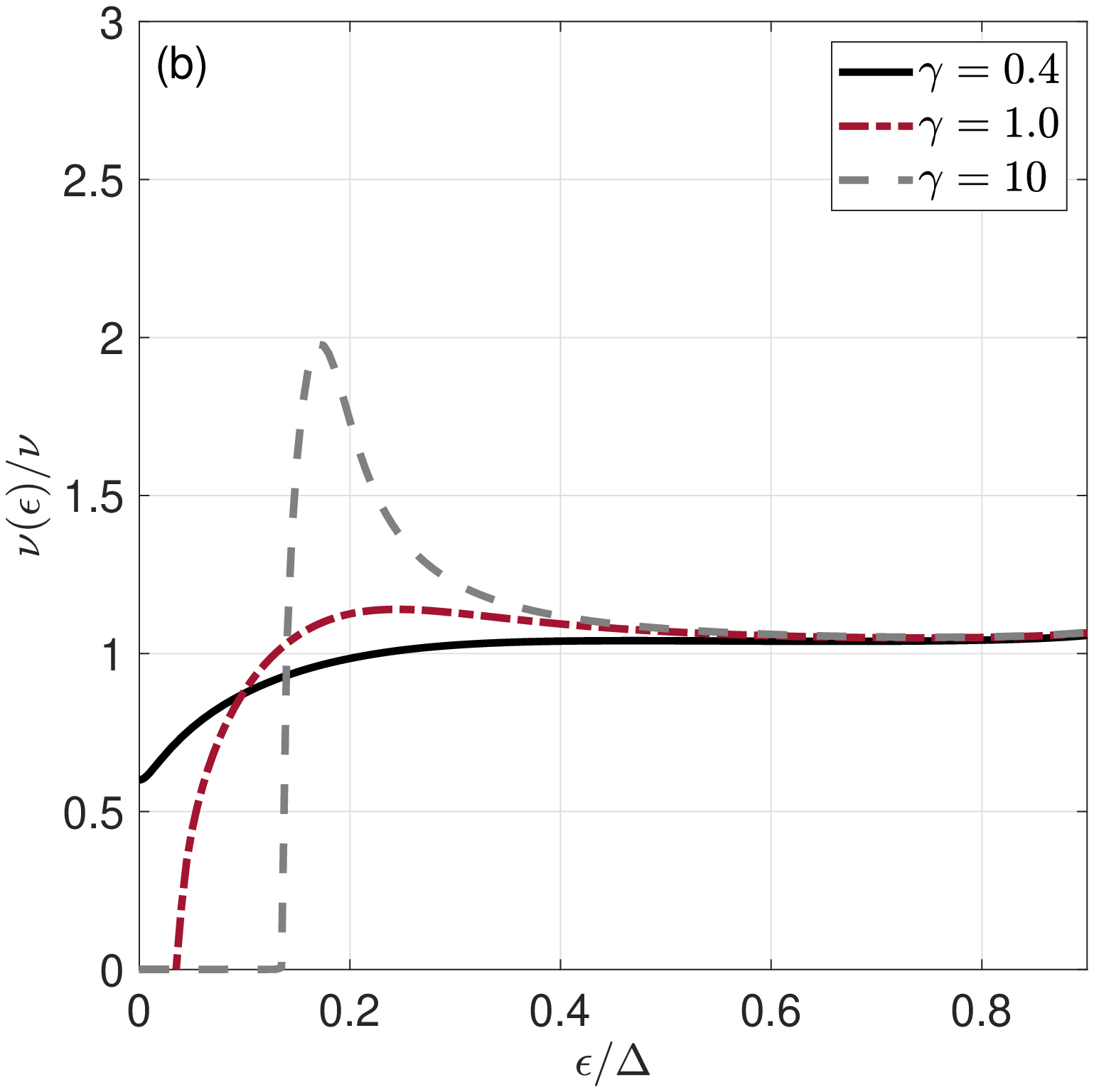}
\includegraphics[scale=0.3]{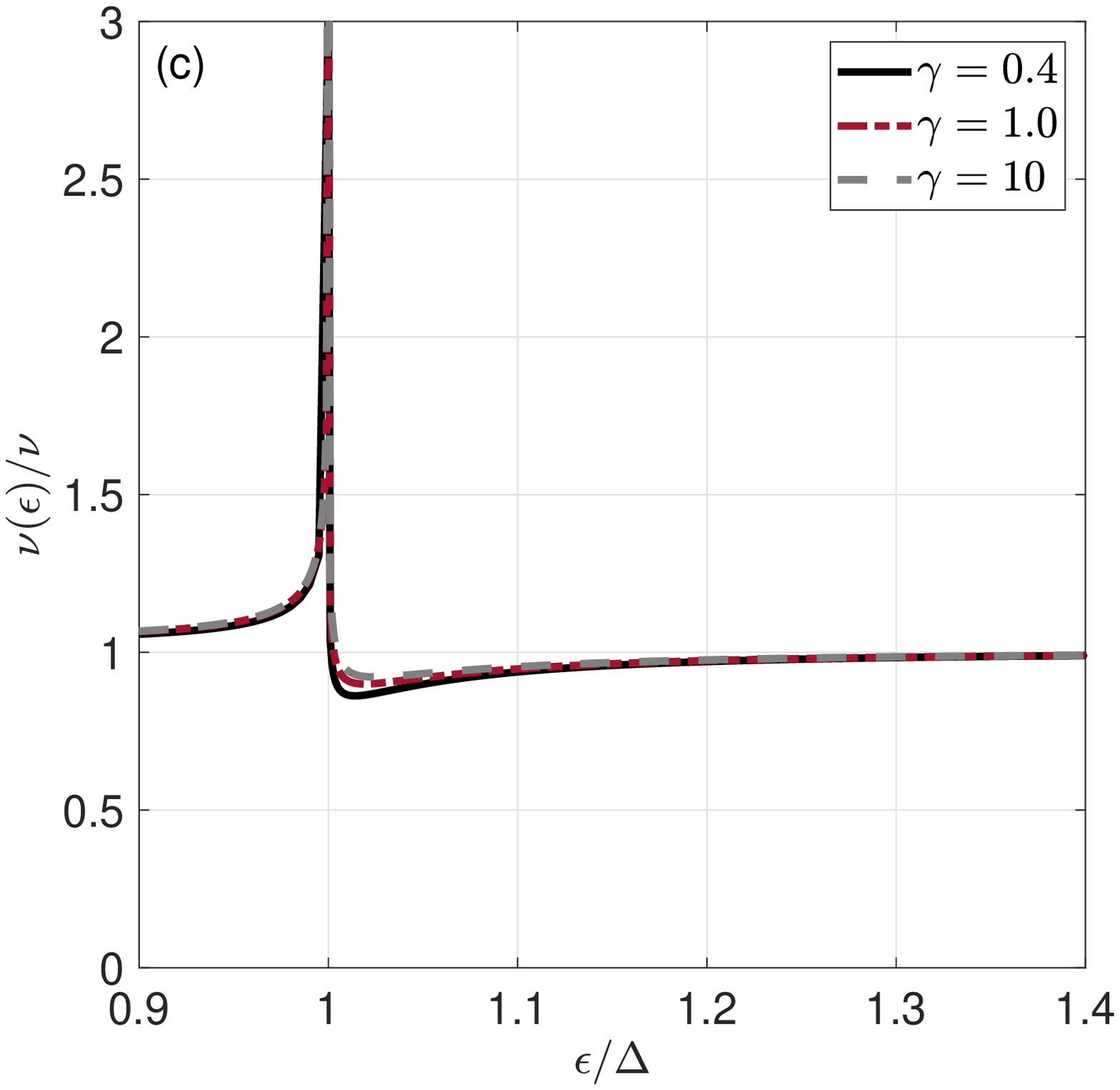} 
\caption{(a) Density of states $\nu(\epsilon)$ for the vanishing magnetic field normalized to the density of states $\nu=\nu(\mu)$ in the absence of superconducting leads for different values of $e_t = E_t/\Delta = 0.1, 1, 10$.
Panels (b) and (c) show density of states of the microbridge at a finite magnetic field as a function of frequency $\epsilon$ for different values of $\gamma = E_t / E_\Phi = (0.1 / 0.01, 0.1 / 0.1, 0.1 / 0.25)$ and fixed phase $\phi = 0$, where $e_\Phi = E_\Phi / \Delta$ .}
\label{Fig-DOS}
\end{figure}

We first consider the DoS in absence of a magnetic field, $B=0$, as shown in Fig.~\ref{Fig-DOS}(a), and recall that the proximity induced minigap $E_g$  
is a function of the ratio $E_t/\Delta$. For the weak coupling limit, $E_t\ll \Delta$, it displays the typical superconductor square-root singularity 
above the minigap $E_g=2E_t\cos\phi$,  $\nu(\epsilon)\sim \theta(\epsilon-E_g)(|\epsilon-E_g|)^{1/2}$, and a weaker singularity
$\nu(\epsilon)\sim 1/(|\epsilon-\Delta|)^{1/4}$ around the superconducting gap. 
In the opposite strong-coupling limit, $\Delta\ll E_t$, there is only a single singularity above the superconducting gap with $E_g\sim\Delta$. In the intermediate case $E_t=\Delta$ we observe two singularities, as for $E_t/\Delta\ll 1$, but now the minigap becomes large $E_g\lesssim \Delta$, as for the case $E_t\gg \Delta$. 

Turning on finite magnetic fields, we focus on the limit $E_t/\Delta\ll1$. In this limit, we can explore the sensitivity of the minigap $E_g$ to the magnetic field while its pair-breaking effect on the superconducting leads is still negligible. Fig.~\ref{Fig-DOS}(b) shows the DoS for $E_t\ll\Delta$ and various values of $E_\Phi$. Increasing the magnetic field from $B=0$, the minigap continuously reduces and closes once $E_\Phi\gtrsim E_g$. At the same time, the square root singularity at $E_g$ is smoothed out and turns into a monotonic function which displays behavior qualitatively similar to that found in the Abrikosov-Gor'kov theory of gapless superconductivity \cite{AG-JETP60}. 
Once the gap closes, the DoS quickly evolves into the nearly constant function $\nu(\epsilon)$. As expected, the singularity at $\epsilon\sim\Delta$ is hardly affected by small magnetic fields $E_\Phi\ll \Delta$, see the right panel of Fig~\ref{Fig-DOS}(c). We observe, however, a small dip above the singularity $\epsilon \gtrsim \Delta$ that develops and becomes more pronounced for 
smaller values of $\gamma$. It should be stressed that the sub- and above the gap features in the DoS are extremely sensitive to the boundary action used in the saddle point analysis of Usadel equation. For instance, in the model of transparent interfaces, that can be captured by the full circuit-theory action \cite{Nazarov}, the DoS in the sub-gap region may display secondary gaps \cite{Reutlinger-1,Reutlinger-2,Whisler}, while a singularity at $\Delta$ may be turned into a vanishing DoS and an unusual structure of the crossover to higher energies arises \cite{AL-DOS,Mazanik}.    

Notice that above results were derived using the exact solution of Eq.~\eqref{eq:mf}. The latter is a rather cumbersome expression and therefore not stated here. Although the mean field solution obtained via power series provides an exceedingly good approximation for the full-fledged solution, 
in both limiting cases, strong and weak magnetic field, it fails to fully capture the structure of the minigap. In the weak magnetic field limit, $E_\Phi < E_t$, it overestimates the size of the minigap and  there is a singularity in the region $\epsilon < \Delta$. In the opposite limit, $E_\Phi > E_t$, there exists a threshold value $E_\Phi^* $ beyond which the minigap closes. The approximated mean field solution fails to reproduce this behavior and always results in a gapless density of states. It is also worth stating that the mean-field analysis of DoS presented in this section misses the sub-gap tails \cite{Frahm,Meyer,Lamacraft,Beloborodov,Ostrovsky,Feigelman}
and zero-bias peaks. The latter include disorder-induced class D peak \cite{Bagrets} and Majorana peak \cite{Ioselevich}. These fine-structure features of the DoS appear at the level of nonperturbative analysis of $Q$-matrix manifold and become resolved at the energy scales of level spacing. This parameter regime is beyond the domain of our assumptions specified earlier by Eq. \eqref{eq:energy-scales}. The results of this section are amenable to scanning-tunneling probes in hybrid S-TI proximity circuits and heterostructures, see e.g. Refs. \cite{Jia-1,Dayton,Tessmer,Zeljkovic,Jia-2}.   


\section{Semiclassical partition function}\label{sec:Z}

In Sec.~\ref{sec:Usadel}, we studied solution of the Usadel equation, which is the saddle point equation of the nonlinear sigma model. This solution allows us to calculate the average current through the Josephson junction. In order to obtain the Josephson current fluctuations, we shall now go one step further and find the semiclassical partition function from the sigma model action \eqref{eq:smodelaction}. We choose the exponential parametrization 
\begin{align}
T=\mbox{e}^W, \quad [W,Q_0]_+=0,\label{eq:para1}
\end{align}
for the fluctuations in the vicinity of the saddle point. 
The symmetry \eqref{eq:symmetry} of the $Q$ field can be accounted for by imposing the constraint $W=\sigma_1^{\mathrm{ph}}W^t\sigma_1^{\mathrm{ph}}$ on the generators $W$. The condition $W^\dagger=-W$ ensures the convergence of integrals in $W$. It is convenient to represent $W$ as the sum of two terms, $W=W_d+W_c$, where we define Diffusons ($d$) and Cooperons ($c$) by the conditions
\begin{align}
&[W_d,\Lambda]_+=0,\quad [W_d,\sigma_3^{\mathrm{ph}}]=0, \quad [W_c,\Lambda]=0,\quad [W_c,\sigma_3^{\mathrm{ph}}]_+=0.\label{eq:para3c}
\end{align}
As we can see, the Diffusons $W_d$ are diagonal in ph space and off-diagonal in Matsubara space, and vice versa for $W_c$.

Relying on the quadratic expansion, the integration over generators leads to the semiclassical partition function
\begin{align}
\mathcal Z(\phi_1, \phi_2)
=
\left(\det {\cal D}\right)^{R^2} \left(\det {\cal C}\right)^{R^2}  e^{-RS_0}.
\label{eq:semiclassicalZ}
\end{align}
The action evaluated at the saddle point is given by
\begin{align}
S_0=&\frac{\pi\nu V}{2}\sum_\epsilon\sum_{j=1,2}[ E_{\Phi} m_2^2(\phi_j)-{2}v_i(\phi_j) m_i(\phi_j)].
\end{align}
The fluctuation determinants are defined through
\begin{align}
\det {\cal D}^{-1}
&=
\prod_q
\prod_{\epsilon_1 > 0}
\prod_{\epsilon_2 > 0}
\left(
\lambda_{\epsilon_1,-\epsilon_2}^{D, +}
\lambda_{\epsilon_1,-\epsilon_2}^{D, -}
\right),\quad 
\det {\cal C}^{-1}
=
\prod_q
\prod_{\epsilon_1 > 0}
\prod_{\epsilon_2 > 0}
\left(
\lambda_{\epsilon_1\epsilon_2}^{C, +}
\lambda_{\epsilon_1\epsilon_2}^{C, -}
\right),
\end{align}
with the eigenvalues 
\begin{align}
&
\lambda^{\pm}_{\epsilon_1,\epsilon_2}
=
Dq^2 
+
\bold{m}(1)\cdot \bold{v}(1)
+
\bold{m}(2)\cdot \bold{v}(2)
+
M_\Phi^{\pm},
\label{eq:lambda}
\\
&
M_\Phi^{\pm}
=
\frac{E_\Phi}{8}\left(
\left[
m_3(1) + m_3(2)\right]^2
-
4\left[
m_2(1) \mp m_2(2)
\right]^2
\right)
,
\end{align}
where we have introduced the multi-index notation $(i) = (\epsilon_i, \phi_i)$ and a mass term, $M_\Phi^{\pm}$, generated by the presence of an external magnetic field. Notice that while these eigenvalues look identical, the Diffusons are only defined 
for $\epsilon_1 > 0$ and $-\epsilon_2 > 0$, whereas the Cooperon modes have positive frequencies only, $\epsilon_1, \epsilon_2 > 0$. The derivation of Eq.~\eqref{eq:semiclassicalZ} is detailed in \ref{app:fluctuations}. In Eq.~\eqref{eq:semiclassicalZ}, we neglected terms that are diagonal in sample space and also discarded $\phi$-independent constants, because such terms cannot contribute to the calculation of the Josephson current fluctuations. 

With the help of Eq.~\eqref{eq:IKbasic}, we arrive at
the general expression for the average Josephson current in the short junction limit
\begin{align}
I(\phi)
&
=
2eT\partial_\phi S_0
=
-
\pi\nu e TV
\sum_\epsilon
\left[
\partial_\phi
\left(
2
v_i m_i
-
E_\Phi m_2^2
\right)
\right].
\label{eq:Iaverage}
\end{align} 
In an analogous way, Eq.~\eqref{eq:IKbasic} results in the following expression for the sample-to-sample current fluctuations
\begin{align}
K(\phi_1,\phi_2)
&=
(2eT)^2
\sum_{s=\pm}
\left(
F^{s}_2(\phi_1,\phi_2)-F^{s}_1(\phi_1,\phi_2)
\right),\label{eq:K12}
\end{align}
with
\begin{align}
F^{s}_1(\phi_1,\phi_2)
&=
\sum_{q}
\sum_{\epsilon_1>0}
\sum_{\epsilon_2}
\frac{\partial^2_{12} \lambda^{s}_{\epsilon_1, \epsilon_2}}
{\lambda^{s}_{\epsilon_1, \epsilon_2} },\quad
F^{s}_2(\phi_1,\phi_2)
=
\sum_{q}
\sum_{\epsilon_1>0}
\sum_{\epsilon_2}
\frac{\partial_1 \lambda^{s}_{\epsilon_1,\epsilon_2} 
\partial_2 \lambda^{s}_{\epsilon_1, \epsilon_2}}
{\left( \lambda^{s}_{\epsilon_1, \epsilon_2} \right)^2},\label{eq:F12}
\end{align}
where positive and negative frequencies $\epsilon_2$ account for the Cooperon and Diffuson contribution, respectively,
and $\partial_{1,2}$ denotes derivatives with respect to the two phase differences $\phi_1$ and $\phi_2$.


\section{Average current and sample-to-sample fluctuations at zero magnetic field}\label{sec:Josephson-Fluctuations}

We next discuss the average Josephson current and its fluctuations at zero magnetic field. We focus on the set-up displayed in Fig.~\ref{Fig-STIS} in the short junction limit, for which $E_{\rm Th}=D/L^2$ is the largest energy scale. We further distinguish the quantum dot geometry with confined transverse direction, $E_{\rm Th}^\perp\gg \{\Delta, E_t\}$, where $E_{\rm Th}^\perp=D/W^2$ 
is the Thouless energy related to the transverse direction, and the quasi-one dimensional geometry with extended transverse direction $E_{\rm Th}^\perp\ll \{\Delta,E_t\}$.

\subsection{Average current}

Building on our discussion in Sec.~\ref{sec:Usadel}, 
the solution of the saddle point equation in the absence of a magnetic field is given by 
$Q_\Delta=\hat{n}_i\sigma_i^{\rm ph}$. 
The average current $I(\phi)$ can therefore be found from Eq.~\eqref{eq:Iaverage} by setting $E_\Phi=0$ and $m_i = n_i$. The mean field vector $\hat{\bf m}$ is then 
parallel to ${\bf v}$, leading to
\begin{align}
I(\phi)
&
=\frac{GE_t}{2e}
J(\phi),
\qquad
J(\phi)
=
4\pi\sin(\phi) T
\sum_{\epsilon > 0}
\frac{\Delta^2}{\omega(\Delta, \epsilon)v(\epsilon, \phi)},
\label{eq:J}
\end{align}
where for the sake of clarity we indicated the dependence of the scalar 
$v(\epsilon, \phi) = |{\bf v}|$ on the phase difference $\phi$ and the Fermionic Matsubara frequency $\epsilon$ and we defined $\omega(\Delta, \epsilon) = \Delta^2 + \epsilon^2$. Here, we used the relation 
$E_t=\delta G/2e^2$ to connect the dwell energy with the normal-state conductance of the junction and $\delta$ represents the mean level spacing. Equation \eqref{eq:J} is valid for arbitrary ratios $e_t\equiv E_t/\Delta$. We notice that the average Josephson current does not depend on the width of the junction. Eq.~\eqref{eq:J} is consistent with previously reported results~\cite{Aslamazov,Kupriyanov,Brouwer}. We will now address the parameter dependence of the average current in the limiting cases of long and short dwell times, $E_t\ll \Delta$ and $E_t\gg \Delta$, respectively. For these cases, simple analytical solutions can be obtained. In Sec.~\ref{subsubsec:arbitrarydwell}, we will then discuss  arbitrary dwell times based on a fully numerical evaluation of Eq.~\eqref{eq:J}.

\subsubsection{Long dwell time: $E_t\ll \Delta$}
Specializing 
on the limit $E_t\ll \Delta$, 
we may approximate $v_3\approx \epsilon$, cf.~Eq.~\eqref{eq:vcomponents}. 
The scale for the average current is then set by $GE_t/2e$, 
and $J$ becomes a function of the dimensionless variables 
$t=T/\Delta$ and 
$e_t=E_t/\Delta$ only. In this approximation and at low temperatures, $T\ll \Delta$, the dimensionless $J(\phi)$ assumes the following asymptotic form~\cite{Houzet-Skvortsov, Brouwer}
\begin{align}
J(\phi)
=
2\sin(\phi)\ln\left[\frac{1}{{\rm max}(t, e_t\cos(\phi/2))}\right].\label{eq:Jphilog}
\end{align}

\subsubsection{Short dwell time: $E_t\gg \Delta$}

In the short dwell time limit, and for zero temperature, the dimensionless function $J$ is proportional to the complete elliptic integral of the first kind $\mathbf{K}$~\cite{Houzet-Skvortsov, Brouwer},
\begin{align}
J(\phi)
=
\frac{1}{e_t}
\sin(\phi)
\mathbf{K}
\left(
\sin^2\frac{\phi}{2}
\right)
=
\frac{1}{e_t}
\sin(\phi)
\int_0^\infty
dy
\frac{1}{\sqrt{\cos^2(\phi/2) + \sinh^2y}}
.
\label{eq:Jshort}
\end{align}
It is worth noting that the scale of the average current in this case is set by the order parameter $\Delta$, compare Eqs.~\eqref{eq:J} and \eqref{eq:Jshort}. 

\subsubsection{Arbitrary dwell time}
\label{subsubsec:arbitrarydwell}

\begin{figure}
\centering
\includegraphics[scale = 0.32]{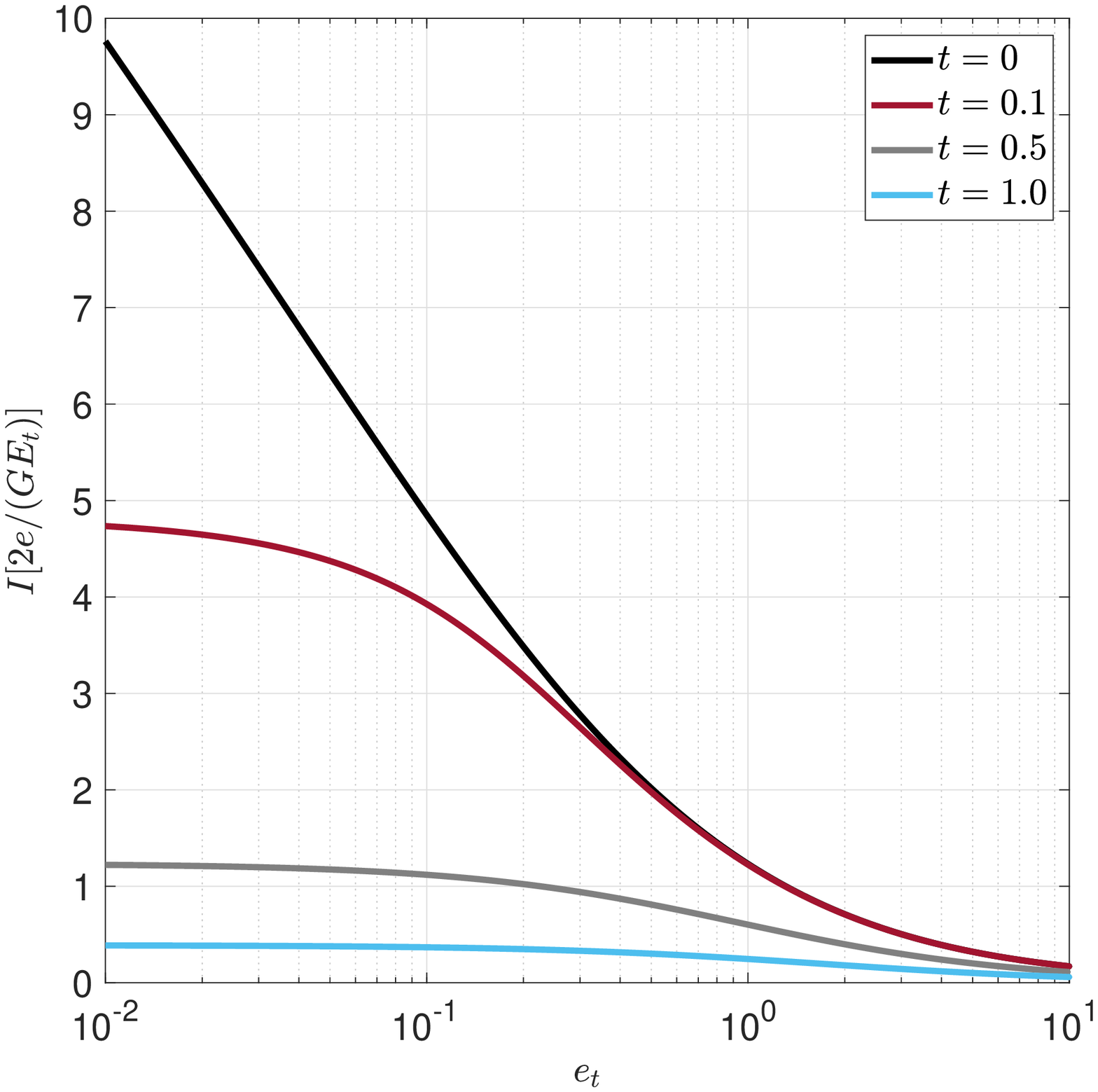}
\includegraphics[scale=0.32]{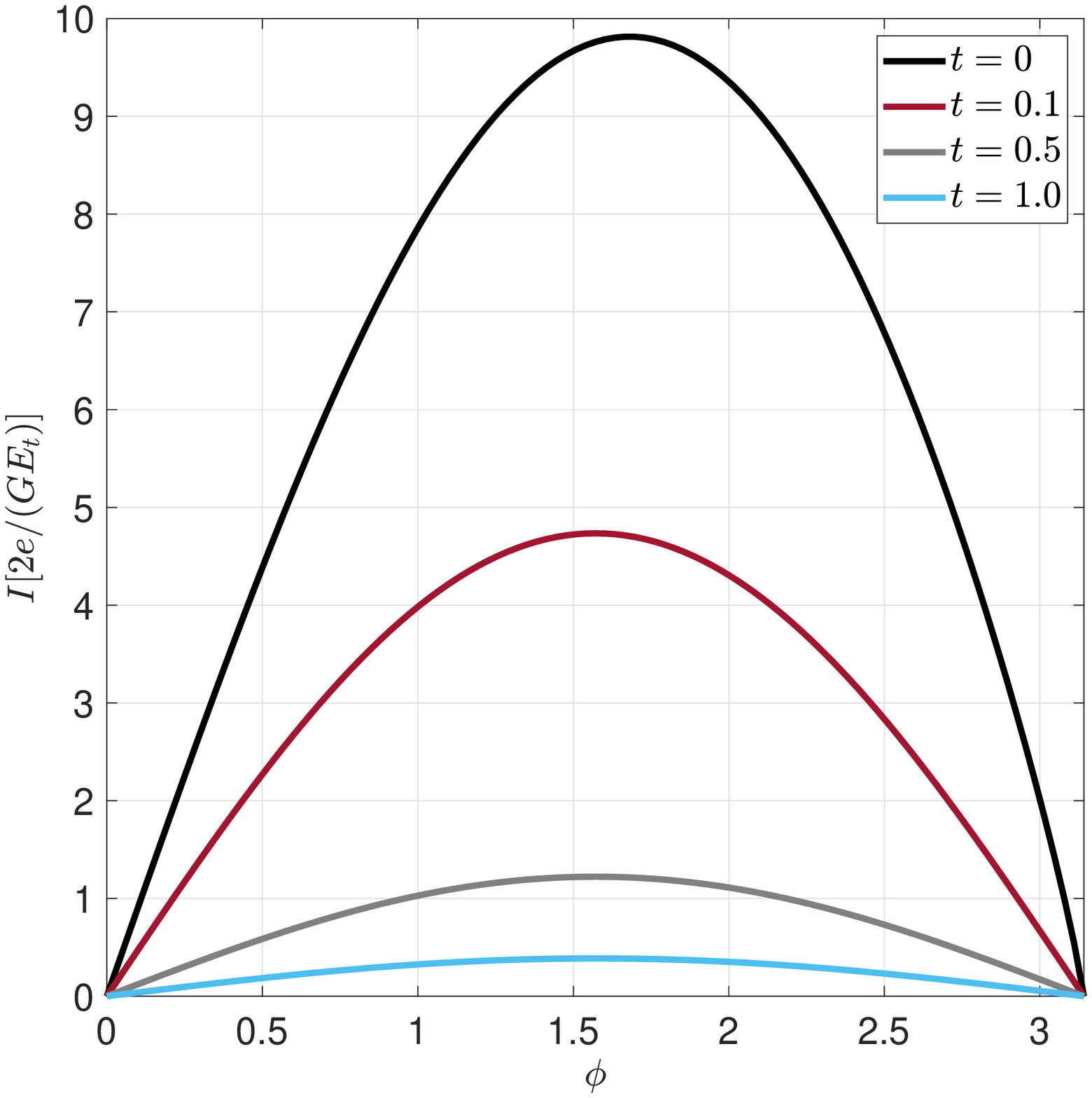}
\caption{The average current $I(\phi)$ at zero magnetic field as a function of the dwell energy $e_t = E_t / \Delta$ for various values of $t = T / \Delta$ and $\phi = \pi / 2$ on the left hand side, and as a function of $\phi$ for various $t$ and $e_t = 0.01$, on the right hand side.}
\label{Fig:J1}
\end{figure}

For the general case of arbitrary dwell times, we employ Eq.~\eqref{eq:J} to perform numerical calculations. We display the dependence of the average current, $I$, on $t$ and $e_t$ in Fig.~\ref{Fig:J1}.

As expected from Eq.~\eqref{eq:Jphilog}, $I$ grows monotonically as the temperature decreases, and the weak low-temperature singularity is cut-off for finite $e_t$. Likewise, $I$ grows with decreasing $e_t$, but the growth is limited for finite $T$. The dependence of $I$ on the phase difference $\phi$ is illustrated in Fig.~\ref{Fig:J1} for fixed $e_t$. At the lowest temperatures, the average current attains its maximum around $\phi = \pi / 2$. In the absence of a phase difference, $\phi = 0$, and at $\phi = \pi$ the average current vanishes. Overall, the $\phi$-dependence of the average current $I$ is dominated by the prefactor $\sin(\phi)$ in Eq.~\eqref{eq:J}.
The average Josephson current does not depend on the width of the junction. This is different for the sample-to-sample fluctuations, as we discuss next.

\subsection{Sample-to-sample fluctuations}

The calculation of the current fluctuations requires the knowledge of the eigenvalues $\lambda^{\pm}_{\epsilon_1,\epsilon_2}$ of the fluctuation determinant stated in Eq.~\eqref{eq:lambda}. For a vanishing magnetic field, the eigenvalues for Diffusons and Cooperons become identical and are given by
\begin{align}
\lambda_{\epsilon_1,\epsilon_2}
&
=
Dq^2
+
v(1)
+
v(2),
\end{align}
where $(i)=(\epsilon_i,\phi_i)$ is a convenient multi-index notation. 
At zero magnetic field, with the help of Eq.~\eqref{eq:K12} and the two-fold derivatives with respect to the phases in Eq.~\eqref{eq:F12}, we obtain the general formula for the current fluctuations,
\begin{align}
\label{eq:varI-general}
\mathrm{var}I(\phi)
&
=
(4eT)^2
\sum_{\epsilon_1, \epsilon_2 > 0}
\sum_q
\frac{\partial_{\phi_1}v(1)\partial_{\phi_2}v(2)}{[Dq^2 + v(1) + v(2)]^2}
.
\end{align}
Compared to the results reported in Refs.~\cite{Houzet-Skvortsov,Micklitz}, the variance in Eq.~\eqref{eq:varI-general} is four times smaller. This is due to the strong spin-orbit coupling in the topological insulator surface, which suppresses fluctuations in the spin triplet channel, while the singlet mode remains effective. We will discuss the current fluctuations in two limits, the quantum dot geometry, for which $E_{\rm Th}^{\perp}\gg E_t$, and the quasi-one-dimensional limit $E_{\rm Th}^{\perp}\ll E_t$.\\

\begin{figure}[!tb]
\centering
\includegraphics[scale=0.32]{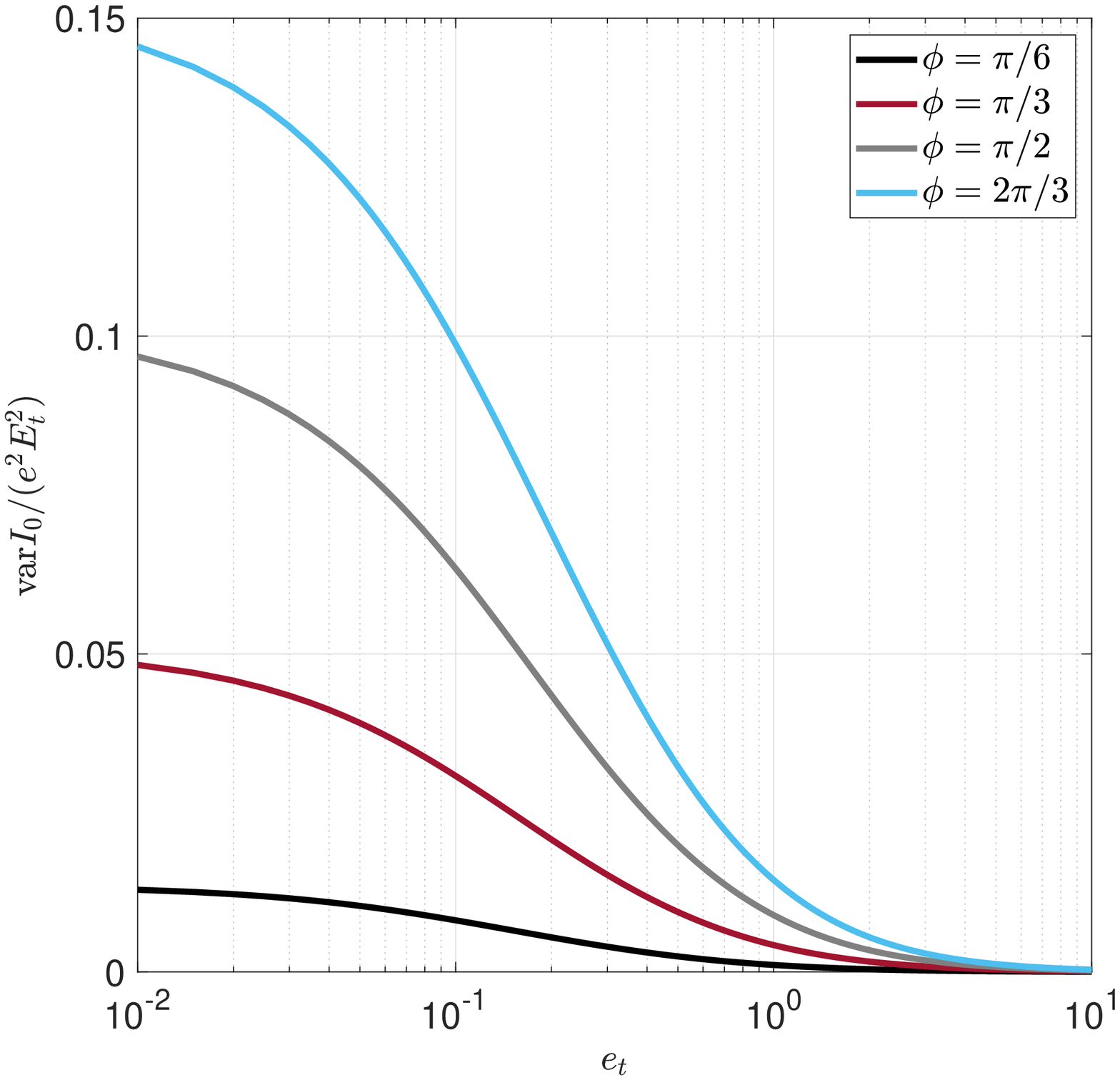}
\includegraphics[scale=0.32]{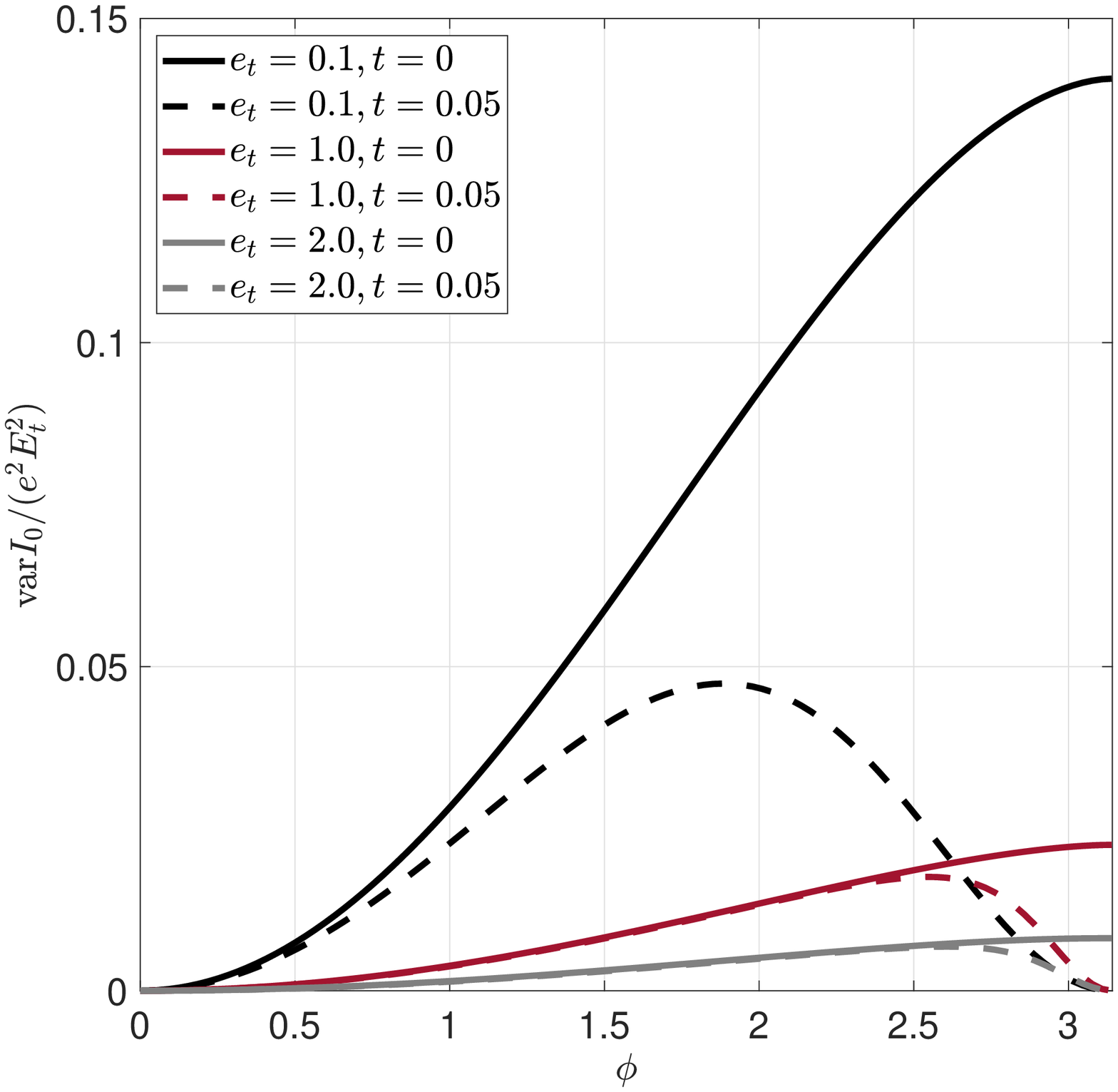}
\caption{On the right hand side, the variance of the Josephson current
as a function of the phase difference $\phi$ in the quantum dot geometry. Solid lines represent the zero temperature limit, whereas dashed lines denote the finite temperature limit. On the left hand side, we display $\mathcal K_0$ as a function of $e_t = E_t/\Delta$ for various fixed phases, $\phi = \pi /6, \pi / 3, \phi / 2, 2\pi / 3$.}
\label{fig:ABCDCurrFlucZeroMag}
\end{figure}

{\it Quantum dot limit, $E_{\rm Th}^{\perp}\gg E_t$:---}In the quantum dot geometry, spatial fluctuations of the Diffuson modes in the transverse direction can be neglected and the current fluctuations are given by~\cite{Houzet-Skvortsov}
\begin{align}
\label{eq:varI0}
&
\mathrm{var}I_0(\phi)
=
e^2E_t^2
\mathcal{K}_0(\phi)
,\qquad
\mathcal{K}_0(\phi)
=
\sin^2(\phi)
T^2
\sum_{\epsilon_1, \epsilon_2 > 0}
\frac
{
16E_t^2\Delta^4
}
{
\omega(\Delta, \epsilon_1)\omega(\Delta, \epsilon_2)v(1)v(2)[v(1) + v(2)]^2
},
\end{align}
where the sub-index in the sample-to-sample fluctuations denotes the effective dimensionality of the system. In Fig.~\ref{fig:ABCDCurrFlucZeroMag}, the variance $\mathrm{var}{I}_0$ is displayed as a function of the ratio $E_t/\Delta$.
\\

\begin{figure}[!tb]
\centering
\includegraphics[scale=0.32]{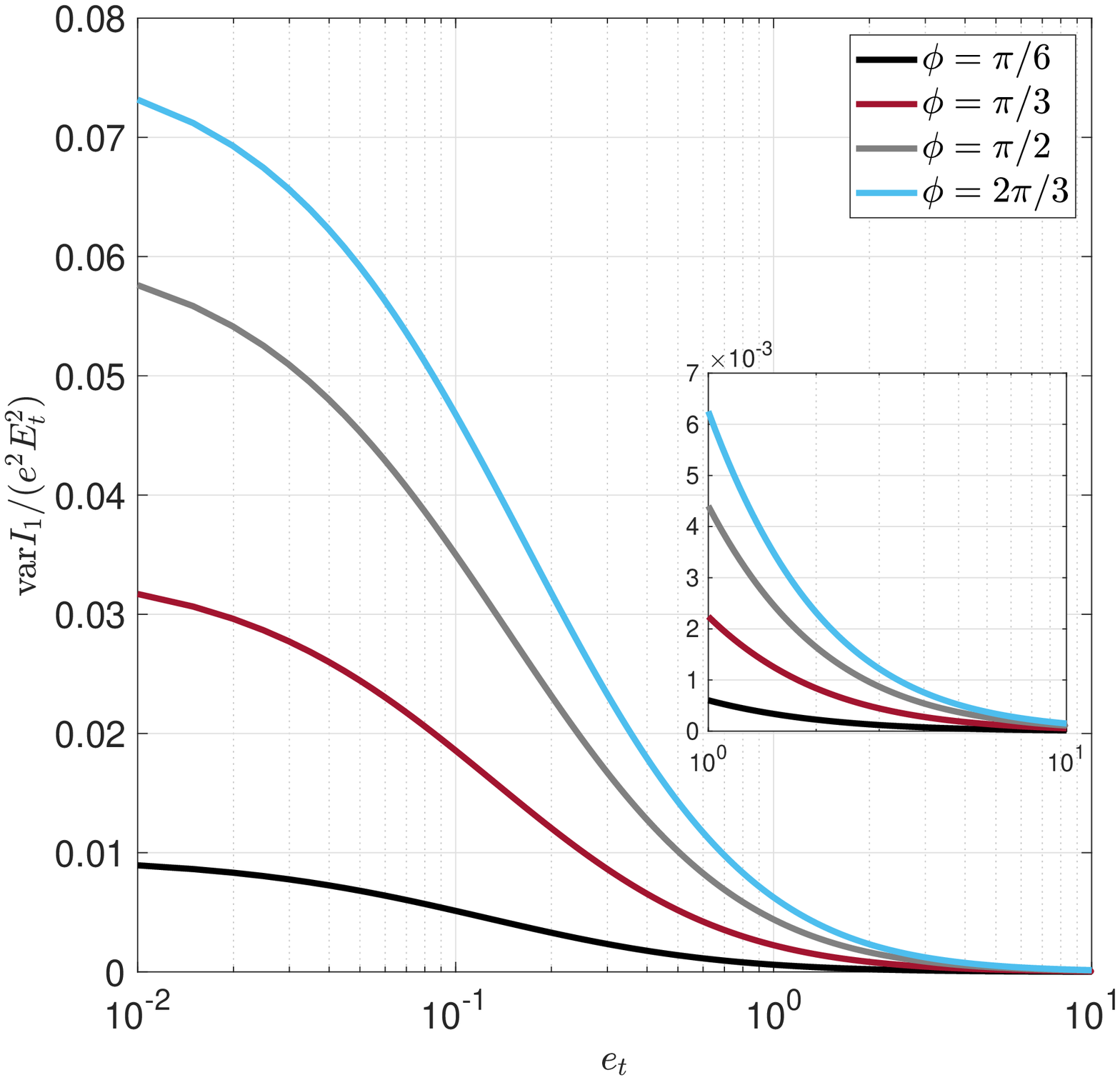}
\includegraphics[scale=0.32]{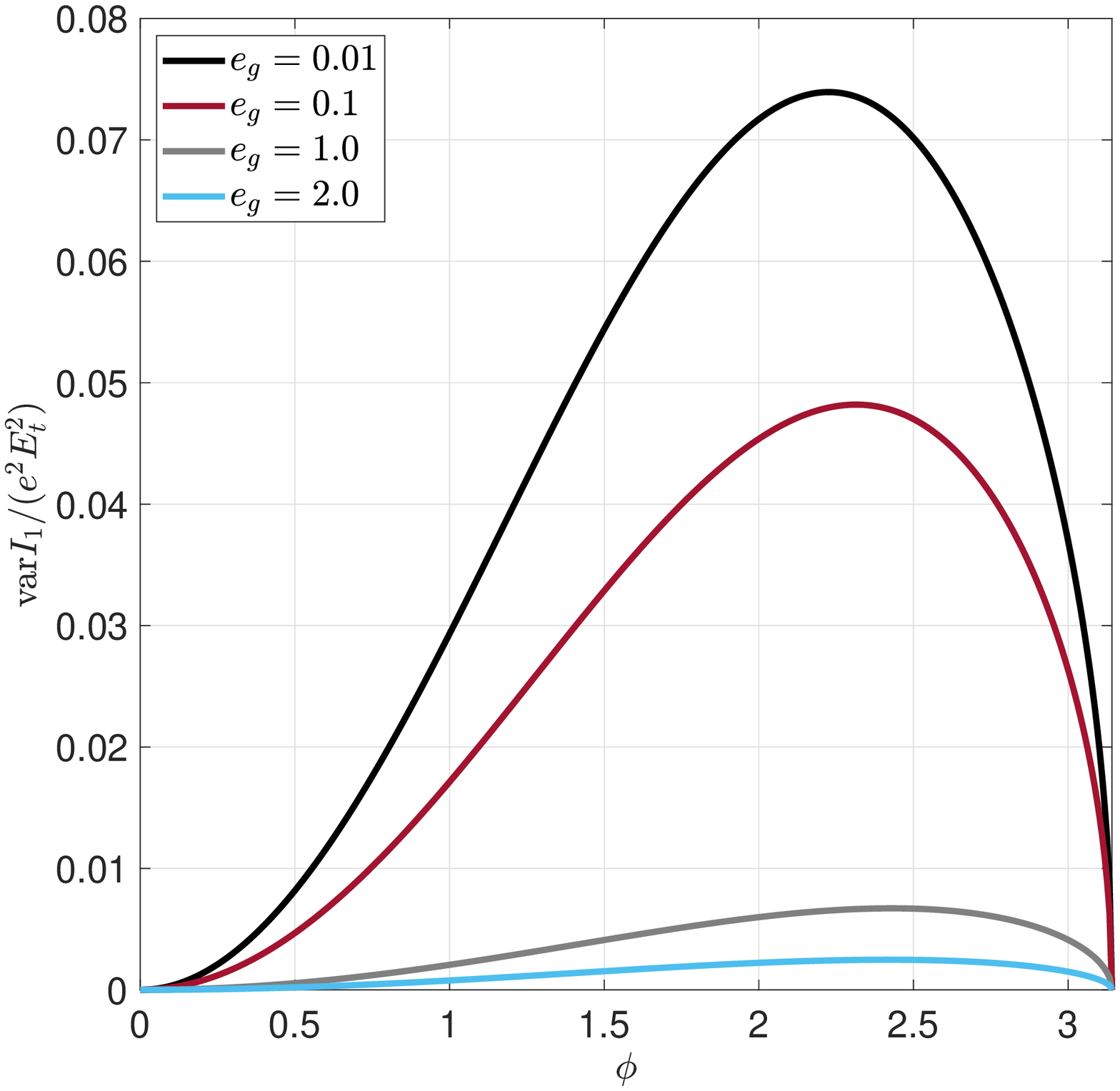}
\caption{On the right hand side, the variance of the Josephson current
as a function of the phase difference $\phi$ in the quasi-one-dimensional geometry. On the left hand side, we display $\mathcal K_1$ as a function of $e_t = E_t/\Delta$ for various fixed phases, $\phi = \pi / 6, \pi / 3, \phi / 2, 2\pi / 3$.}
\label{fig:ABCDCurrFlucZeroMagQ1D}
\end{figure}

{\it Quasi-one-dimensional limit, $E^{\perp}_{\rm Th}\ll E_t$:---}
In the quasi-one-dimensional geometry spatial fluctuations of the Diffuson modes in the transverse direction have to be taken into account. Employing the same equations as in the previous limit, Eq.~\eqref{eq:K12} and Eq.~\eqref{eq:F12}, and performing the sum over momenta $q$, we obtain the following expression for the variance of the Josephson current
\begin{align}
\label{eq:varI1}
\mathrm{var}I_1(\phi)
&
=
e^2E_t^2
\sqrt{\frac{{E}_t}{E^{\perp}_{\rm Th}}}
\mathcal{K}_1(\phi),\qquad
\mathcal{K}_1(\phi)
=
4\sin^2 (\phi) T^2
\sum_{\epsilon_1,\epsilon_2>0}
\frac
{
\Delta^4E_t^{3/2}
}
{
\omega(\Delta, \epsilon_1)\omega(\Delta, \epsilon_2)
v(1)v(2)[v(1)+v(2)]^{3/2}
}.
\end{align}
The plot for the current fluctuations $\mathrm{var}I_1(\phi)$ is shown in Fig.~\ref{fig:ABCDCurrFlucZeroMagQ1D}. Next, we move on to discuss the current fluctuations specifically in the limit of long and short dwell times.

\subsubsection{Long dwell time: $E_t\ll \Delta$}
{\it Quantum dot limit, $E^{\perp}_{\rm Th}\gg E_t$:---}In the long dwell time limit and at zero temperature, the scale for the variance of the current is set by $E_t^2$, and we obtain an analytical expression
\begin{align}
&
\label{eq:longdwell-varI0}
\mathrm{var}I_0(\phi)
=
e^2E_t^2
\mathcal{K}_0(\phi)
,\qquad
\mathcal{K}_0(\phi)
=
\frac{\sin ^2(\phi )}{\pi^2}
\iint_0^\infty
\frac
{
dx_1 dx_2
}
{
\sqrt
{
X_1(\phi)X_2(\phi)
}
\left(
\sqrt
{
X_1(\phi)
}
+
\sqrt
{
X_2(\phi)
}
\right)^2
}
,
\end{align}
where $X_i(\phi)=\cos^2(\phi/2) + x_i^2$ and $x_i = \epsilon_i / \Delta$. Investigating the behavior of the function $\mathcal K_0$, we observe that its dependence on the phase difference $\phi$ can be described by a simple power-law in $\cos\phi/2$, $\mathcal K_0\approx \sin^2(\phi)[\cos(\phi/2)]^{-2}$. At $\phi = \pi$, we expect that both the average current and the current fluctuations vanish. However, it is clear that $\mathcal{K}_0$ does not reproduce this behavior as the phase $\phi$ approaches $\pi$.
Such failure has to do with the violation of the criterion of validity for our Gaussian approximation, whose existence hinges on the small parameter $\lambda / \delta\gg 1$, $\lambda$ being an eigenvalue of the Gaussian action and $\delta$ the mean level spacing. In this approximation, the mass of the system is proportional to $\cos(\phi/2)$ and as a consequence when the phase becomes close to $\pi$ the criterion of validity for our approximation is no longer satisfied.
A more detailed analysis of the action reveals the correct result in this limit~\cite{Micklitz}. As already suggested in Ref.~\cite{Houzet-Skvortsov}, the vanishing of the average current and the current fluctuations for $\phi\rightarrow \pi$ is restored at finite temperatures even in the Gaussian approximation, as can be seen in Fig.~\ref{fig:ABCDCurrFlucZeroMag}.

{\it Quasi-one-dimensional limit, $E^{\perp}_{\rm Th}\ll E_t$:---}Focusing on the zero-temperature limit, we  transform summations over Matsubara frequencies into integrations again, and express the latter in terms of dimensionless quantities to find 
\begin{align}
\label{eq:longdwell-varI1}
\mathrm{var}I_1(\phi)
=
e^2E_t^2
\sqrt{\frac{{E}_t}{E^{\perp}_{\rm Th}}}
\mathcal K_1(\phi)
,
\qquad
\mathcal{K}_1(\phi)
&=
\frac{\sqrt{2}\sin^2 (\phi)}{4\pi^2}
\iint_0^\infty
\frac{dx_1dx_2}
{\sqrt{X_1(\phi)X_2(\phi)}\left(\sqrt{X_1(\phi)}+\sqrt{X_2(\phi)}\right)^{\frac{3}{2}}}.
\end{align}
The scale of the fluctuations is now set not only by the squared dwell energy but also by the parameter $\sqrt{E_t / E_{\mathrm{Th}}^\perp}$. The result of the integrations in $x_1$ and $x_2$ can be approximated by a power law in $\cos\phi/2$ and, as a consequence, the phase dependence of $\mathcal{K}_1(\phi)$ is governed by the function
$\sin^2(\phi)[\cos(\phi/2)]^{-3/2}$, which monotonically vanishes as $\phi$ approaches $\pi$. The presence of a momentum structure in the Gaussian action leads to this significant difference in comparison to the quantum dot geometry, for which finite temperatures had to be invoked in order to reproduce this behavior in the Gaussian approximation.

\subsubsection{Short dwell time: $E_t\gg \Delta$}
{\it Quantum dot limit, $E^{\perp}_{\rm Th}\gg E_t$:---}At zero temperature, the current fluctuations read as~\cite{Beenakker93}
\begin{align}
\label{eq:shortdwell-varI0}
&
\mathrm{var}I_0(\phi)
=
e^2E_t^2
\mathcal{K}_0(\phi)
,
\quad
\mathcal{K}_0(\phi)
=
\frac{\Delta^2}{E_t^2}
\mathcal{K}^S_0(\phi),
\\
&
\mathcal{K}^S_0(\phi)
=
\frac{\sin^2(\phi)}{4\pi^2}
\iint_0^\infty
\frac{\sqrt{X_1(0)}\sqrt{X_2(0)}dx_1
dx_2}
{\sqrt{X_1(\phi)}\sqrt{X_2(\phi)}
\left[
\sqrt{X_2(0)}
\sqrt
{
X_1(\phi)
}
+
\sqrt{X_1(0)}
\sqrt
{
X_2(\phi)
}
\right]^2}
.
\end{align}
A quick inspection of this expression reveals that in this regime the scale is now set by $\Delta^2$. In this limit, when $\phi$ approaches $\pi$ the product between $\sin^2(\phi)$ and the dimensionless function $\mathcal K_0^S$ yields
a non-zero result, which clearly violates the condition $\mathrm{var}I_0(\pi) = 0$. As already found in the long dwell time limit, finite temperatures restore the correct behavior in our formalism, see Fig. \ref{fig:ABCDCurrFlucZeroMag}.

\begin{figure}[H]
\centering
\includegraphics[scale=0.32]{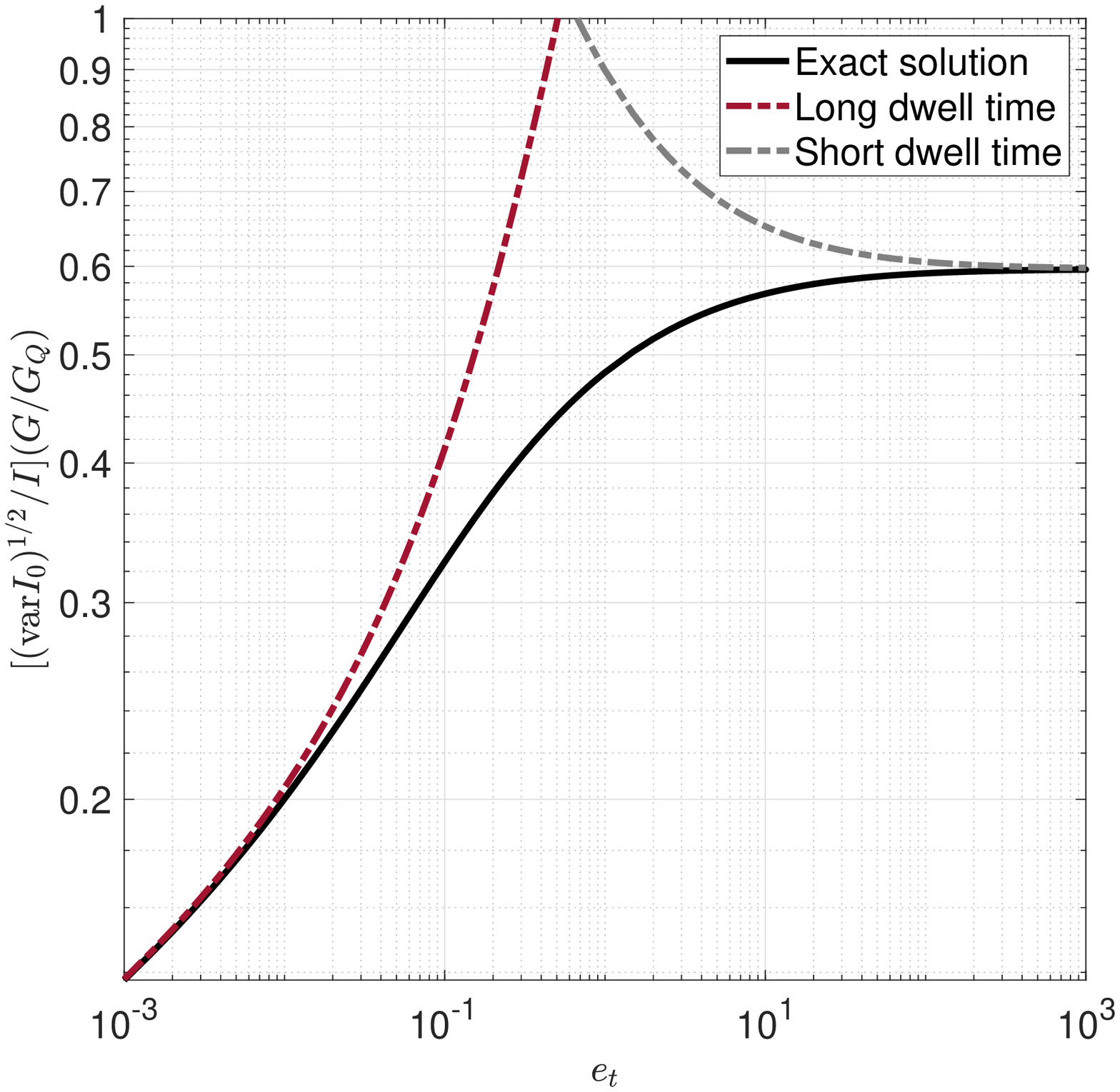}
\includegraphics[scale=0.32]{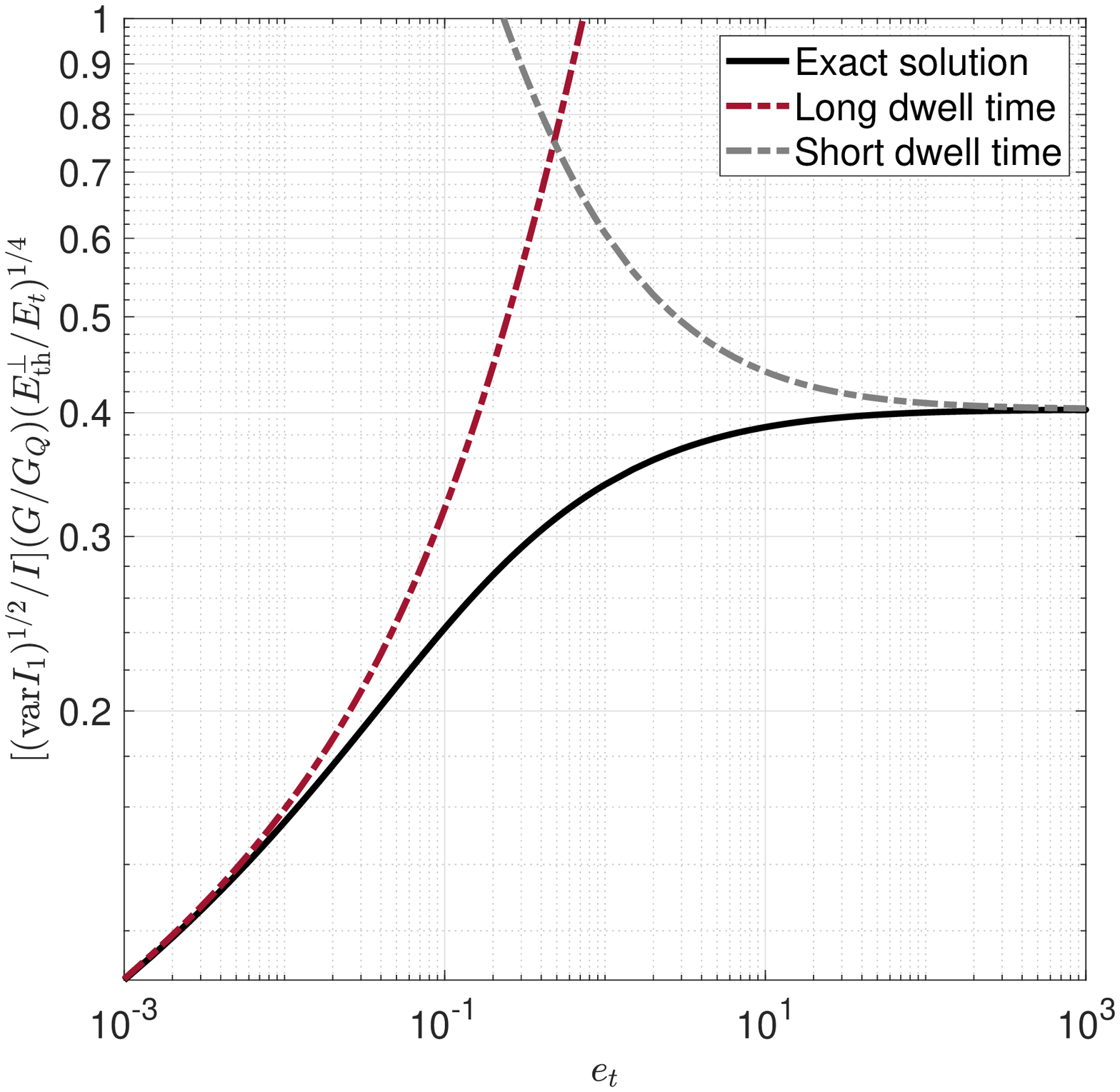}
\caption{The ratio between the current fluctuations and the average current as a function of $e_t$ in the absence of a magnetic field and at zero temperature. On the left hand side, we show this ratio for the quantum dot geometry, on the right hand side for the quasi-one-dimensional case.}
\label{fig:ZM-Ratio-varI-CPR}
\end{figure}

{\it Quasi-one-dimensional limit, $E^{\perp}_{\rm Th}\ll E_t$:---}Considering the zero temperature limit, the current fluctuations yield
\begin{align}
\label{eq:shortdwell-varI1}
&
\mathrm{var}I_1(\phi)
=
e^2E_t^2
\sqrt{\frac{{E}_t}{E^{\perp}_{\rm th}}}
\mathcal K_1(\phi)
,
\quad
\mathcal K_1(\phi)
=
\frac{\Delta^2}{E_t^2}
\mathcal K^S_1(\phi),
\\
&
\mathcal K^S_1(\phi)
=
\frac{\sqrt{2}\sin^2(\phi)}{16\pi^2}
\iint_0^\infty
\frac
{[X_1(0)]^{1/4}[X_2(0)]^{1/4}dx_1
dx_2}
{
\sqrt{X_1(\phi)}
\sqrt{X_2(\phi)}
\left(
\sqrt{X_2(0)}\sqrt{X_1(\phi)}
+
\sqrt{X_1(0)}\sqrt{X_2(\phi)}
\right)^{3/2}
}
.
\end{align}
In analogy to the 
long dwell-time limit, in a quasi-one-dimensional geometry the scale is set by $\Delta^2$, and also by the parameter $\sqrt{E_t / E^\perp_{\mathrm{Th}}}$. In addition to that, as in the previous cases, we find that the quasi-one-dimensional geometry restores the correct result for the fluctuations at $\phi = \pi$, $\mathrm{var}I_1(\pi) = 0$, see details in Fig. \ref{fig:ABCDCurrFlucZeroMagQ1D}.

\subsubsection{Arbitrary dwell time}
We can now compare the magnitudes of fluctuations and the average current for the quantum dot and the quasi-one-dimensional geometry. For the quantum dot geometry, we find
\begin{equation}
\label{eq:ratio-QD}
\frac{[\mathrm{var}I_0(\phi)]^{1/2}}{I(\phi)}
=
\frac{G_Q}{G}
\frac{[4\pi^2\mathcal K_0(\phi)]^{1/2}}{J(\phi)}
.
\end{equation}
In the quasi-one-dimensional geometry, we obtain the following expression
\begin{equation}
\label{eq:ratio-1D}
\frac{[\mathrm{var}I_1(\phi)]^{1/2}}{I(\phi)}
=
\frac{G_Q}{G}
\left(\frac{E_t}{E^\perp_{\mathrm{Th}}}\right)^{1/4}
\frac{[4\pi^2\mathcal K_1(\phi)]^{1/2}}{J(\phi)}
.
\end{equation}
With the help of Eqs.~\eqref{eq:Jphilog},~\eqref{eq:Jshort},~\eqref{eq:longdwell-varI0},~\eqref{eq:longdwell-varI1},~\eqref{eq:shortdwell-varI0} and~\eqref{eq:shortdwell-varI1}, we can estimate that the ratios in Eqs.~\eqref{eq:ratio-QD} and~\eqref{eq:ratio-1D} are of the order of $G_Q/ G$, where $G_Q = e^2 / \pi$ is the conductance quantum. Furthermore, as a result of the hierarchy of energy scales, $\delta \ll E_t\ll \Delta\ll E_{\rm th}$, for a quasi-one-dimensional system, the ratio is proportional to the parameter $E_t/E^\perp_{\mathrm{Th}}$. As we observe in
Fig.~\ref{fig:ZM-Ratio-varI-CPR}, for both geometries the approximate analytical results obtained in this section are in good agreement with numerical results. In the long dwell-time limit, the fluctuation-to-average current ratio behaves as $1/\log(1/e_t)$ and in the short dwell-time limit the dwell energy dependence is completely absent, hence the ratio tends to a constant value. Next we discuss how these findings are changed in presence of a magnetic field.


\section{Average current and sample-to-sample fluctuations at finite magnetic field}\label{sec:Josephson-Fluctuations-FiniteB}

As previously discussed, we continue to explore the weak coupling regime $E_t \ll \Delta$ 
where the mini-gap is set by the dwell energy $E_t$.  The magnetic field then allows to tune the population of sub-gap states,
with mini-gap closure at $E_\Phi \sim E_t$,
while pair-breaking effects on the superconducting leads can be  neglected. 
 We focus on the sensitivity of the average Josephson current and its fluctuations to the mini-gap closure at strong magnetic fields, where an analytical solution of the mean field equation is available. These analytical calculations are complemented by calculations building on the numerical solution of the mean field equation and allowing to describe the crossover into the weak magnetic field regime.

\subsection{Average current}
\begin{figure}[H]
\centering
\includegraphics[scale=0.23]{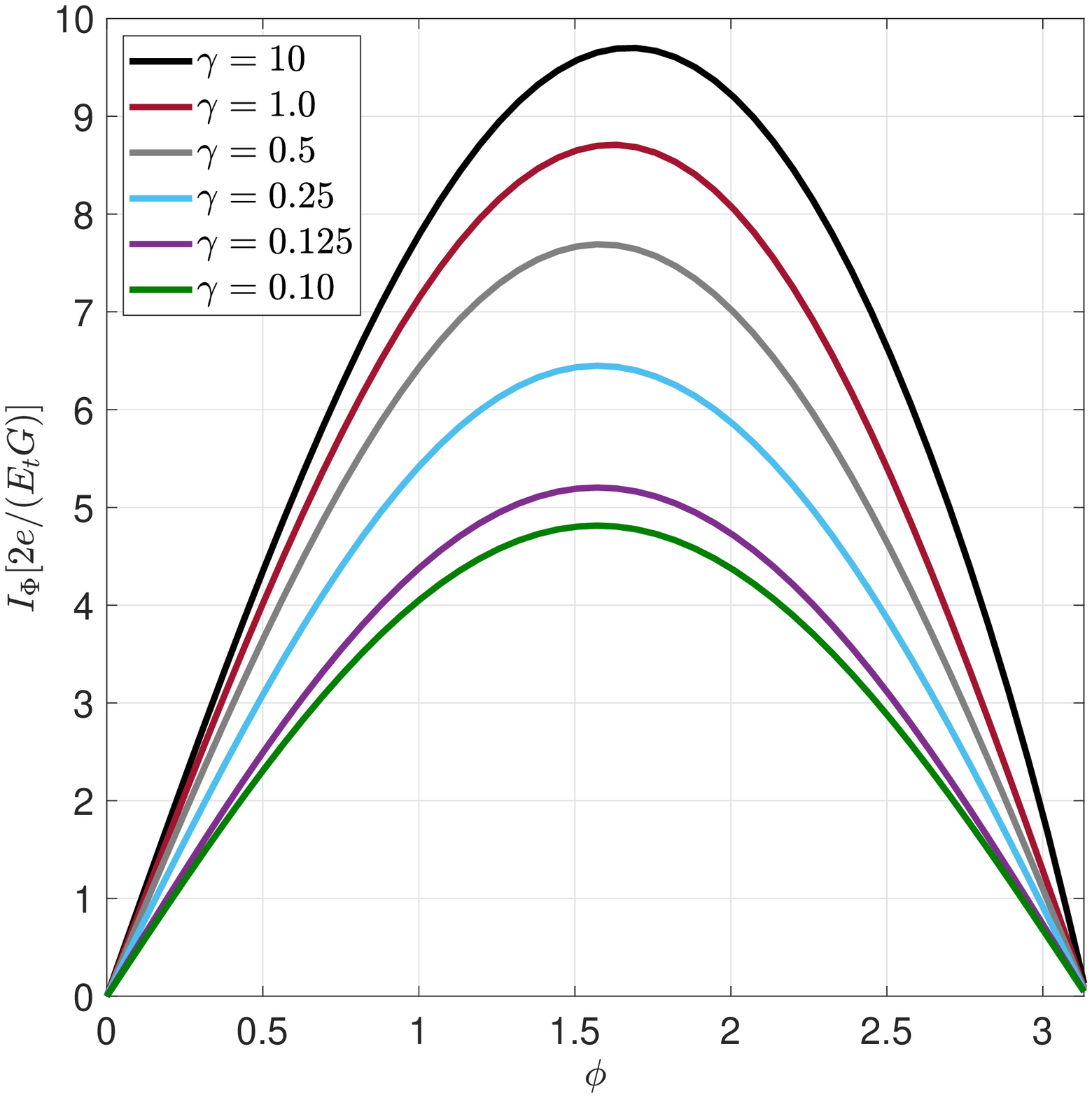}
\includegraphics[scale=0.23]{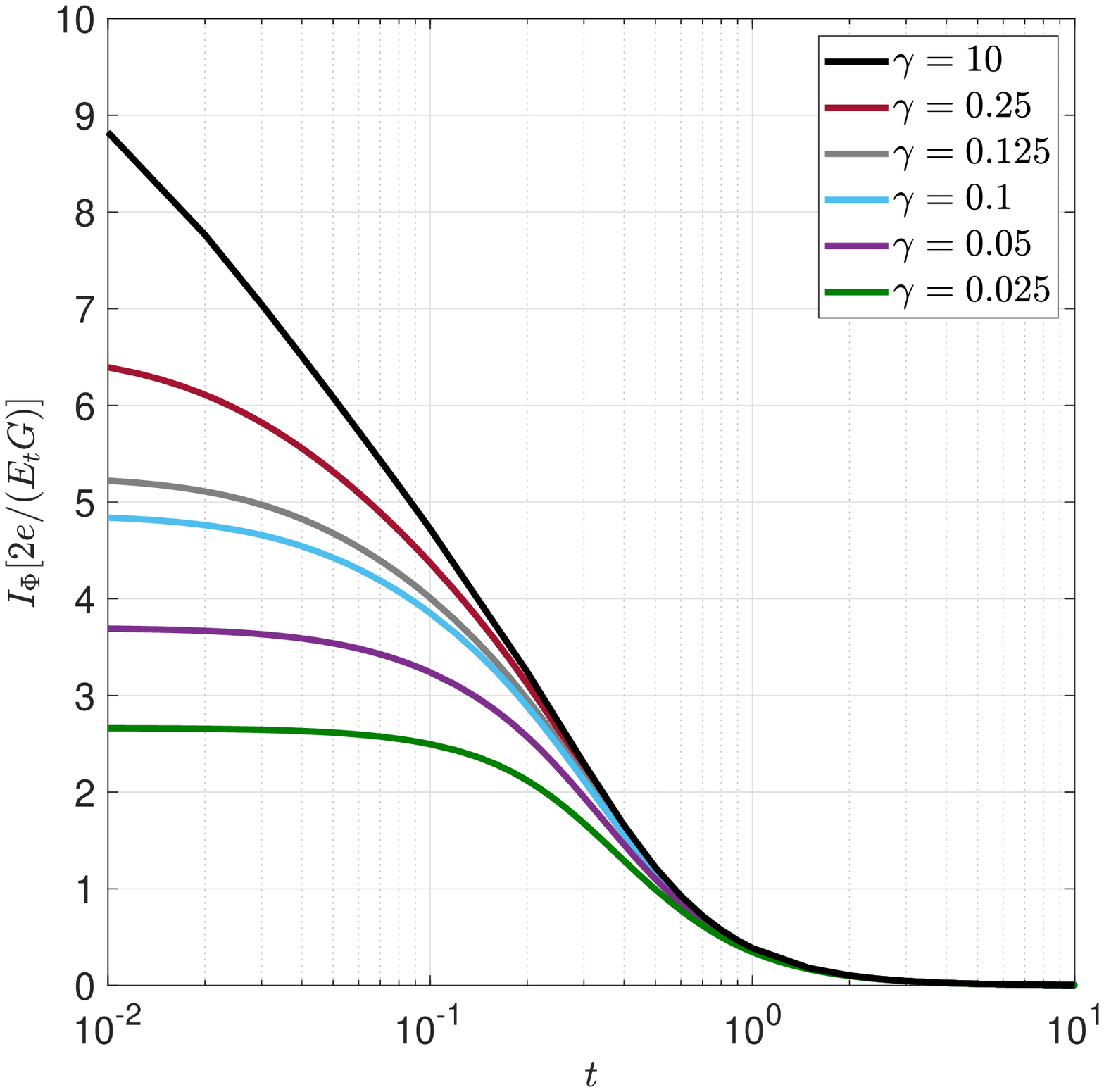}
\includegraphics[scale=0.23]{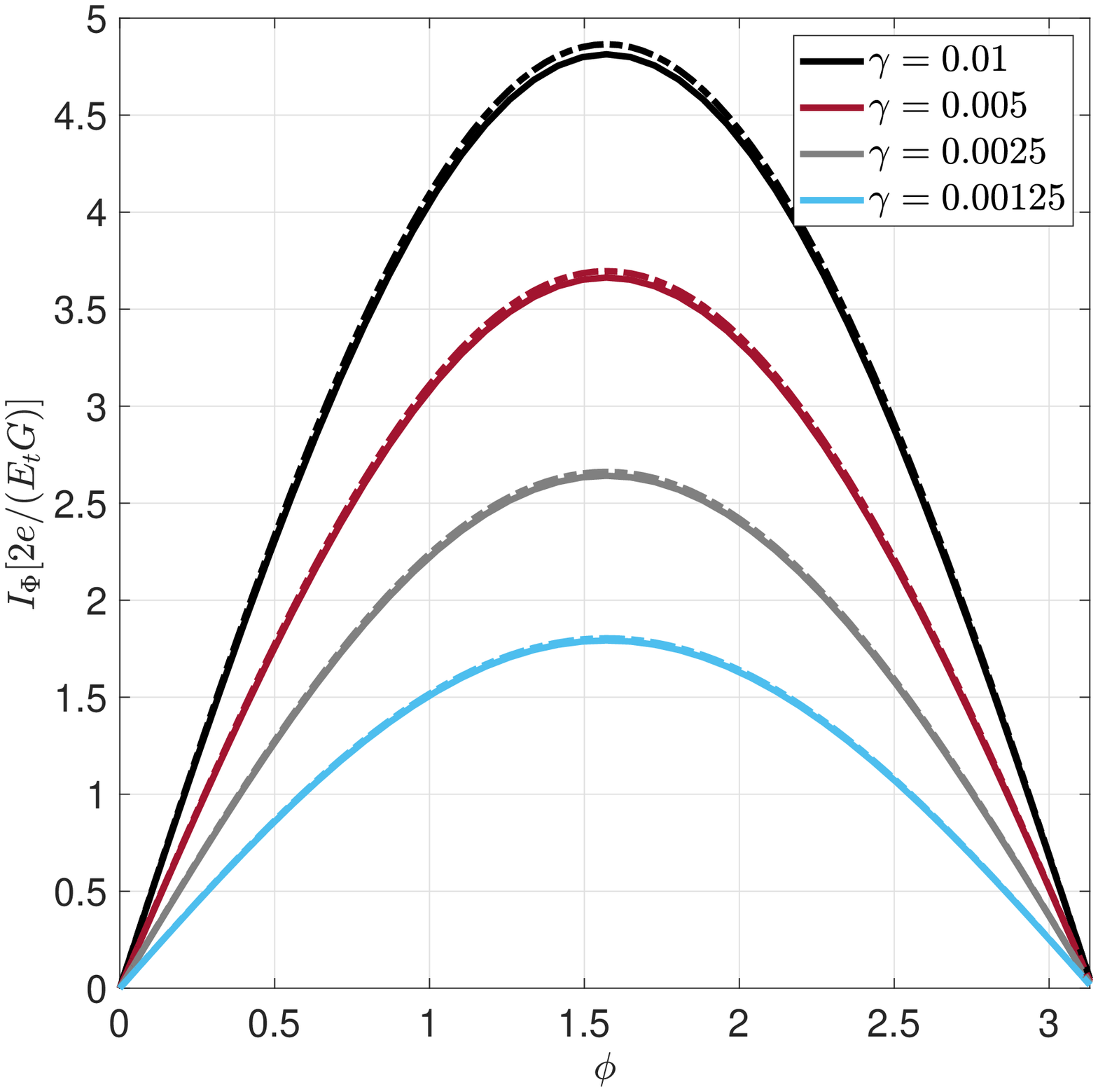}
\caption{Left panel: $J_\Phi$ at zero temperature as a function of $\phi$ and for various values of $\gamma= E_t/E_\Phi$. Middle panel: $J_\Phi$ as a function of temperature $t=T/\Delta$ for various values of  $\gamma$, cf. Eq.~\eqref{eq:Jphi}. Here the dwell energy is chosen as $e_t =E_t/\Delta= 1/100$ and the phase difference as $\phi = \pi/2$. Right panel: The average current at zero temperature for various values of $\gamma = E_t / E_\Phi$, where we fixed $e_t = E_t/\Delta=1/1000$. The solid lines represent the exact numerical solution and dash-dotted lines the analytical approximation.}
\label{fig:JPhiMag}
\end{figure}

From the mean field solution
$Q_\Delta =\hat m_i\hat{\sigma}_i$,
with $\hat m_i$ in the limit $E_\Phi \gg E_t$, 
 and Eq.~\eqref{eq:Iaverage}
we find 
the average Josephson current at
strong magnetic fields
\begin{align}
&
I_\Phi(\phi)
=
\frac{GE_t}{2e}
J_\Phi,\qquad 
J_\Phi
=
4\pi
\sin(\phi)
T
\sum_{\epsilon > 0}
\frac
{
\Delta^2
}
{
(\epsilon + E_\Phi)\Delta^2
+
\epsilon
(
\epsilon^2 + 2E_t\sqrt{\Delta^2 + \epsilon^2} + \epsilon E_\Phi
)
}
.\label{eq:Jphi}
\end{align}
Using that $E_t\ll\Delta$, we can neglect terms involving the dwell energy in  $J_\Phi$, and perform the summation  arriving at an expression for the average Josephson current in terms of polygamma functions, see \ref{app:average-current} for details.
The result is shown in Fig.~\ref{fig:JPhiMag}.

The scale for the current is set by $G E_t / ( 2 e )$, 
similar to the zero magnetic field case $B=0$. In contrast 
to the latter, 
the phase dependence of $J_\Phi$ in the strong magnetic field limit is, however, fully governed by the sine function, Eq.~\eqref{eq:Jphi}. Technically, 
corrections to the mean field solution  Eq.~\eqref{eq:SM-mean-field2}
are suppressed in $E_t / E_\Phi\ll 1$ giving only  insignificant contributions, and deviations from a sinusoidal behavior are therefore strongly suppressed. As evident from Eq.~\eqref{eq:Jphi}, increasing the external magnetic field  monotonically suppresses the average Josephson current. At low temperatures $T\ll \Delta$,
the dimensionless function $J_\Phi$  shares the logarithmic asymptotic form
of the zero magnetic field expression
\begin{align}
J_\Phi=2\sin(\phi)\ln\left[\frac{1}{{\rm max}(t, e_\Phi)}\right],
\label{J_log_t_e_phi}
\end{align}
where now $E_\Phi\gg E_t$ replaces the dwell energy $E_t$ found at $B=0$.

From the numerical solution of the mean field equation, we can calculate the average current for arbitrary ratios of   $E_t / E_\Phi$. The result is shown 
 in Fig.~\ref{fig:JPhiMag}. The average current as a function of the phase (left panel) shows a dominant sinusoidal behavior for all ratios $E_t/E_\Phi$, attaining its maximum at $\pi / 2$ in the strong magnetic field limit, which is slightly shifted to larger values with increasing ratio $E_t / E_\Phi$. The phase-dependence of the current does not show any signs of Fraunhofer patterns, in agreement with the discussions in  references~\cite{Hammer},~\cite{Bergeret} and~\cite{Barone}. The average current as a function of temperature is shown in the middle panel of Fig. \ref{fig:JPhiMag}. Since the weak logarithmic divergence of Eq.~\eqref{J_log_t_e_phi}  is cut off by the larger of $T$ and $E_\Phi$, the average current at low temperatures $T\ll \Delta$
 decreases with increasing magnetic field, and all curves for different values  $E_t/E_\Phi$ then collapse into a single curve at high temperatures $T\gg \Delta$.

 Finally, we compare in right panel of Fig.~\ref{fig:JPhiMag} 
 the average current from the analytical mean field solution at strong magnetic fields to the exact current obtained from the numerical solution of the mean field equation, here at zero temperature and $e_t = E_t/\Delta=1/1000$.
 As expected, the analytical solution describes the average current very well for these small values  $E_t/E_\Phi\leq 0.01$. 
 
 \begin{figure}[H]
\centering
\includegraphics[scale=0.32]{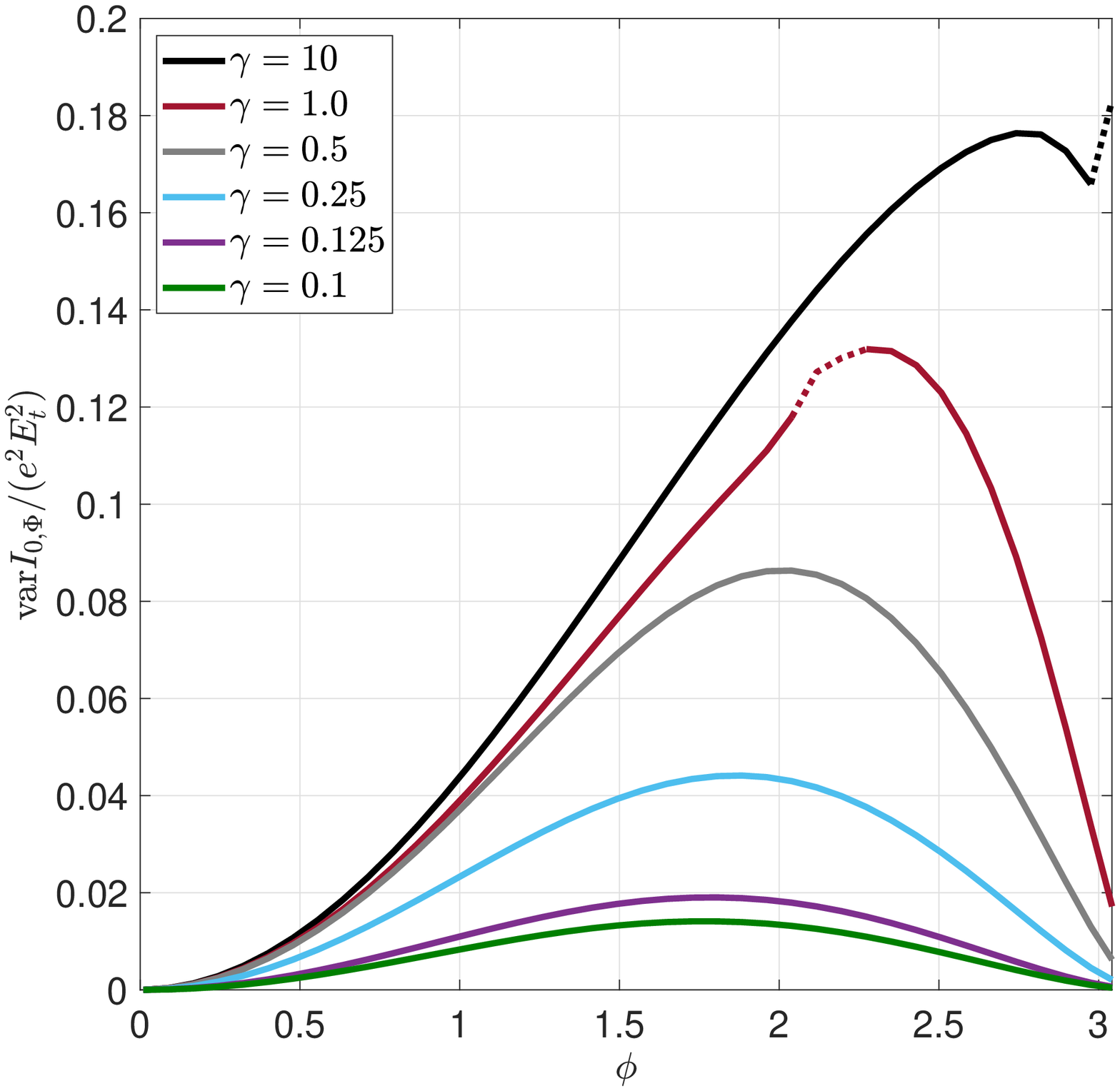}
\includegraphics[scale=0.32]{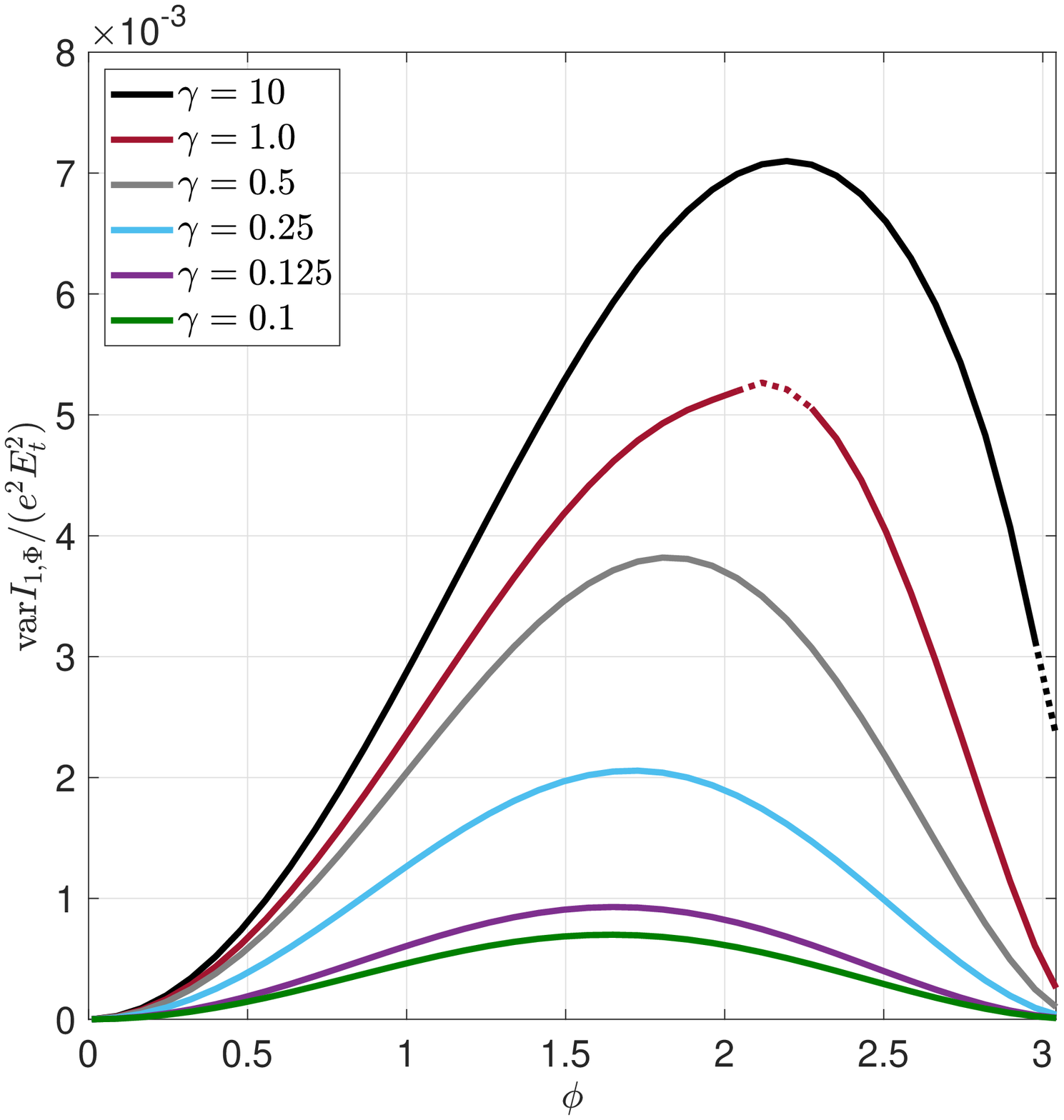}
\caption{Left panel: $\mathcal K_{0, \Phi}$  (quantum dot) as a function of phase angle $\phi$ for various values of $\gamma = E_t/E_\Phi$. Right panel: $\mathcal K_{1, \Phi}$ (quasi-one-dimensional geometry) as a function of phase difference $\phi$ for various values of $\gamma = E_t/E_\Phi$. The dashed lines indicate the parameter region for which the semiclassical approximation becomes uncontrolled. In all figures we fixed $e_t = E_t/\Delta = 1/100$ and varied $E_\Phi$.}
\label{fig:Kphi}
\end{figure}

\subsection{Sample-to-sample fluctuations}

To prepare the calculation of current fluctuations, we first notice that eigenvalues of Diffuson and Cooperon modes (X=D,C)
at strong magnetic fields become
\begin{align}
&
\lambda^{X, \pm}_{\epsilon_1, -\epsilon_2}
=
Dq^2
+
\epsilon_1
+
\epsilon_2
+
M_X^\pm(\epsilon_1, \epsilon_2),
\end{align}
with Diffuson masses $M_D^s$ 
\begin{align}
&
M_D^\pm(\epsilon_1, \epsilon_2)
=
2E_t^2\Delta^2\cos^2\left(\frac{\phi}{2}\right)
\left[
\sum_{i = 1}^2
\frac
{
1
}
{
(E_\Phi + \epsilon_i)
\sqrt{\omega(\Delta, \epsilon_i)}
}
\mp
\frac
{2E_\Phi}
{
\sqrt{(E_\Phi + \epsilon_1)(E_\Phi + \epsilon_2)\omega(\Delta, \epsilon_1)\omega(\Delta, \epsilon_2)}
}
\right],
\end{align}
and Cooperon masses $M_C^\pm=M_D^\pm+E_\Phi/2$. Notice that the magnetic field lifts previous degeneracies at $B=0$, 
and all four modes $(D/C,\pm)$ now  
contribute differently to the current fluctuations. Then, starting out from the general expression for current fluctuations
\begin{align}
\mathrm{var} I_\Phi(\phi)
&=
(2eT)^2
\sum_{s=\pm}
\left[
F^{s}_2(\phi)
-
F^{s}_1(\phi)
\right]
,
\end{align}
we employ that in the limit of strong magnetic fields the functions $F_1^s$ and $F_2^s$ are given by
\begin{align}
&
F^s_1(\phi)
=
\sum_{X = D,C}
\sum_{\epsilon_1, \epsilon_2 > 0}
\sum_q
\frac
{
\partial_1
[
M_X^s(\epsilon_1, \epsilon_2)
]
\partial_2
[
M_X^s(\epsilon_1, \epsilon_2)
]
}
{
[
Dq^2
+
\epsilon_1
+
\epsilon_2
+
M_X^s(\epsilon_1, \epsilon_2)
]^2
}
,
\\
&
F^s_2(\phi)
=
\sum_{X = D,C}
\sum_{\epsilon_1, \epsilon_2 > 0}
\sum_q
\frac
{
\partial^2_{12}M_X^s(\epsilon_1, \epsilon_2)
}
{
Dq^2
+
\epsilon_1
+
\epsilon_2
+
M_X^s(\epsilon_1, \epsilon_2)
}
.
\end{align}
We next explore these general expression for the 
two geometries of interest, that is, the quantum dot and quasi-one-dimensional structure, defined by  $E^\perp_{\rm Th}\gg E_t$ and $E^\perp_{\rm Th}\ll E_t$, respectively.

\subsubsection{Quantum dot limit: $E^\perp_{\rm Th}\gg E_t$} 

Current fluctuations for the quantum dot geometry in the zero temperature limit can be simplified to
\begin{align}
\label{eq:varI0Phi}
&
\mathrm{var}I_{0, \Phi}(\phi)
=
e^2E_t^2
\mathcal{K}_{0, \Phi}\left(\phi\right)
,
\quad
\mathcal{K}_{0, \Phi}(\phi)
=
\left(
\frac{E_t}{E_\Phi}
\right)^2
\mathcal{F}_{0, \Phi}(\phi),
\\
&
\mathcal{F}_{0, \Phi}\left(\phi\right)
=
\frac{\sin^2\left(\phi\right)}{\pi^2}
\sum_{X = C/D}
\sum_{s = \pm 1}
\left[
f_{0, 1}^{X, s}(\phi, \gamma)
+
f_{0,2}^{X,s}(\phi, \gamma)
\right]
,
\end{align}
with functions $f_{0, 1}$ and $f_{0, 2}$ defined as
\begin{align}
&
f_{0,1}^{X, 1}(\phi, \gamma)
+
f_{0,1}^{X, -1}(\phi, \gamma)
=
2
\iint_0^\infty
dxdx'
\frac
{
\omega(1, e_\Phi x)\omega(1, e_\Phi x^\prime)
}
{
\Omega_X^+(x, x^\prime)
\Omega_X^-(x, x^\prime)
}
,
\\
&
f_{0,2}^{X,s}(\phi, \gamma)
=
\iint_0^\infty
dxdx'
\frac
{
\eta_X^s(x, x^\prime)
}
{
\left[
\Omega_X^s(x, x^\prime)
\right]^2
}
.
\end{align}
To write the equations in a compact manner, we used $\omega(\Delta, \epsilon) = \Delta^2 + \epsilon^2$ and introduced
\begin{align*}
&
\eta_X^s(x_1, x_2)
=
\prod_{i\neq j, i, j = 1}^2
\left[\omega(1, e_\Phi x_i) - s b_X(x_i) (1 + x_j) \sqrt{\omega(1, e_\Phi x_i)\omega(1, e_\Phi x_j)}\right]
\\
&
\Omega_{X}^s(x_1, x_2)
=
\left(
a_X + x_1 + x_2
\right)
\prod_{i = 1}^2
(1 + x_i)
\omega(1, e_\Phi x_i)
+
2
\gamma^2
\cos^2\left(\frac{\phi}{2}\right)
\times\nonumber\\
&\qquad \qquad \qquad \times
\left[
\sum_{i = 1}^2(1 + x_i)\omega(1, e_\Phi x_i)
-
2s
\sqrt{\omega(1, e_\Phi x_1)\omega(1, e_\Phi x_2)}
\right]
.
\end{align*}
Here, the numerical constant $a_X$ is zero for diffusons and $1/2$ for Cooperons, and $b_D(\epsilon) = 1$ for diffusons, respectively,  
$b_C(\epsilon) = (1/2)[1 + \epsilon/(1 + \epsilon)]$
for Cooperons.

While fluctuations in the absence of magnetic fields are set by the (squared) dwell energy, they are suppressed by the additional factor $(E_t / E_\Phi)^2$ in the strong magnetic field limit. The left panel of Fig.~\ref{fig:Kphi} shows the current fluctuations  $\mathrm{var}I_{\Phi,0}$ as a function of $\phi$ for 
different values $\gamma=E_t/E_\Phi$. The increase of fluctuations with $\gamma$ is clearly visible and 
we also observe a shift of the maximum from close to $\pi$ at weak magnetic fields to smaller values as the magnetic field increases. We caution again that the semiclassical approximation loses its validity once the 
action takes values ${\cal O}(1)$. The corresponding regions are  close to the maximum value of fluctuations and indicated by the dashed lines. For $E_t/E_\Phi\gtrsim 1$, the action  becomes large enough to justify the semiclassical approximation for all values of $\phi$.

\begin{figure}[H]
\centering
\includegraphics[scale=0.32]{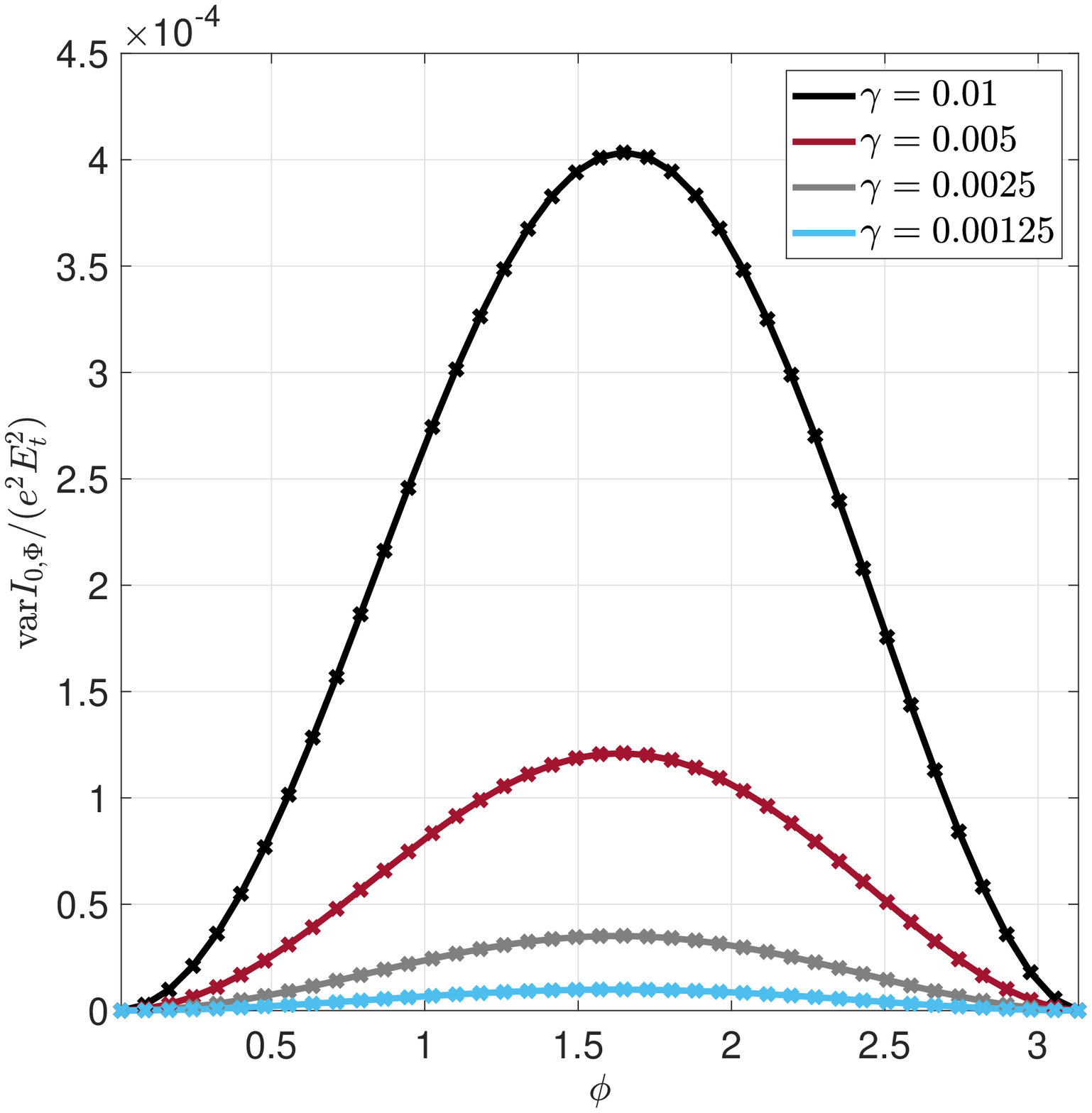}
\includegraphics[scale=0.32]{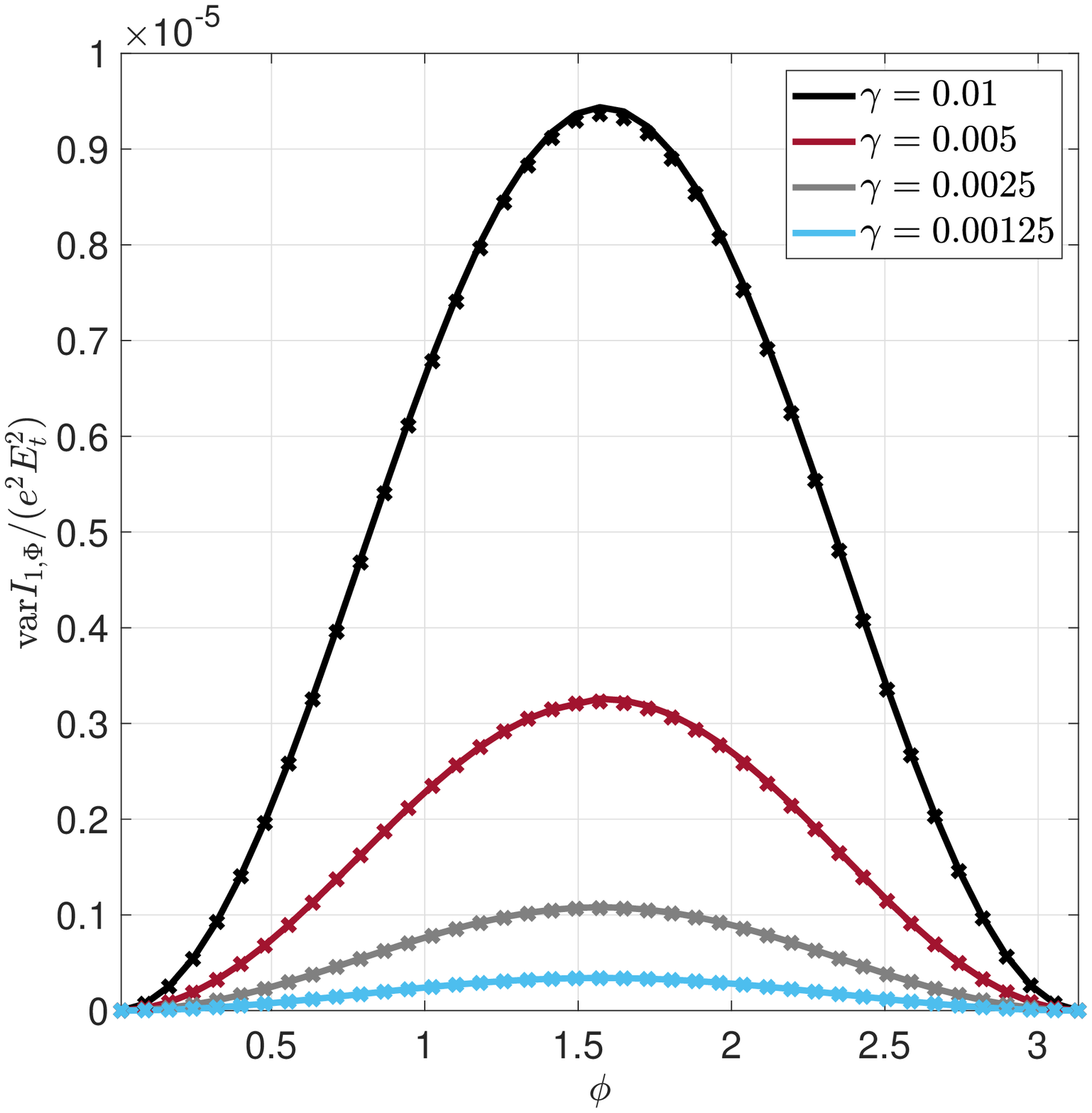}
\caption{Current fluctuations in the strong magnetic field limit as a function of phase for 
various values $\gamma=E_t/E_\Phi$
(we here fixed $e_t = E_t/\Delta=1/1000$ and varied $E_\Phi$). 
Solid lines and markers denote the analytical result employing the approximate solution of the mean field equation and the result 
 building on the numerical solution of the mean field equation, respectively.
 Left panel: quantum dot geometry. Right panel: quasi-one-dimensional geometry.}
\label{fig:KphiApp-Exact}
\end{figure}

In the left panel of Fig.~\ref{fig:KphiApp-Exact}, we compare the analytical solution based on the analytical mean field solution at large magnetic fields to the 
fluctuations calculated using the 
exact numerical solution of the mean field equation. 
Again we find very good agreement for all values $\gamma < 0.01$. 

Finally, we show in the left panel of Fig.~\ref{fig:SM-ratio} the ratio between the square root of current fluctuations and average current for the  quantum dot geometry in the strong magnetic field regime,
\begin{align}
\label{eq:SM-ratio-QD}
\frac
{
\sqrt{\mathrm{var}I_{0, \Phi}(\phi)}
}
{
I_\Phi(\phi)
}
=
\left(
\frac{G_Q}{G}
\right)
\left(
\frac{E_t}{E_\Phi}
\right)
\frac
{
\sqrt{4\pi^2\mathcal{F}_{0, \Phi}(\phi)}
}
{
J_\Phi
}
.
\end{align}
As previously noted, 
large magnetic fields suppresses the relative size of fluctuations by an additional factor 
$E_t/E_\Phi$ compared to the zero magnetic field limit $B=0$.\\

\subsubsection{Quasi-one-dimensional limit: $E^\perp_{\rm Th}\ll E_t$}

For the quasi-one-dimensional geometry 
 current fluctuations at zero temperature read
\begin{align}
\label{eq:varI1Phi}
&
\mathrm{var}I_{1, \Phi}(\phi)
=
e^2E_t^2
\sqrt{\frac{E_t}{E^\perp_{\rm Th}}}
{\cal K}_{1, \Phi}(\phi)
,
\quad 
{\cal K}_{1, \Phi}(\phi)
=
\sqrt
{
\frac{E_t^3}{E_\Phi^3}
}
{\cal F}_{1, \Phi}(\phi)
,
\\
&
{\cal F}_{1, \Phi}(\phi)
=
\frac{\sin^2(\phi)}{\pi^2}
\sum_{s=\pm}\sum_{X=C,D}
\left(
f_{1, 1}^{X,s} 
+
f_{1, 2}^{X,s}
\right).
\end{align}
Here, the functions $f_{1, i}^{X,s}$ depend on the ratio  $\gamma = E_t/E_\Phi$ and are defined as 
\begin{align}
&
\sum_{s = \pm}
f_{1, 1}^{X,s}(\gamma)
=
\iint_0^\infty
dxdx^\prime
\frac
{
1
}
{
\sqrt
{
\left( 1 + x\right)
\left( 1 + x^\prime\right)
\Omega^{+}_X(x, x^\prime)
\Omega^{-}_X(x, x^\prime)
}
\left[
\sqrt
{
\Omega^{+}_X(x, x^\prime)
}
+
\sqrt
{
\Omega^{-}_X(x, x^\prime)
}
\right]
},
\\
&
f_{1, 2}^{X, s}(\gamma)
=
\frac{1}{4}
\iint_0^\infty
dxdx^\prime
\frac
{
\eta_X^{s}(x, x^\prime)
}
{
\sqrt
{
(1 + x)
(1 + x^\prime)
[\Omega^{s}_X(x, x^\prime)]^{3}
}
}
.
\end{align}
As compared to the zero magnetic field limit, 
fluctuations at strong magnetic fields in the quasi-one-dimensional geometry are suppressed
by an additional factor $(E_t/E_\Phi)^{3/2}$.
In terms of this small parameter, the one-dimensional integration over momenta leads to a mildly weaker suppression of fluctuations compared to the quantum dot geometry.

\begin{figure}[H]
\centering
\includegraphics[scale=0.32]{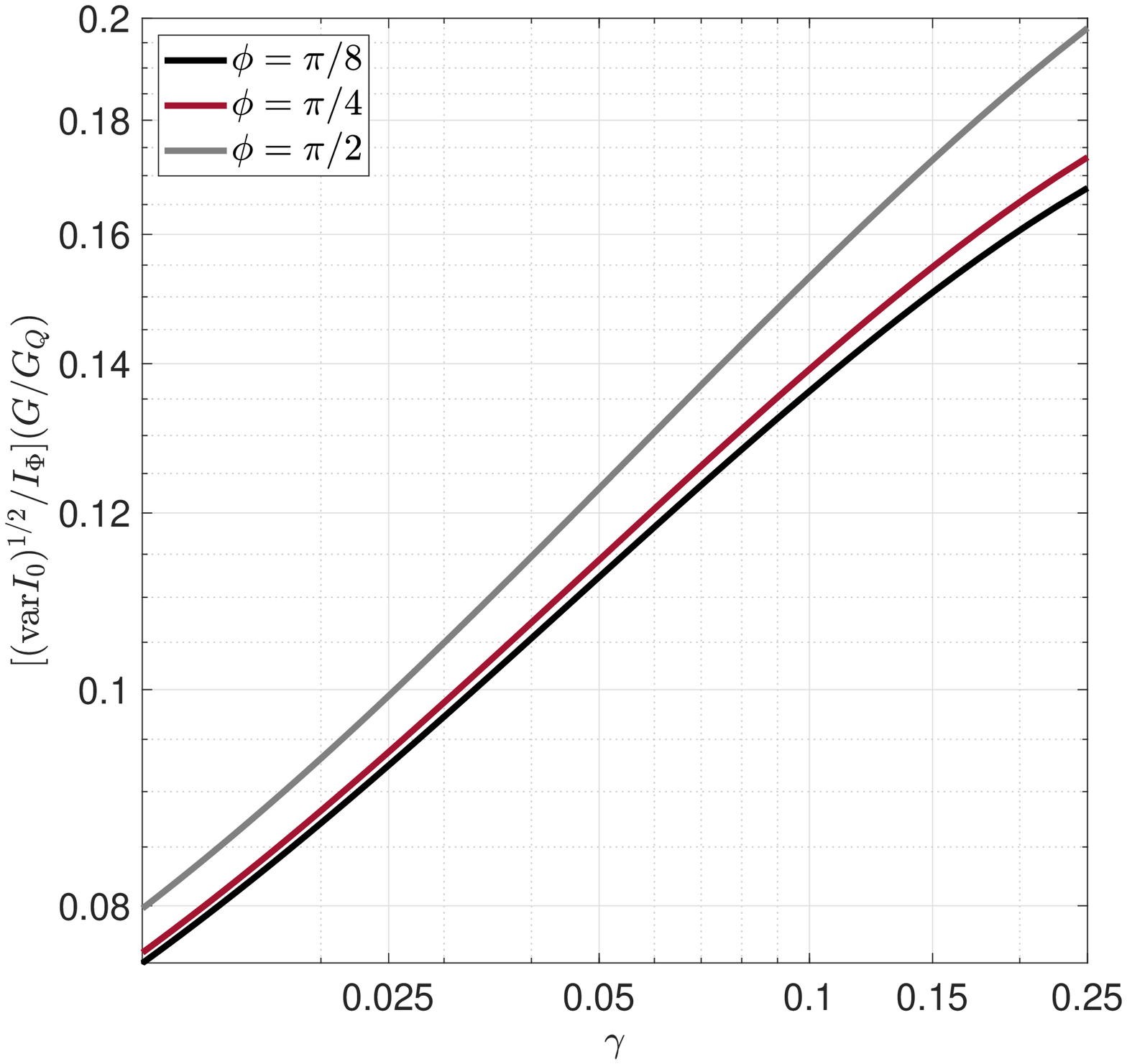}
\includegraphics[scale=0.32]{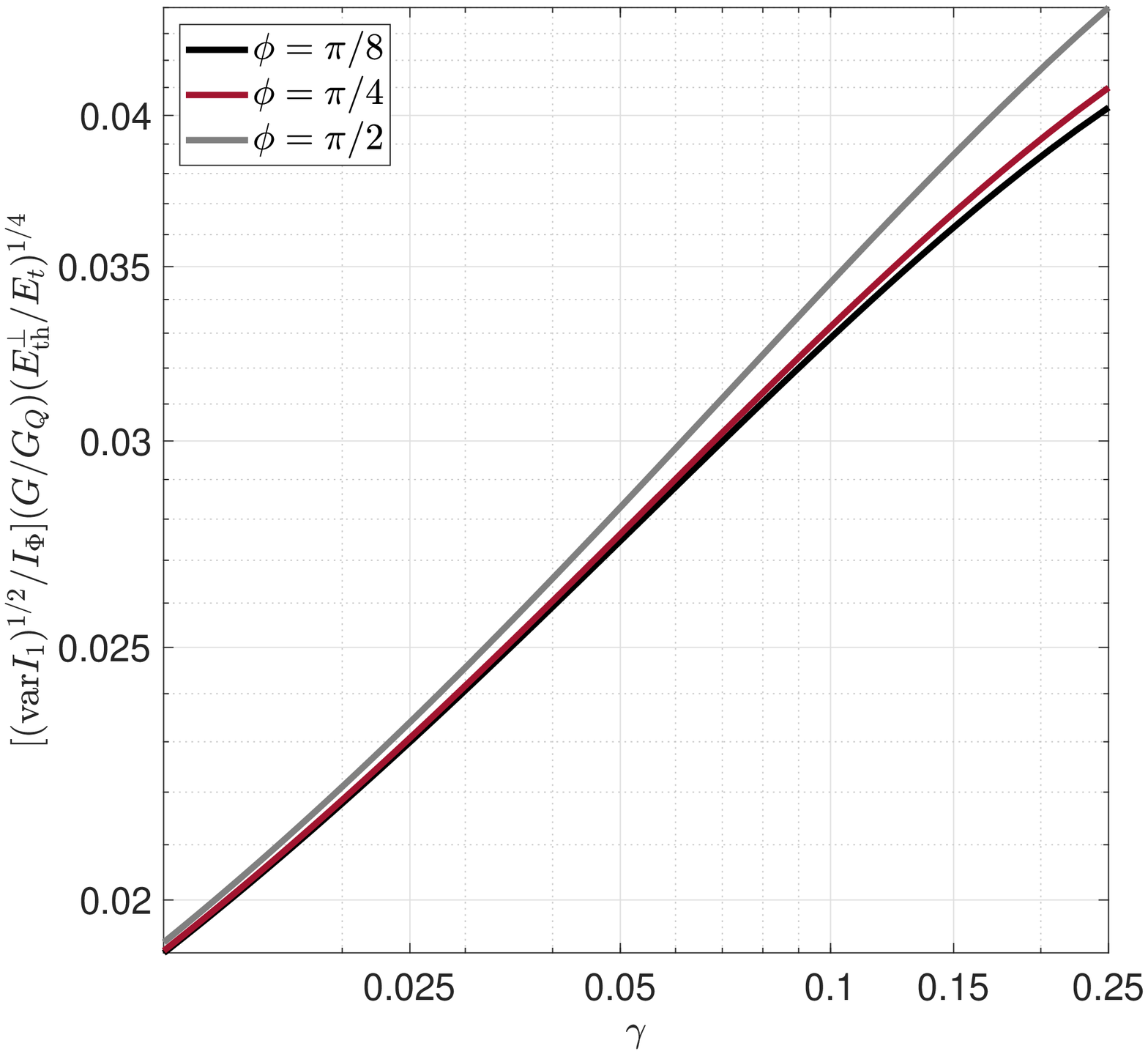}
\caption{Relative size of current fluctuations 
$\sqrt{{\rm var}I_\Phi}/I_\Phi$
in the strong magnetic field limit as a function of $\gamma = E_t / E_\Phi$ and various values of $\phi$.
Left panel: quantum dot geometry. Right panel: quasi-one-dimensional geometry.
}
\label{fig:SM-ratio}
\end{figure}

The relative scale of current fluctuations for the 
 the quasi-one-dimensional geometry
  then reads
\begin{align}
\label{eq:SM-ratio-1D}
\frac
{
\sqrt{\mathrm{var}I_{1, \Phi}(\phi)}
}
{
I_\Phi(\phi)
}
=
\left(
\frac{G_Q}{G}
\right)
\left(
\frac{E_t}{E_\Phi}
\right)^{3/4}
\left(
\frac{E_t}{E^\perp_{\mathrm{th}}}
\right)^{1/4}
\frac
{
\sqrt{4\pi^2\mathcal{F}_{1, \Phi}(\phi)}
}
{
J_\Phi
},
\end{align}
with an 
additional suppression $(E_t/E_\Phi)^{3/4}$
compared to the corresponding zero magnetic field 
expression.
The right panels of Figs.~\ref{fig:Kphi}, ~\ref{fig:KphiApp-Exact}, and ~\ref{fig:SM-ratio} compare the corresponding results for the quantum dot and quasi-one-dimensional geometries. 
Specifically, we observe in Fig.~\ref{fig:SM-ratio} that in both geometries the relative size of current fluctuations monotonically increases as a function of $E_t/E_\Phi$ in a nearly power-law fashion.

\section{Summary}
\label{sec:discussion}
In Table~\ref{table1}, we summarize the parametric dependence of the average current and current fluctuations on the four energy scales $\Delta$, $E_t$, $E_\Phi$ and $E^\perp_{\mathrm{Th}}$ for a topological insulator contacted to identical superconducting leads, at zero temperature. The current-phase relation in the absence of an external magnetic field displays a typical Ambegaokar-Baratoff relation confirming previous findings already reported in the literature, where the scale of the current is set by $\text{min}(E_t, \Delta)$. In the long dwell time limit, $E_t\ll \Delta$, the current-phase relation does not display a sinusoidal behavior and its scale is set by the dwell energy $E_t$~\cite{Brouwer}. In the opposite limit, short dwell time $E_t\gg \Delta$, the average current as a function of the phase only mildly deviates from a sinusoidal form and the scale is set by the superconducting gap $\Delta$~\cite{Kupriyanov},~\cite{Brouwer}. In the limit of a strong magnetic field, $E_t\ll E_\Phi$, the scale of the average current is set by $E_t$, similarly to the long dwell time limit, but in here the current depends logarithmically on $E_\Phi$.

In the limit of a zero magnetic field and in the quantum dot geometry, the current fluctuations are also separated into two groups: the long and short dwell times. In the former limit, the scale of the fluctuations is set by the (squared) dwell energy~\cite{Houzet-Skvortsov}, similar to the average current. The scale of the fluctuations in the latter limit is also set by the energy that determines the corresponding average current, in this case the (squared) superconducting gap~\cite{Houzet-Skvortsov},~\cite{Micklitz},~\cite{Beenakker93}. Turning on the strong magnetic field, the scale of the fluctuations is still set by the dwell energy, in analogy to the long dwell time limit. However, the presence of a strong magnetic field generates an additional suppression of the fluctuations in terms of the small parameter $E_t / E_\Phi$.

Finally, for a quasi-one-dimensional geometry, the scales setting the magnitude of the current fluctuations are identical to the quantum dot geometry. Notwithstanding, the integration over the momenta generates an additional energy dependence via the parameter $E_t/E^\perp_{\mathrm{Th}}$. In the absence of an external magnetic field, the qualitative discussion remains unchanged, but quantitatively the fluctuations are considerably smaller in comparison to the quantum dot geometry. In the presence of a strong magnetic field, the most striking difference between this present case and the quantum dot geometry is the magnitude of the current fluctuations. As a consequence of the integration over the momenta the fluctuations are smaller, though the suppression caused by the small parameter $E_t/E_\Phi$ in here is mildly weaker.

\begin{table}[!htbp]
\centering
\begin{tabular}{|c|c|c|c|}
\hline
&$E_t\ll\Delta, E_\Phi = 0$ & $E_t\gg\Delta, E_\Phi = 0$ & $E_t\lesssim E_\Phi \ll \Delta$\\ 
\hline\rule[-2ex]{0pt}{5.5ex}
$\frac{e}{G}I(\phi)$
&
$
\quad
E_t
\ln\left(\frac{\Delta}{E_t}\right)\sin(\phi)
\quad
$
&
$
\quad
\frac{1}{2}
\Delta
\mathbf{K}\left(\sin^2\frac{\phi}{2}\right)\sin(\phi)
\quad
$
&
$
\quad
E_t
\ln\left(\frac{\Delta}{E_\Phi}\right)
\sin(\phi)
\quad
$
\\
\rule[-2ex]{0pt}{5.5ex}
$\frac{1}{e^2}\mathrm{var}I_0(\phi)$
&
\quad
$E_t^2
\mathcal{K}_{0}(\phi)
\quad
$
&
$
\quad
\Delta^2
\mathcal{K}_0^S(\phi)
\quad
$
&
$
\quad
E_t^2
\frac{E_t^2}{E_\Phi^2}
\mathcal{F}_{0,\Phi}(\phi)
\quad
$
\\ 
\rule[-2ex]{0pt}{5.5ex}
$\frac{1}{e^2}\mathrm{var}I_1(\phi)$
&
$
\quad
E_t^2
\sqrt{\frac{E_t}{E^{\perp}_{\mathrm{Th}}}}
\mathcal{K}_{1}(\phi)
\quad
$
&
$
\quad
\Delta^2
\sqrt{\frac{E_t}{E^{\perp}_{\mathrm{Th}}}}
\mathcal{K}_1^S(\phi)
\quad
$
&
$
\quad
E_t^2
\sqrt{\frac{E_t}{E^\perp_{\mathrm{Th}}}}
\sqrt{\frac{E_t^3}{E_\Phi^3}}
\mathcal{F}_{1, \Phi}(\phi)
\quad
$
\\
\hline
\end{tabular}
\caption{
At zero temperature, the average current, $I(\phi)$, and the current fluctuations in $d-$dimensions, $\mathrm{var} I_d(\phi)$: the second row is described by Eqs~\eqref{eq:longdwell-varI0}, \eqref{eq:shortdwell-varI1}, and \eqref{eq:varI0Phi} in each column, respectively, and the third row by Eqs.~\eqref{eq:longdwell-varI1}, \eqref{eq:shortdwell-varI1}, and \eqref{eq:varI1Phi}. 
}
\label{table1}
\hfill{}
\end{table}

Experimentally, sample-to-sample fluctuations of supercurrents are not easily observed. Instead, fluctuations in a given sample as a function of the chemical potential are more accessible. Josephson junctions consisting of a TI surface states in contact with superconducting contacts allow for variation of $\mu$ by means of a gate voltage control. In addition, narrow constrictions and point-contact junctions can be defined lithographically or electrostatically using split gates. For such systems one would expect that when
the chemical potential is varied on the scale of Thouless energy, the low-temperature critical current will fluctuate universally by an amount of order $\sim e\Delta/h$, independent of the properties of the junction. The critical current noise in topological junctions was observed in Ref. \cite{Kurter-PRB} but thus far interpreted in terms of the charge noise and relocation of topological surface states induced by the gate control.


\section{Acknowledgments}

We thank Dale Van Harlingen, Nicholas Sedlmayr, Stuart Tessmer, and Smitha Vishveshwara for valuable discussions on the broad range of topics related to properties of superconductor--topological insulator interfaces. Support for this work at the University of Wisconsin-Madison was provided by the National Science Foundation, Quantum Leap Challenge Institute for Hybrid Quantum Architectures and Networks, Grant No. 2016136 (A.L.). At the University of Alabama, this work was supported by the National Science Foundation under Grant No. DMR-1742752 (G.S.) and by the MINT summer internship program (G. V.). This work was performed in part at Aspen Center for Physics, which is supported by National Science Foundation Grant No. PHY-1607611. 
M.~M. acknowledge financial support by CNPq (164500/2018-9), G.~V., and T.~M.~acknowledge financial support by Brazilian agencies CAPES, CNPq 
and FAPERJ.


\begin{appendix}
\section{Replica field theory}
\label{app:replica}

\subsection{Replica trick}

Employing the replica trick, we can write the free energy as 
\begin{align}
F&=T\lim_{R\to 0} \frac{1}{R}(Z^R-1). 
\end{align}
The replicated partition function $Z^R=\int {\cal D}[\bar{\psi}\psi]\,
e^{-S[\bar{\psi},\psi,V]}$ is described by the action
\begin{align}
\label{app:fermion_action}
S[\bar{\psi},\psi,V]
&=\sum_{r=1}^R\sum_n\int d^2x\,\bar{\psi}_n^r
\left(
- i\epsilon_n
+ 
\cal H
-
\mu \sigma_3^{\rm ph}
\right)
\psi_n^r.
\end{align}
Here $\mathcal{H} = H_S + H_\Gamma$ is the effective junction Hamiltonian, introduced in Eqs.~\eqref{eq:HS} and~\eqref{H_Gam} in the main text. It describes the TI surface states subjected to a specific realization $V$ of the random disorder potential, 
and accounts for the coupling to the superconducting leads via the boundary 
Hamiltonian $H_\Gamma$. The spinors $\bar\psi_n$, $\psi_n$ in Eq.~\eqref{app:fermion_action} 
are $4\times R$ dimensional fields, 
living in the direct product of spin, particle-hole (Nambu) and replica space, 
and the sum is over fermionic Matsubara frequencies $\epsilon_n=(2n+1)\pi T$.

\subsection{Sample-space}

The calculation of sample-to-sample fluctuations is 
simplified by introducing two copies of the system. 
That is, doubling once more spinor components, and introducing the 
(block) diagonal matrices
\begin{align}
\mathcal H
&\mapsto
\mathcal H(\phi_1,\phi_2)
\equiv
{\rm diag}
(\mathcal H(\phi_1), \mathcal H(\phi_2)),\quad
\epsilon
\mapsto 
\epsilon
\otimes \mathbbm{1}_2,\quad 
\mu
\mapsto 
\mu 
\otimes \mathbbm{1}_2.
\end{align}
This two dimensional extension is referred to 
as `sample space' in the following. We are thus working with the replicated partition function in enlarged space,
$Z^R
=
\int {\cal D}[\bar{\psi}\psi]\,
e^{-S[\bar{\psi},\psi,V]}$, 
with action
\begin{align}
S[\bar{\psi},\psi,V]
&=
\int d^2x\,
\bar{\psi}
\left(
-i\epsilon
+ 
\mathcal H(\phi_1,\phi_2)
-
\mu \sigma_3^{\rm ph}
\right)
\psi,
\end{align}
where $\bar\psi$, $\psi$ are now $2\times 2 \times 2 \times M\times R$ 
dimensional fields living in the direct product of 
spin, particle-hole (Nambu), 
sample, Matsubara and replica space, respectively. 
We did not write out explicitly scalar products in Matsubara and replica spaces, 
and to compactify notation also introduced the matrix 
of Matsubara frequencies $(\hat \epsilon)_n=\epsilon_n$ 
operating in an $M$ dimensional space of Matsubara frequencies
($M$ is here some irrelevant cut off for frequencies, 
e.g. set by the largest energy scale $1/T\tau$). The partition function $Z^R$ allows the calculation of the average Josephson current and its fluctuations as described in Eq.~\eqref{eq:IKbasic} in the main text. 
Finally, recalling the Nambu spinor structure,  
one can verify the following symmetry relation for fields,
\begin{align}
\sigma_2^{\rm S}\otimes\sigma_2^{\rm ph} \bar{\psi}^t(\bold{x},\tau) 
&=-\psi(\bold{x},\tau).
\end{align}

\subsection{Disorder average}

The replica trick allows to readily perform the average over the random disorder potential. We then arrive 
 at the four-fermion contribution 
\begin{align}
\ln\left\langle 
\exp\left[\int dx \bar{\psi} V(\bold{x})\sigma_3^{\rm ph}\psi\right]
\right\rangle
=
\frac{1}{2\pi\nu\tau} 
\int dx \left[\bar{\psi} \sigma_3^{\rm ph} \psi \bar{\psi} \sigma_3^{\rm ph} \psi\right],
\end{align}
which can be further organized by separating the two low-momentum channels that represent the slow diffusion modes 
in a disordered single-particle system with time-reversal symmetry
(`Diffusons' and `Cooperons').
Proceeding with the Hubbard-Stratonovich transformation,
we introduce the $8RM$-dimensional matrix
$Q$ with entries in 
 spin, Nambu, sample, Matsubara and replica space.  
The latter satisfies the symmetry constraint
\begin{align}
Q(\bold{x},\tau\tau') 
&=
\sigma_2^{\rm S}\otimes\sigma_1^{\rm ph}
Q^t(\bold{x},\tau'\tau)
\sigma_2^{\rm S}\otimes\sigma_1^{\rm ph},
\end{align}
inherited from the Nambu spinors. It makes it possible to decouple both slow modes via the transformation 
\begin{align}
\exp\left[-\frac{1}{\pi\nu\tau} 
\int d^2x\,
{\rm tr}(\sigma_3^{\rm ph}\Psi\bar{\psi}\sigma_3^{\rm ph}\Psi\bar{\psi})\right]= \int {\cal D}Q
\, 
\exp\left[-\frac{\pi \nu}{16\tau}\Tr Q^2
+
\frac{i}{2\tau}\int d^2x\bar{\psi}(\bold{x})
Q(\bold{x}) 
\sigma_3^{\rm ph}
\psi(\bold{x})
\right].
\end{align}
It makes the system's action to be quadratic in fermionic fields that can be explicitly integrated out leading to the determinant of the corresponding matrix Green's function operator. Using the celebrated formula for the determinant to the trace-log transformation, $\det{O}=\exp(\Tr\ln O)$, we arrive at the disorder averaged generating functional,
$\langle 
Z^R
\rangle_V
=
\int {\cal D}Q \,
e^{-S[Q]}$,
with the action
\begin{align}
\label{app:Q_action}
S[Q]
&=
\frac{\pi \nu}{16\tau}\Tr Q^2 - \frac{1}{2}\Tr\ln(G^{-1}_Q).
\end{align}
Here, we defined the Greens' function 
\begin{align}
G^{-1}_Q
&=
i\hat \epsilon - \left( v\bold{k}\cdot\sigma -\mu \right)\sigma_3^{\rm ph } - H_\Gamma + \frac{i}{2\tau}
Q\sigma_3^{\rm ph},
\end{align}
with $\hat \epsilon$ and $\hat\Delta=$ and $\hat \phi$ being 
diagonal matrices in sample space. 
Eq.~\eqref{app:Q_action} is still an exact representation of the original replica partition function. It defines the starting point for a derivation of the low energy effective action.
The latter describes the soft rotations around saddle points of 
Eq.~\eqref{app:Q_action}, as discussed next.

\subsection{Mean field equation}

The variation of the action~\eqref{app:Q_action} leads to the saddle point equation
\begin{align}
\label{app:meanfield_eqn}
Q_0
&=
\frac{2i}{\pi\nu}
\int \frac{d^2k}{(2\pi)^2}
\frac{\mu + \frac{i}{2\tau}Q_0 + v\bold{k}\cdot\sigma}{\left(\mu + \frac{i}{2\tau}Q_0\right)^2 - v^2\bold{k}^2},
\end{align}
discussed in the main text.
Referring to the latter for further details, 
we here only recall its solution
\begin{align}
Q_0
&=\sigma_3^{\rm ph}\otimes \Lambda,
\end{align} 
in accordance 
with the causal structure of the model.
Here, $\Lambda$ is the diagonal matrix in Matsubara space with elements 
$(\Lambda)_n={\rm sgn}(\epsilon_n)$.  
The parametrization 
$T(\bold{x})Q_0T^{-1}(\bold{x})$ includes
soft fluctuations around the saddle point which
 leave the first term invariant, namely the high energy contribution to 
 Eq.~\eqref{app:Q_action}.  
The final soft mode action then is found from a low energy
 expansion of the remaining `trace log'.

\subsection{Trace-log expansion}

We then organize the second contribution to Eq.~\eqref{app:Q_action} as follows 
\begin{align}
S_{\rm eff}
&=
- \frac{1}{2}{\rm Tr}\ln(G^{-1}_Q)
\equiv
- \frac{1}{2}{\rm Tr}\ln\left( 1
- G_0 O_T
\right),
\end{align}
 where we dropped an inessential constant that vanishes in the replica limit, 
 and introduced
\begin{align}
G_0^{-1}
&=
-v\bold{k}\cdot\sigma +\mu + \frac{i}{2\tau}
Q_0,
\\
O_T
&=
T^{-1}\left(
- i\hat \epsilon \sigma_3^{\rm ph}  
 - T[v\bold{k}\cdot\sigma, T^{-1}]
+  \hat H_\Gamma 
\sigma_3^{\rm ph} 
\right) 
T.
\end{align}
Expanding in the small energies $\{\epsilon, E_t,\Delta\}\ll 1/\tau$, 
and gradients $\partial_\bold{x}T(\bold{x})$ 
of the slowly fluctuating field, we arrive at
\begin{align}
S_{\rm eff}
&\simeq
 \frac{1}{2}{\rm Tr}\left( G_0 O_T
\right)
+ \frac{1}{4}{\rm Tr}\left( G_0 O_T G_0 O_T
\right)
\equiv 
S_1 + S_2.
\end{align}

\subsection{Spin singlet mode}

We notice that only 
homogeneous modes $T(\bold{x})\equiv T$ lacking any structure in spin-space 
have vanishing commutator 
$[v\bold{k}\cdot\sigma, T^{-1}]$ in $O_T$.
That is, only spin singlet matrices are soft modes. 
Indeed, a brief estimate shows that spin triplet modes 
have masses $\sim\nu/\tau$ which constitutes a large energy in our problem. 
Neglecting the latter, we project onto the spin singlet mode, and
 find from the linear order `trace log' expansion
\begin{align}
S_1
&=
-\frac{\pi\nu}{2}
\int d^x
{\rm tr}\left( \epsilon\sigma_3^{\rm ph}  Q +  iH_\Gamma \sigma_3^{\rm ph}  Q
\right),
\end{align}
(the spin space is now traced out in the action above) where $Q=TQ_0T^{-1}$. 
Similarly, we find from the second order `trace log' expansion 
\begin{align}
S_2 
&=
-\frac{ v^2}{4}
\sum_{i,k=0}^2\sum_{j,l=1}^2  \,
{\rm Tr}\left(
g_0^i \sigma_i   A_j\sigma_j g_0^k\sigma_k A_l\sigma_l  
\right)=
-\frac{ v^2}{2}
\sum_{i=0}^2 \,
{\rm tr}
\left[
\Xi(Q_0,A_i)
\right],
\end{align}
where 
\begin{align}
\Xi(Q_0,A_i)
\equiv D_R A_i  A_i
-D_I A_iQ_0 A_iQ_0
+
2iD'
 A_iA_iQ_0
 \end{align}
and we have decomposed the Green's functions in terms of a linear combination of Pauli matrices and the identity matrix
\begin{align}
G_0
=
\sum_{i = 0}^2
g_0^i
\sigma_i
=
\frac{1}{2}
\sum_{i = 0}^2
\sum_{j = \pm}
g_0^{ij}
(
1
+
j
Q_0
)
\sigma_i
.
\end{align}

Here we introduced $A_i=T\partial_i T^{-1}$, the trace  
${\rm tr}$ which excludes the trace over spin-space. Fixing $\alpha = \mu + i/(2\tau)$ we defined the Green's functions above as
\begin{align}
&
g_0^{0+}
=
\frac
{
\mu + \frac{i}{2\tau}
}
{
\left(
\mu + \frac{i}{2\tau}
\right)^2
-
v^2k^2
}
=
\left(
g_0^{0-}
\right)^*
,\quad
g_0^{i+}
=
\frac
{
vk_i
}
{
\left(
\mu + \frac{i}{2\tau}
\right)^2
-
v^2k^2
}
=
\left(
g_0^{i-}
\right)^*
\end{align}
and the constants
\begin{align}
D_R
&\equiv 
\int \frac{d^2k}{(2\pi)^2}
[{\rm Re} g_0^{0+}(\bold{k})]
[{\rm Re} g_0^{0+}(\bold{k})]
,
D_I 
\equiv 
\int \frac{d^2k}{(2\pi)^2}
[
{\rm Im}
g_0^{0+}(\bold{k})
]
[
{\rm Im}
g_0^{0+}(\bold{k})
]
.
\end{align}
Converting the summations into integrations, we obtain
\begin{align}
D_R
&
=
\frac{1}{2\pi v^2}
\int
d\epsilon
\epsilon
[
{\rm Re}
g_0^{0+}(\epsilon/v)
]
[
{\rm Re}
g_0^{0+}(\epsilon/v)
]=
\frac{\pi\nu\tau}{4}
,
\\
D_I
&
=
\frac{1}{2\pi v^2}
\int
d\epsilon
\epsilon
[
{\rm Im}
g_0^{0+}(\epsilon/v)
]
[
{\rm Im}
g_0^{0+}(\epsilon/v)
]
=
\frac{\pi\nu\tau}{4}
,
\end{align}
where we have performed a change of variables using $\epsilon = vk$. While these constants provide the dominant contributions in $1/(\mu\tau)$, the contribution from the region we neglected, $D'
\equiv
\sum_{\bold k}[{\rm Re} g_0^{0+}(\bold{k})]
[{\rm Im} g_0^{0+}(\bold{k})]
$ is only subleading in $1/\mu\tau$.
Employing the identity 
${\rm tr}(A_iA_i-Q_0A_iQ_0A_i)=-{\rm tr}(\frac{1}{2}(\partial_i Q_0)^2)$, 
the above action can be rewritten as $S_2=\frac{1}{8}\pi\nu D_0\int d^2x \;{\rm tr}\left(\partial_i Q\partial_i Q\right)$, where $Q=TQ_0T^{-1}$, $D_0=v^2\tau/2$ and the density of states per spin direction is defined as $\nu=\mu/(2\pi v^2)$. In the derivation presented so far, the massive spin fluctuations were neglected entirely. In fact, it is known that these modes can renormalize the diffusion coefficient for the singlet modes. Here, we will take a pragmatic approach and account for this effect by introducing the renormalized diffusion coefficient $D_0\rightarrow D=v^2\tau_{tr}/2$ into the action, so that finally
\begin{align}
S_2=\frac{\pi\nu D}{8}\int d^2x \;{\rm tr}\left(\partial_i Q\partial_i Q\right).
\end{align}
Notice that after 
projection onto the spin singlet mode,
the matrix field satisfies the symmetry constraint 
\begin{align}
Q = \sigma_1^{\rm ph} Q^t \sigma_1^{\rm ph}.
\end{align}
Finally, adding both contributions $S_1+S_2$ we arrive at action Eq.~\eqref{eq:firstaction} 
in the main text. 

\section{Average Current}
\label{app:average-current}
Starting out from the general expression for the current phase relation
\begin{equation}
I(\phi)
=
-
\pi\nu e TV
\sum_\epsilon
\left[
\partial_\phi
\left(
2
v_i m_i
-
E_\Phi m_2^2
\right)
\right],
\end{equation}
we find the average current in the strong magnetic field limit,
\begin{align}
I_\Phi(\phi)
=
\left(
\frac{GE_t}{e}
\right)
\mathrm{Re}
\left[
\sum_{\epsilon>0}
\frac{2\pi t\sin\phi}{(e_\Phi + |\epsilon|)(1 + i\epsilon)}
\right]
.
\end{align}
The summation is then readily done using the identity
\begin{equation}
\sum_{n = 0}^\infty
\frac{1}{(n + a)(n + b)}
=
\frac{\psi(a) - \psi(b)}{a - b},
\end{equation}
resulting in
\begin{equation}
I_\Phi(\phi)
=
\left(
\frac{GE_t}{e}
\right)
\mathrm{Re}
\left[
\frac
{
(1 + ie_\Phi)
\psi\left(\frac{1}{2} - \frac{i}{2\pi t}\right)
-
(1 + ie_\Phi)
\psi\left(\frac{1}{2} + \frac{e_\Phi}{2\pi t}\right)
}
{
1 + e_\Phi^2
}
\right]
\sin\phi,
\end{equation}
where $\psi$ is the polygamma function.

\section{Fluctuations}
\label{app:fluctuations}
In this appendix, we provide some details on the derivation of the semiclassical partition function $\mathcal{Z}$, Eq.~\eqref{eq:semiclassicalZ}, from the sigma model action~\eqref{eq:smodelaction}. To this end, we expand the matrix field $Q$ in terms of generators $W$, c.f. \eqref{eq:para1}, up to second order and find the fluctuation determinant. At zeroth order, simply replacing $Q\rightarrow Q_0$, we obtain the saddle point action $S^{(0)}=RS_0$, where
\begin{align}
S_0
=
&
\frac{\pi\nu V}{2}
{\rm tr}'
(
E_{\Phi} m_2^2
-
2v_i m_i
).
\end{align}
Here, we traced out the Nambu and replica spaces, so that the trace operation ${\rm tr}'$ only comprises summations over Matsubara frequencies and the sample space.

At linear order in the generators $W$, the action vanishes in the geometry we study. The saddle point equation eliminates terms containing no spatial derivatives. The remaining term in the action is an irrelevant boundary contribution.

Fluctuations are determined by the second order expansion in $W$. For this term, we obtain the expression
\begin{align}
S^{(2)}
=
&
\pi\nu \int d^2 x De^2{\bf A}^2{\rm tr}[(m_2\sigma^{\rm ph}_1 W)^2 + m_2^2 W^2]
\\
+
&\frac{\pi\nu}{2}\int d^2 x\;{\rm tr}[D(Q_0\partial_y W+ie{\bf A}[m_3,W]_+)^2-2v_i m_i W^2]\nonumber 
\end{align}
In order to perform the Gaussian integration in $W$ and find the fluctuation determinant, we need to account for the constraints discussed below Eq.~\eqref{eq:para1}. In accordance with these constraints, we parametrize the Diffuson and Cooperon contributions to $W$ as
\begin{align}
W_d=\left(\begin{array}{cc} P_{d}&0\\0&P_{d}^t\end{array}\right)_{\rm ph},\quad W_c&=\left(\begin{array}{cc} 0&P_c\\ -P_c^*&0\end{array}\right)_{\rm ph},
\end{align}
where the Diffuson and Cooperon matrices $P_d$ and $P_c$ fulfill the additional constraints $P_d^\dagger=-P_d$, $P_c^t=P_c$. All fields in these equation are functions of two imaginary time arguments. The fields $P_c$ and $P_d$ are further constraint by the relations $[P_d,\Lambda]_+=0$, and $[P_c,\Lambda]=0$. Since $\Lambda$ takes a particularly simple form in Matsubara frequency space, $\Lambda_{\epsilon_n}=\mbox{sgn}({\epsilon_n})$, the constraints are conveniently resolved in frequency space as well,
\begin{align}
P_d(\epsilon_1,\epsilon_2)&=d_{\epsilon_1,\epsilon_2}\theta_{\epsilon_1}\theta_{-\epsilon_2}-d^\dagger_{\epsilon_1,\epsilon_2}\theta_{-\epsilon_1}\theta_{\epsilon_2},\quad P_c(\epsilon_1,\epsilon_2)=c_{\epsilon_1,\epsilon_2}\theta_{\epsilon_1}\theta_{\epsilon_2}+c^t_{-\epsilon_1,-\epsilon_2}\theta_{-\epsilon_1}\theta_{-\epsilon_2},
\end{align}
where $\theta$ is the Heaviside step function.

Using the parametrization introduced above, we obtain the quadratic forms for Diffuson and Cooperon modes as
\begin{align}
\frac{S_{D}^{(2)}}{ 2\pi\nu L}=&\sum_{\epsilon_1>0,\epsilon_2<0}\sum_{ab,mn} 
\int dy \Big(D\left|\mathcal{D}_yd_\alpha\right|^2+\frac{1}{8}E_\Phi[m_3^a(\epsilon_1)+m_3^b(\epsilon_2)]^2|d_\alpha|^2+[m_i^a(\epsilon_1)v_i^a(\epsilon_1)+m_i^b(\epsilon_2)v_i^b(\epsilon_2)]|d_\alpha|^2 \nonumber\\
&-\frac{1}{2}E_\Phi[m_2^a(\epsilon_1)^2+m_2^b(\epsilon_2)]|d_\alpha|^2-E_\Phi[m_2^a(\epsilon_1)m_2^b(\epsilon_2){\rm Re}(d^*_\alpha d^t_{\bar{\alpha}})]\Big),
\label{eq:diff}
\end{align}
where we defined $\mathcal{D}_y=\partial_y+ieBL[m_3^a(\epsilon_1)+m_3^b(\epsilon_2)]/2$ and used the multi-index notation $\alpha = \epsilon_1\epsilon_2, ab, mn$, $\bar{\alpha}=(-\epsilon_1)(-\epsilon_2),ab,mn$, and
\begin{align}
\frac{S_{C}^{(2)}}{ 2\pi\nu L}
=&\sum_{\epsilon_1>0,\epsilon_2>0}\sum_{ab,nm}\int dy\Big(
D
|\mathcal{D}_y c_\alpha|^2+
\frac{1}{8}E_\Phi
[
m_3^a(\epsilon_1)
+
m_3^b(\epsilon_2)
]^2
|c_\alpha|^2+
[
m_i^a(\epsilon_1)v_i^a(\epsilon_1) + m_i^b(\epsilon_2)v_i^b(\epsilon_2)
]
|c_\alpha|^2\nonumber\\
&-
\frac{1}{2}E_\Phi
[
m_2^a(\epsilon_1)^2 + m_2^b(\epsilon_2)^2
]
|c_\alpha|^2+
E_\Phi
[
m_2^a(\epsilon_1)m_2^b(\epsilon_2)\mathrm{Re}(c_\alpha c_\alpha^t)
]\Big).
\label{eq:coop}
\end{align}

Our next goal will be to integrate out the $d$ and $c$ modes and to find the fluctuation determinant. We cannot immediately read off the eigenvalues due to the presence of the derivative $\mathcal{D}_y$ and due the nontrivial structure present in the last line of Eqs.~\eqref{eq:diff} and \eqref{eq:coop}. As far as the derivative is concerned, due to the translational invariance in the $y$ direction we can effectively replace $\mathcal{D}_y\rightarrow \partial_y$ for the calculation of the fluctuation determinant. 

We will discuss the diagonalization of the quadratic form for the case of the Cooperon. The Diffuson contribution can be treated by analogy. It is convenient to write down the Cooperon field in the form $c_\alpha = c_\alpha^\prime + ic_\alpha^{\prime\prime}$, where $c'$ and $c''$ are the real and imaginary parts of $c$, respectively. Then, the Cooperon contribution to the quadratic action reads as
\begin{align}
&\frac{S_{C}^{(2)}}{2\pi\nu L}
=\sum_{\epsilon_1>0,\epsilon_2>0}\sum_{ab,mn}\int \frac{dq}{2\pi}
\big(c_{\alpha,q}^{\prime}
\mathcal O_{\epsilon_1\epsilon_2}^{ab}
c_{\alpha,-q}^{\prime}+
c_{\alpha,q}^\prime \mathcal{N}_{\epsilon_1\epsilon_2}^{ab} c_{\alpha,-q}^{\prime T}
+
c_{\alpha,q}^{\prime\prime}
\mathcal O_{\epsilon_1\epsilon_2}^{ab}
c_{\alpha,-q}^{\prime\prime}
-
c_{\alpha,q}^{\prime\prime} \mathcal{N}_{\epsilon_1\epsilon_2}^{ab}c_{\alpha,-q}^{\prime\prime T}\big)
,
\end{align}
where we introduced the notation
\begin{align}
\mathcal O_{\epsilon_1\epsilon_2}^{ab}
=&
Dq^2
+
\frac{1}{8}E_\Phi
[
\hat{m}_3^a(\epsilon_1) + \hat{m}_3^b(\epsilon_2)
]^2
+
\hat{m}_i^a(\epsilon_1)v_i^a(\epsilon_1)
+
\hat{m}_i^b(\epsilon_2)v_i^b(\epsilon_2)-
\frac{1}{2}E_\Phi
[
\hat{m}_2^a(\epsilon_1)^2 + \hat{m}_2^b(\epsilon_2)^2
]\nonumber
\\
\mathcal{N}_{\epsilon_1\epsilon_2}^{ab}
=&
E_\Phi
\hat{m}_2^a(\epsilon_1)\hat{m}_2^b(\epsilon_2)
.
\end{align}
For the fully diagonal terms in the action, for which $\epsilon_1=\epsilon_2$, $a=b$, $m=n$, we readily read off the two eigenvalues $\mathcal{O}^{aa}_{\epsilon_1\epsilon_1}\pm \mathcal{N}^{aa}_{\epsilon_1\epsilon_1}$. In order to find the remaining eigenvalues, we will first arrange the variables $c'$ and $c''$ into vectors so that the quadratic form is represented by a block-diagonal matrix. For $(\epsilon_1\epsilon_2,ab,mn)\neq(\epsilon_2\epsilon_1,ba,mn)$, we introduce four-component vectors $\chi$ as
\begin{align}
\chi=(c',[c']^t,c'',[c'']^t)^T\label{eq:chi}
\end{align}
where the transposition $T$ indicates that we view this object as a column vector. For a given $\alpha$, the contribution to the quadratic form can then be represented as
\begin{align}
\chi^T_{\alpha,q}\left(\begin{array}{cc} \mathcal{M}_{\epsilon_1\epsilon_2}^{ab}&0\\0&\bar{\mathcal{M}}_{\epsilon_1\epsilon_2}^{ab}\end{array}\right)\chi_{\alpha,-q},
\end{align}
where we introduced the block matrices
\begin{align}
\mathcal{M}=\left(\begin{array}{cc} \mathcal{O}&\mathcal{N}\\ \mathcal{N}&\mathcal{O}\end{array}\right),\quad \bar{\mathcal{M}}=\left(\begin{array}{cc} \mathcal{O}&-\mathcal{N}\\ -\mathcal{N}&\mathcal{O}\end{array}\right).
\end{align}
We used the symmetries $\mathcal{O}=\mathcal{O}^t$ and $\mathcal{N}=\mathcal{N}^t$ to cast the expression in this form. Both $\mathcal{M}$ and $\bar{\mathcal{M}}$ have the two eigenvalues
\begin{align}
\tilde{\lambda}^{C,\pm}=\mathcal{O}\pm \mathcal{N}.
\end{align} 
Having identified all eigenvalues of the fully diagonal and the off-diagonal parts of the quadratic form, what remains is to find their multiplicity. When grouping $c'$, $[c']^t$ (and $c''$, $[c'']^t$) into vector $\chi$, the summation in $\alpha$ needs to be constraint to cover only half of the degrees of freedom in order to avoid overcounting. This is compensated by the degeneracy of eigenvalues from the $c'$ and $c''$ sectors. We can summarize the above discussion by stating that for each $\alpha=(\epsilon_1,\epsilon_2,ab,mn)$ we find two associated eigenvalues $(\tilde{\lambda}^{C,+})_{\epsilon_1\epsilon_2}^{ab}$ and $(\tilde{\lambda}^{C,-})_{\epsilon_1\epsilon_2}^{ab}$. These eigenvalues are  independent of the replica indices $m$ and $n$. In order to make contact with the notation used in the main text, we define
\begin{align}
\lambda^{C,\pm}_{\epsilon_1\epsilon_2}=(\tilde{\lambda}^{C,\pm})^{12}_{\epsilon_1\epsilon_2}=(\tilde{\lambda}^{C,\pm})_{\epsilon_2\epsilon_1}^{21}.
\end{align}
Upon integration in $c'$ and $c''$, and discarding irrelevant $\phi$-independent constants and sample-space diagonal terms (which are not relevant for the calculation of the current fluctuations) we arrive at the expression for $\mathcal{Z}$ stated in Eq.~\eqref{eq:semiclassicalZ}.


\end{appendix}
 


\begin{thebibliography}{10}
\expandafter\ifx\csname url\endcsname\relax
  \def\url#1{\texttt{#1}}\fi
\expandafter\ifx\csname urlprefix\endcsname\relax\def\urlprefix{URL }\fi
\expandafter\ifx\csname href\endcsname\relax
  \def\href#1#2{#2} \def\path#1{#1}\fi


\bibitem{BLA-UCF}
B. L. Al'tshuler, \textit{Fluctuations in the extrinsic conductivity of disordered conductors}, JETP Lett. \textbf{41}, 648 (1985) [Pis'ma Zh. Eksp. Teor. Fiz. \textbf{41}, 530 (1985)].

\bibitem{Lee-Stone}
P. A. Lee and A. Douglas Stone, \textit{Universal Conductance Fluctuations in Metals}, Phys. Rev. Lett.\textbf{55}, 1622 (1985).

\bibitem{Lee-Stone-Fukuyama}
P. A. Lee, A. Douglas Stone, and H. Fukuyama, \textit{Universal conductance fluctuations in metals: Effects of finite temperature, interactions, and magnetic field}, Phys. Rev. B \textbf{35}, 1039 (1987).

\bibitem{Beenakker-RMP}
C. W. J. Beenakker, \textit{Random-matrix theory of quantum transport}, 
Rev. Mod. Phys. \textbf{69}, 731 (1997).

\bibitem{Mehta}
M. L. Mehta, \textit{Random Matrices}, (Elsevier Ltd. 2004, 3rd edition).

\bibitem{Aleiner-Blanter}
I. L. Aleiner, Ya. M. Blanter, 
\textit{Inelastic Scattering Time for Conductance Fluctuations}, 
Phys. Rev. B \textbf{65}, 115317 (2002).

\bibitem{Imry}
Y. Imry, \textit{Active Transmission Channels and Universal Conductance Fluctuations}, 
Europhys. Lett. \textbf{1}, 249 (1986).

\bibitem{BLA-BIS}
B. L. Al'tshuler, and B. I. Shklovskii, \textit{Repulsion of energy levels and conductivity of small metal samples}, Sov. Phys. JETP \textbf{64}, 127 (1986) [Zh. Eksp. Teor. Fiz. \textbf{91}, 220 (1986)].

\bibitem{Dorokhov}
O. N. Dorokhov, \textit{On the coexistence of localized and extended states in the metallic phase}, Solid State Commun. \textbf{51}, 381 (1984).

\bibitem{Marmorkos}
I. K. Marmorkos, C. W. J. Beenakker, and R. A. Jalabert, 
\textit{Three signatures of phase-coherent Andreev reflection}, 
Phys. Rev. B \textbf{48}, 2811(R) (1993).

\bibitem{Brunn}
J. Bruun, V. C. Hui, and C. J. Lambert, \textit{Coherence-length dependence of fluctuations in the conductance of normal-superconducting interfaces}, 
Phys. Rev. B \textbf{49}, 4010 (1994).

\bibitem{Takane-Ebisawa}
Y. Takane and H. Ebisawa, \textit{Conductance Fluctuations in Small Normal and Superconductor Composite Wire Systems}, J. Phys. Soc. Jpn. \textbf{60}, 3130 (1991).

\bibitem{Exp-UCF-SN-1}
S. G. den Hartog, C. M. A. Kapteyn, B. J. van Wees, T. M. Klapwijk, W. van der Graaf, and G. Borghs, \textit{Sample-Specific Conductance Fluctuations Modulated by the Superconducting Phase}, Phys. Rev. Lett. \textbf{76}, 4592 (1996).

\bibitem{Exp-UCF-SN-2}
K. Hecker, H. Hegger, A. Altland, and K. Fiegle, \textit{Conductance Fluctuations in Mesoscopic Normal-Metal/Superconductor Samples}, Phys. Rev. Lett. \textbf{79}, 1547 (1997).

\bibitem{Brouwer-Beenakker}
P. W. Brouwer and C. W. J. Beenakker, \textit{Insensitivity to time-reversal symmetry breaking of universal conductance fluctuations with Andreev reflection}, Phys. Rev. B \textbf{52}, 16772 (1995).

\bibitem{Chalker-Macedo}
J. T. Chalker and A. M. S. Mac\^edo, \textit{Complete Characterization
of Universal Fluctuations in Quasi-One-Dimensional Mesoscopic
Conductors}, Phys. Rev. Lett. \textbf{71}, 3693 (1993).

\bibitem{Beenakker-PRL91}
C. W. J. Beenakker, \textit{Universal limit of critical-current fluctuations in mesoscopic Josephson junctions}, Phys. Rev. Lett. \textbf{67}, 3836 (1991); Erratum Phys. Rev. Lett. \textbf{68}, 1442 (1992). 

\bibitem{BLA-BZS}
B. L. Al'tshuler, B. Z. Spivak, 
\textit{Mesoscopic fluctuations in a superconductor-normal metal-superconductor junction},
ZhETF \textbf{92}, 609 (1987) [Sov. Phys. JETP \textbf{65}, 343 (1987)]  

\bibitem{Houzet-Skvortsov}
M. Houzet, M. A. Skvortsov, \textit{Mesoscopic fluctuations of the supercurrent in diffusive Josephson junctions}, 
Phys. Rev. B \textbf{77}, 057002 (2008). 

\bibitem{Micklitz}
T. Micklitz, \textit{Interface dependence of the Josephson-current fluctuations in short mesoscopic superconductor/normal-conductor/superconductor junctions}, 
Phys. Rev. B \textbf{75}, 144509 (2007).

\bibitem{HXAL}
Hong-Yi Xie and Alex Levchenko, \textit{Topological supercurrents interaction and fluctuations in the multiterminal Josephson effect}, 
Phys. Rev. B \textbf{99}, 094519 (2019). 

\bibitem{BBS}
W. Belzig, C. Bruder, and Gerd Sch\"on, \textit{Local density of states in a dirty normal metal connected to a superconductor}, 
Phys. Rev. B \textbf{54}, 9443 (1996). 

\bibitem{Hammer}
J. C. Hammer, J. C. Cuevas, F. S. Bergeret, and W. Belzig, \textit{Density of states and supercurrent in diffusive SNS junctions: Roles of nonideal interfaces and spin-flip scattering}, 
Phys. Rev. B \textbf{76}, 064514 (2007).

\bibitem{AL-DOS}
Alex Levchenko, \textit{Crossover in the local density of states of mesoscopic superconductor/normal-metal/superconductor junctions}
Phys. Rev. B \textbf{77}, 180503(R) (2008).

\bibitem{Reutlinger-1}
J. Reutlinger, L. Glazman, Yu. V. Nazarov, and W. Belzig, 
\textit{``Smile" Gap in the Density of States of a Cavity between Superconductors}, Phys. Rev. Lett. \textbf{112}, 067001 (2014).

\bibitem{Reutlinger-2}
J. Reutlinger, L. Glazman, Yu. V. Nazarov, and W. Belzig, 
\textit{Secondary ``smile"-gap in the density of states of a diffusive Josephson junction for a wide range of contact types}, 
Phys. Rev. B \textbf{90}, 014521 (2014).

\bibitem{Reutlinger-3}
Johannes Reutlinger, Leonid I. Glazman, Yuli V. Nazarov, Wolfgang Belzig, 
\textit{Universal Properties of Mesoscopic Fluctuations of the Secondary ``Smile Gap}, preprint arXiv:2109.03055 [cond-mat.supr-con]. 

\bibitem{Vavilov}
M. G. Vavilov, P. W. Brouwer, V. Ambegaokar, and C. W. J. Beenakker, \textit{Universal Gap Fluctuations in the Superconductor Proximity Effect}, 
Phys. Rev. Lett. \textbf{86}, 874 (2001). 

\bibitem{LKG}
Alex Levchenko, Alex Kamenev, and Leonid Glazman, \textit{Singular length dependence of critical current in superconductor/normal-metal/superconductor bridges}, Phys. Rev. B \textbf{74}, 212509 (2006).

\bibitem{Whisler}
Colin M. Whisler, Maxim G. Vavilov, and Alex Levchenko, \textit{Josephson currents in chaotic quantum dots}, 
Phys. Rev. B \textbf{97}, 224515 (2018).

\bibitem{Sacepe}
B. Sacepe, J. B. Oostinga, J. Li, A. Ubaldini, N. J. G.
Couto, E. Giannini, and A. F. Morpurgo, \textit{Gate-tuned normal and superconducting transport at the surface of a topological insulator}, Nat. Commun. \textbf{2}, 575 (2011). 

\bibitem{Brinkman}
M. Veldhorst, M. Snelder, M. Hoek, T. Gang, V. K. Guduru, X. L. Wang, U. Zeitler, W. G. Van der wiel, A. A. Golubov, H. Hilgenkamp, and A. Brinkman, \textit{Josephson supercurrent through a topological insulator surface state}, Nat. Mater. \textbf{11}, 417 (2012). 

\bibitem{Lu}
F. M. Qu, F. Yang, J. Shen, Y. Ding, J. Chen, Z. Q. Ji, G. G. Liu, J. Fan, X. N. Jing, C. L. Yang, and L. Lu, \textit{Strong Superconducting Proximity Effect in Pb-Bi$_2$Te$_3$ Hybrid Structures}, Sci. Rep. \textbf{2}, 339 (2012).

\bibitem{Goldhaber-Gordon}
J. R. Williams, A. J. Bestwick, P. Gallagher, Seung Sae 39 Hong, Y. Cui, Andrew S. Bleich, J. G. Analytis, I. R. Fisher, and D. Goldhaber-Gordon, \textit{Unconventional Josephson Effect in Hybrid Superconductor-Topological Insulator Devices}, Phys. Rev. Lett. \textbf{109}, 056803 (2012).

\bibitem{Mason}
S. Cho, B. Dellabetta, A. Yang, J. Schneeloch, Z. Xu, T. Valla, G. Gu, M. J. Gilbert, and N. Mason, \textit{Symmetry protected Josephson supercurrents in three-dimensional topological insulators}, Nat. Commun. \textbf{4}, 1689 (2013).

\bibitem{Kurter-PRB}
C. Kurter, A. D. K. Finck, P. Ghaemi, Y. S. Hor, and D. J. Van Harlingen, \textit{Dynamical gate-tunable supercurrents in topological Josephson junctions}, Phys. Rev. B \textbf{90}, 014501 (2014).

\bibitem{Kurter-NC}
C. Kurter, A. D. K. Finck, Y. S. Hor, D. J. Van Harlingen, \textit{Evidence for an anomalous current–phase relation in topological insulator Josephson junctions}, Nat. Commun. \textbf{6}, 7130 (2015).

\bibitem{Sochnikov}
I. Sochnikov, L. Maier, C. A. Watson, J. R. Kirtley, C. Gould, G. Tkachov, E. M. Hankiewicz, C. Br\"une, H. Buhmann, L. W. Molenkamp, and K. A. Moler, \textit{Nonsinusoidal Current-Phase Relationship in Josephson Junctions from the 3D Topological Insulator HgTe}, Phys. Rev. Lett. \textbf{114}, 066801 (2015).

\bibitem{Stehno}
M. P. Stehno, V. Orlyanchik, C. D. Nugroho, P. Ghaemi, M. Brahlek, N. Koirala, S. Oh, and D. J. Van Harlingen, \textit{Signature of a topological phase transition in the Josephson supercurrent through a topological insulator}, 
Phys. Rev. B \textbf{93}, 035307 (2016). 

\bibitem{Finck}
A.D.K. Finck, C. Kurter, Y.S. Hor, and D.J. Van Harlingen, \textit{Phase Coherence and Andreev Reflection in Topological Insulator Devices}, 
Phys. Rev. X \textbf{4}, 041022 (2014).

\bibitem{Bobkova16}
I. V. Bobkova, 
A. M. Bobkov,  
A. A. Zyuzin, 
M. Alidoust,
\textit{Magnetoelectrics in disordered topological insulator Josephson junctions},
Phys. Rev. \textbf{B} 94, 134506 (2016).

\bibitem{Alidoust17}
M. Alidoust, 
H. Hamzehpour, 
\textit{Spontaneous supercurrent and $\phi_0$ phase shift parallel to magnetized topological insulator interfaces}, 
Phys. Rev. B \textbf{96}, 165422 (2017).

\bibitem{Zyuzin18}
A. Zyuzin, 
M. Alidoust, 
D. Loss, 
\textit{Josephson junction through a disordered topological insulator with helical magnetization}, 
Phys. Rev. B \textbf{93}, 214502 (2016).

\bibitem{Vishveshwara}
Suraj S. Hegde, Guang Yue, Yuxuan Wang, Erik Huemiller, D. J. Van Harlingen, Smitha Vishveshwara, \textit{A topological Josephson junction platform for creating, manipulating, and braiding Majorana bound states}, 
Annals of Physics \textbf{423}, 168326 (2020). 


\bibitem{Sedlmayr}
Nicholas Sedlmayr, Alex Levchenko, 
\textit{Hybridization mechanism of the dual proximity effect in
superconductor–topological insulator interfaces},
Solid State Communications \textbf{327}, 114221 (2021). 

\bibitem{Golubov}
A. A. Golubov, M. Yu. Kupriyanov, and E. Il'ichev, \textit{The current-phase relation in Josephson junctions}, Rev. Mod. Phys. \textbf{76}, 411 (2004).

\bibitem{Altland97}
A. Altland and M. R. Zirnbauer, \textit{Nonstandard symmetry classes in mesoscopic normal-superconducting hybrid structures}, Phys. Rev. B \textbf{55}, 1142 (1997).


\bibitem{Wegner}
F. J. Wegner, \textit{The mobility edge problem: Continuous symmetry and a conjecture}, Z. Phys. B \textbf{35}, 207 (1979).

\bibitem{Efetov}
K. B. Efetov, A. I. Larkin, and D.E. Khmelnitskii, \textit{Interaction of diffusion modes in the theory of localization}, Zh. Eksp. Teor. Fiz. \textbf{79}, 1120 (1980) [Sov. Phys. JETP \textbf{52}, 568 (1980)].

\bibitem{Altland10}
A. Altland and B. Simons, \textit{Condensed Matter Field Theory}, (Cambridge University Press, 2010, 2nd edition).

\bibitem{Aleiner06}
I. L. Aleiner and K. B. Efetov, \textit{Effect of Disorder on Transport in Graphene}, Phys. Rev. Lett. \textbf{97}, 236801 (2006).

\bibitem{Ostrovsky06}
P. M. Ostrovsky, I. V. Gornyi, and A. D. Mirlin, \textit{Electron transport in disordered graphene}, Phys. Rev. B \textbf{74}, 235443 (2006).

\bibitem{Garate}
I. Garate and L. Glazman, \textit{Weak localization and antilocalization in topological insulator thin films with coherent bulk-surface coupling}, Phys. Rev. B \textbf{86}, 035422 (2012).

\bibitem{Velkov}
H. Velkov, G. N. Bremm, T. Micklitz, and G. Schwiete, \textit{Transport in topological insulators with bulk-surface coupling: Interference corrections and conductance fluctuations}, Phys. Rev. B \textbf{98}, 165408 (2018).

\bibitem{Usadel70}
K. D. Usadel, \textit{Generalized Diffusion Equation for Superconducting Alloys}, Phys. Rev. Lett. \textbf{25}, 507 (1970).

\bibitem{Bergeret}
F. S. Bergeret, J. C. Cuevas, 
\textit{The Vortex State and Josephson Critical Current of a Diffusive SNS Junction},
J Low Temp Phys \textbf{153}, 304 (2008).



\bibitem{AG-JETP60}
A. A. Abrikosov and L. P. Gor'kov, Zh. Eksp. Teor. Fiz. \textbf{39}, 1781
(1960) [Sov. Phys. JETP \textbf{12}, 1243 (1961)].

\bibitem{Nazarov} 
Yu. V. Nazarov, \textit{Novel circuit theory of Andreev reflection}, 
Superlattices and Microstructures \textbf{25}, 1221 (1999).

\bibitem{Mazanik}
A. A. Mazanik, Ya. V. Fominov, \textit{Peculiarities of the density of states in SN bilayers}, preprint arXiv:2205.06171 [cond-mat.supr-con].  

\bibitem{Frahm}
K. M. Frahm, P. W. Brouwer, J. A. Melsen, and C. W. J. Beenakker, \textit{Effect of the Coupling to a Superconductor on the Level Statistics of a Metal Grain in a Magnetic Field}, Phys. Rev. Lett. \textbf{76}, 2981 (1996). 

\bibitem{Meyer}
J. S. Meyer and B. D. Simons, \textit{Gap fluctuations in inhomogeneous superconductors}, Phys. Rev. B \textbf{64}, 134516 (2001). 

\bibitem{Lamacraft}
A. Lamacraft and B. D. Simons, \textit{Tail States in a Superconductor with Magnetic Impurities}, Phys. Rev. Lett. \textbf{85}, 4783 (2000).

\bibitem{Beloborodov}
I. S. Beloborodov, B. N. Narozhny, and I. L. Aleiner, \textit{Effect of Time Reversal Symmetry Breaking on the Density of States in Small Superconducting Grains}, Phys. Rev. Lett. \textbf{85}, 816 (2000). 

\bibitem{Ostrovsky}
P. M. Ostrovsky, M. A. Skvortsov, and M. V. Feigel'man, \textit{Density of States below the Thouless Gap in a Mesoscopic SNS Junction}, Phys. Rev. Lett. \textbf{87}, 027002 (2001). 

\bibitem{Feigelman}
M. V. Feigel'man and M. A. Skvortsov, \textit{Universal Broadening of the Bardeen-Cooper-Schrieffer Coherence Peak of Disordered Superconducting Films}, 
Phys. Rev. Lett. \textbf{109}, 147002 (2012). 

\bibitem{Bagrets}
Dmitry Bagrets and Alexander Altland, \textit{Class 
D Spectral Peak in Majorana Quantum Wires}, 
Phys. Rev. Lett. \textbf{109}, 227005 (2012). 

\bibitem{Ioselevich}
P. A. Ioselevich, P. M. Ostrovsky, and M. V. Feigel'man, \textit{Majorana state on the surface of a disordered three-dimensional topological insulator}, 
Phys. Rev. B \textbf{86}, 035441 (2012). 

\bibitem{Jia-1}
Jin-Peng Xu, Canhua Liu, Mei-Xiao Wang, Jianfeng Ge, Zhi-Long Liu, Xiaojun Yang, Yan Chen, Ying Liu, Zhu-An Xu, Chun-Lei Gao, Dong Qian, Fu-Chun Zhang, and Jin-Feng Jia, \textit{Artificial Topological Superconductor by the Proximity Effect}, Phys. Rev. Lett. \textbf{112}, 217001 (2014). 

\bibitem{Dayton}
Ian M. Dayton, Nicholas Sedlmayr, Victor Ramirez, Thomas C. Chasapis, Reza Loloee, Mercouri G. Kanatzidis, Alex Levchenko, and Stuart H. Tessmer, \textit{Scanning tunneling microscopy of superconducting topological surface states in Bi$_2$Se$_3$}, Phys. Rev. B \textbf{93}, 220506(R) (2016).

\bibitem{Tessmer}
Nicholas Sedlmayr, E. W. Goodwin, Michael Gottschalk, Ian M. Dayton, Can Zhang, Erik Huemiller, Reza Loloee, Thomas C. Chasapis, Maryam Salehi, Nikesh Koirala, Mercouri G. Kanatzidis, Seongshik Oh, D. J. Van Harlingen, Alex Levchenko, S. H. Tessmer, \textit{Dirac surface states in superconductors: a dual topological proximity effect}, arXiv:1805.12330 [cond-mat.supr-con]. 

\bibitem{Zeljkovic}
Bryan Rachmilowitz, He Zhao, Hong Li, Alexander LaFleur, J. Schneeloch, Ruidan Zhong, Genda Gu, and Ilija Zeljkovic, \textit{Proximity-induced superconductivity in a topological crystalline insulator}, 
Phys. Rev. B \textbf{100}, 241402(R) (2019).

\bibitem{Jia-2}
Hao Yang, Yao-Yi Li, Teng-Teng Liu, Dan-Dan Guan, Shi-Yong Wang, Hao Zheng, Canhua Liu, Liang Fu, and Jin-Feng Jia, \textit{Multiple In-Gap States Induced by Topological Surface States in the Superconducting Topological Crystalline Insulator Heterostructure Sn$_{1-x}$Pb$_x$Te-Pb}, 
Phys. Rev. Lett. \textbf{125}, 136802 (2020). 



\bibitem{Aslamazov}
 L. G. Aslamazov, A. I. Larkin, Yu. N. Ovchinnikov, \textit{Josephson Effect in Superconductors Separated by a Normal Metal},
Zh. Eksp. Teor. Fiz. \textbf{55}, 323 (1968) [Sov. Phys. JETP \textbf{28}, 171 (1969)].

\bibitem{Kupriyanov}
M. Yu. Kupriyanov, V. F. Lukichev, 
\textit{Influence of boundary transparency on the critical current of "dirty" SS'S
structures}, 
Zh. Eksp. Teor. Fiz. \textbf{94} (6), 139 (1988) [Sov. Phys. JETP \textbf{67}, 1163 (1988)].

\bibitem{Brouwer}
P. W. Brouwer and C. W. J. Beenakker, 
\textit{Anomalous Temperature Dependence of the Supercurrent Through a Chaotic Josephson Junction}, 
Chaos, Solitons and Fractals \textbf{8}, 1249 (1997).

\bibitem{Beenakker93}
C. W. J. Beenakker, 
\textit{Random-matrix theory of mesoscopic fluctuations in conductors and superconductors}, Phys. Rev. B \textbf{47}, 15763 (1993).

\bibitem{Barone}
A. Barone, G. Paterno, 
\textit{Physics and Applications of the Josephson Effect}, 
Wiley, New York, 1982.


\end{thebibliography}
\end{document}